\documentclass{jpp}
\usepackage[utf8]{inputenc}
\usepackage{graphicx}
\usepackage[T1]{fontenc}
\usepackage{amsmath}
\usepackage{hyperref}

\usepackage{tikz}
\usetikzlibrary{shapes.geometric}
\usetikzlibrary{decorations.pathreplacing,calligraphy}
\usepackage{amsfonts}
\usepackage[caption=false]{subfig}
\usepackage{pdfpages}
\usepackage{changepage}
\usepackage[normalem]{ulem}

\DeclareMathOperator{\sech}{sech}

\newcommand{\edit}[1]{{\color{black}#1}}
\newcommand{\tofix}[1]{{\color{black}#1}}

\shorttitle{QTTs for the Vlasov-Maxwell equations}
\shortauthor{E. Ye and N. Loureiro}

\title{Quantized tensor networks for solving the Vlasov-Maxwell equations}

\author{Erika Ye \aff{1}
  \corresp{\email{erikaye@lbl.gov}}
  \and Nuno F. Loureiro \aff{2}
  }

\affiliation{
\aff{1} Applied Mathematics and Computational Research Division, Lawrence Berkeley National Laboratory, Berkeley, CA 94720, USA
\aff{2} Plasma Science and Fusion Center, Massachusetts Institute of Technology, Cambridge, MA 02139, USA
}

\begin{document}

\maketitle

\begin{abstract}
The Vlasov-Maxwell equations provide an \textit{ab-initio} description of collisionless plasmas, but solving them is often impractical because of the wide range of spatial and temporal scales that must be resolved and the high dimensionality of the problem. In this work, we present a quantum-inspired semi-implicit Vlasov-Maxwell solver that utilizes the quantized tensor network (QTN) framework. With this QTN solver, the cost of grid-based numerical simulation of size $N$ is reduced from $\mathcal{O}(N)$ to $\mathcal{O}(\text{poly}(D))$, where $D$ is the ``rank'' or ``bond dimension'' of the QTN and is typically set to be much smaller than $N$. We find that for the five-dimensional test problems considered here, a modest $D=64$ appears to be sufficient for capturing the expected physics despite the simulations using a total of $N=2^{36}$ grid points, \edit{which would require $D=2^{18}$ for full-rank calculations}. 
Additionally, we observe that a QTN time evolution scheme based on the Dirac-Frenkel variational principle allows one to use \tofix{somewhat} larger time steps than prescribed by the Courant-Friedrichs-Lewy (CFL) constraint.
As such, this work demonstrates that the QTN format is a promising means of approximately solving the Vlasov-Maxwell equations with significantly reduced cost.

\end{abstract}

\section{Introduction}

\edit{The term ``quantum-inspired algorithms''} loosely describes a set of classical algorithms that utilize some concepts from their quantum counterparts. They often obtain comparable speed-ups so long as some error with respect to the true solution can be tolerated and the original problem of interest has some (potentially unknown) underlying structure that can be capitalized upon. 
% ``Quantum-inspired" algorithms loosely describe a set of classical algorithms that utilize some concepts from their quantum counterparts, often obtaining comparable speed-ups at the cost of accuracy. 
% Often, at the cost of accuracy, the quantum-inspired algorithms exhibit comparable speed-up compared to the original classical algorithm. % In the context of solving partial differential equations, quantum-inspired algorithms utilize a low-rank approximation of the quantum representation of the classical dataset. 
% In the context of solving partial differential equations (PDEs), quantum-inspired algorithms utilize a quantum-like representation of the classical dataset. Using tensor networks, a computational framework in which one can efficiently represent and manipulate low-rank approximations of the original quantum state. They have also been used with much success in quantum chemistry and condensed matter physics\cite{}. 
In the context of solving partial differential equations (PDEs), quantum-inspired algorithms utilize a quantum-like representation of the classical dataset. One can then use classical methods for quantum simulation to approximately solve the PDE. One natural choice is the tensor network framework, a popular tool in the quantum-chemistry and condensed-matter communities \citep{Schollwock_MPS, Vidal2003efficient, White1993density, Chan2002highly, White2005density} 
since it may allow for efficient representation and manipulation of the quantum-like dataset, so long as the dataset is low rank. 
% However, as presented thus far, there is no strong justification that the dataset would be low-rank.
% The justification so far is largely intuitive: the data set is expected to be low-rank when interactions between the quantum particles are local.  Using a similar argument, one might expect tensor networks to efficiently represent classical data that exhibits only local interactions between different length scales. It hypothesized that even turbulent dynamics exhibits scale locality. 

% A low-rank approximation to the quantum-like dataset can be efficiently represented and manipulated using the tensor network  Tensor networks are a classical computational framework that allows one to efficiently represent and manipulate low-rank approximations of the quantum-like data set. A low-rank approximation of the representation can be efficiently expressed and manipulated using the tensor network framework 
%
% One justification is largely intuitive: tensor networks have been widely used in quantum chemistry and condensed matter, utilizing the argument that quantum systems with only local interactions can be well-approximated by low-rank tensor networks. Using a similar argument, one might expect tensor networks to efficiently represent classical data that exhibits only local interactions between different length scales. It hypothesized that even turbulent dynamics exhibits scale locality. 

This concept was first introduced several years ago, and is referred to as \textit{quantics} or \textit{quantized} tensor trains (QTTs) because the data is artificially folded into a high-dimensional tensor representing many small ``quants'' of data \citep{Khoromskij2011quantics, Dolgov2012fast, Oseledets2009approximationLogparams, Oseledets2010approximationQTT}. (Tensor trains are tensor networks with a 1-D geometry.)
It was also proven that several one-dimensional functions (exponential, trigonometric, polynomial) and common operators (e.g., differential operators) on a uniform grid can be efficiently represented in the QTT format \citep{Khoromskij2011quantics, kazeev_low-rank_2012, Kazeev2013lowrank}. It follows that one can often obtain a loose bound on the QTT rank for other smooth functions by expanding them in terms of exponential or polynomial basis sets \citep{Dolgov2012fast}, though tighter bounds can be obtained \citep{Lindsey2023multiscale}.
% It then follows that the QTT rank for smooth functions can be bounded by the number of Fourier coefficients \cite{Dolgov2012fast}. 
However, non-smooth functions can have efficient representations too; the Heaviside step function can be represented with just rank two, triangle functions with rank three \citep{Gourianov2021}. 
The underlying intuition is that functions with some degree of scale locality, meaning that features on long length scales are relatively independent of behavior at small length scales, can be represented as a tensor network with low rank. Since it is believed that turbulent systems also exhibit some degree of scale locality, tensor networks may even be able to efficiently capture turbulent dynamics \citep{gourianov2022exploiting}.  
% Additionally, it is hypothesized that turbulent systems exhibit scale locality, . This locality suggests that the PDE solution may have some internal structure that the tensor network representation can take advantage of.

Several recent works solve the Navier-Stokes equations using the quantized tensor train (QTT) format \citep{Gourianov2021, kiffner_tensor_2023, kornev_numerical_2023, peddinti_complete_2023} with reasonable success. 
In this work, we aim to solve the Vlasov equation, a (6+1)-dimensional PDE that provides an \textit{ab initio} description of collisionless plasma dynamics in the presence of self-consistent electromagnetic fields computed via coupling to Maxwell's equations. Being able to perform these calculations would greatly advance our understanding of many astrophysical phenomena and fusion energy systems. 
However, while the coupled Vlasov-Maxwell system is the gold standard for plasma simulation, \edit{its use is limited} due to prohibitively high computational costs. %---both grid-based and particle-based conventional solvers can only tackle a limited set of problems.
% While one could use a Lagrangian or Monte-Carlo solvers (particle-in-cell) can be and currently are used, they face challenges in managing statistical noise and can only cover a limited range of physical scales.  

% \textcolor{blue}
{There is a class of low-rank grid-based solvers that are closely related to the proposed QTN solver, except that only low-rank approximations with respect to physical dimensions (i.e., the spatial and velocity coordinates) are considered~\citep{Kormann2015, Einkemmer2018, Rahn2022parallel6D, einkemmer_accelerating_2023}. These studies show that\edit{,} often, only modest rank is required to perform common test problems with sufficient accuracy, leading to significant reduction in computational cost. While the mechanics of these solvers are similar to what will be discussed here, they are fundamentally different in that they are not quantized and low-rank approximations within each dimension cannot be made.
} 
% Note that while the mechanics of these non-quantized solvers are similar to what will be discussed here, their investigations are fundamentally different in that they do not quantize the system and thus cannot make low-rank approximations within each dimension.

% While the coupled Vlasov-Maxwell system is the gold standard for plasma simulation, there are a limited but growing number of Eulerian \citep{Valentini2007hybrid, Cerri2017kinetic, Juno2018discontinuous, Hakim2020alias, Hakim2020conservative}, spectral \citep{manzini2016legendre, delzanno2015multi}, or semi-Lagrangian kinetic \citep{Cheng1976integration, sonnendrucker1999semilagrangian, crouseilles2010conservative, liu2010conservative, liu2021conservativeFEM,  vonAlfthan2014vlasiator, Einkemmer2020semilagrangian, Kormann2019massively} solvers --- its high dimensionality generally means a prohibitive amount of resources are required. (Lagrangian or Monte-Carlo solvers (particle-in-cell) are more common ~\citep{Dawson1983particle, Fonseca2002osiris, Franci2018camelia, Franci2018solar, Qin2015canonical, Xiao2021expliciit, Xiao2021symplectic, Sun2015explicit}, but have difficulties managing statistical noise.) There are a few low-rank solvers, which show that one only requires modest rank to perform common test problems with sufficient accuracy \citep{Kormann2015, Einkemmer2018, Rahn2022parallel6D, einkemmer_accelerating_2023}. Note that while algorithmically similar, these works are fundamentally different from the QTN methods investigated here in that they do not make low-rank approximations within each dimension.

Prior work on solving the Vlasov equation with QTTs assumed electrostatic fields, and only considered calculations with one axis in real space and one axis in velocity space (1D1V) \citep{Ye2022}. Here, we consider higher\edit{-}dimensional plasmas in 2D3V (though the algorithm \edit{can readily be extended to} 3D3V calculations) in the presence of electromagnetic fields. % investigating the properties of an alternative quantized tensor network (QTN) geometry beyond the 1-D QTT ansatz, as well as QTN-based time evolution schemes that allow one to take larger time steps than can be used in traditional finite-difference methods. 
We first start with a brief overview of the Vlasov equation and QTTs, and then introduce a comb-like quantized tree tensor network (QTC) geometry that we expect to be a more efficient representation for high-dimensional data than the standard tensor train geometry. In the following section, we detail our semi-implicit finite-difference Vlasov solver in the QTN format. Most notably, we discuss how one can modify the problem such that positivity of the distribution function is preserved despite the low-rank approximation, and introduce a QTN time-evolution scheme that allows one to take larger time steps than allowed by traditional grid-based methods. Finally, we close with numerical simulations of the Orszag-Tang vortex and the GEM magnetic reconnection problem in 2D3V, and a discussion on our observations. % Additional details regarding tensor networks and the Vlasov-Maxwell solver, as well as additional results are provided in the Supplemental Information (SI).

\section{Quantized Tensor Networks}
\edit{Quantized tensor networks are a multi-scale ansatz built from a set of interconnected high-dimensional matrices typically referred to as tensors \citep{Oseledets2009approximationLogparams, Khoromskij2011quantics, dolgov_tensor_2014}.
For more background information on QTNs, we refer readers to the previous references as well as \citet{Ripoll2021quantuminspired} and \citet{Ye2022}.

Suppose a 1-\edit{D} function ${h}(x)$ is discretized into $N=d^L$ grid points. One can \textit{quantize} the representation, by which one defines artificial dimensions to describe the function. This effectively appears as reshaping the vector into an $L$-dimensional tensor with size $d$ along each dimension, $h(x) \cong \textsf{h}(i_1, i_2, \hdots, i_L)$,
where the indices $i_1, i_2, ... i_L \in \{ 0, \hdots, d-1\}$ encode the grid points in $x$. For example, if $d=2$, one possible encoding is a binary mapping, such that the $i^\text{th}$ grid point is indexed by its binary string $(i_1, i_2, ..., i_L)_2$. In this case, the indices sequentially correspond to finer grid resolutions.
Another possible encoding is a signed register, in which one takes the two's complement of the physical indices conditioned on the first index, such that $i \cong (i_1, i_1 + i_2, ..., i_1 + i_L)_2$. This mirror-like mapping may be more efficient for symmetric functions, since the second half of the data is flipped with respect to the original \citep{Ripoll2021quantuminspired}. It is used later in Section 5 when representing Gaussian functions}. 

% With this mapping, each tensor sequentially correspond to finer grid resolutions. 

To obtain a \edit{QTT} representation, the $L$-dimensional tensor is decomposed into a 1-D chain of $L$ tensors: 
\begin{align}
    % \edit{h}(x_i) \cong 
    \textsf{h}(i_1, i_2, ..., i_L) = % \nonumber \\
    \sum_{\alpha_1=1}^{r_1} ... \sum_{\alpha_{L-1}=1}^{r_{L-1}} % \nonumber \\
    \textsf{M}^{(1)}_{\alpha_1} (i_1) \textsf{M}^{(2)}_{\alpha_1, \alpha_2}(i_2) \hdots \textsf{M}^{(L)}_{\alpha_{L-1}} (i_L),
    \label{eq:tt_ansatz}
\end{align}
where $\textsf{M}^{(p)}$ is the $p^\text{th}$ tensor in the chain with one ``physical'' bond ($i_p$) of size $d$ that is used to access the actual data, and two ``virtual'' bonds ($\alpha_{p-1}, \alpha_{p}$) of size $r_{p-1}$ and $r_{p}$ that connect the tensor to its two neighbors. \edit{Upon contraction of the tensor network, the sums in the above equation are performed and the virtual bonds no longer exist.} (The exceptions are $\textsf{M}^{(1)}$, which only has one virtual bond ($\alpha_1$) of size $r_1$, and $\textsf{M}^{(L)}$, which also only has one virtual bond ($\alpha_{L-1}$) of size $r_{L-1}$.)  
\edit{The simplest way to obtain the tensors in Eq.~\eqref{eq:tt_ansatz} is to perform a series of singular value decompositions \citep{Oseledets2010approximationQTT}, isolating each dimension $i_p$ one at a time. If no approximation is made and the vector is full-rank, then rank $r_p$ is $\text{min}(d^p, d^{L-p})$. The rank is largest at the middle of the tensor train, with $r_{L/2}=d^{L/2}=\sqrt{N}$. To obtain a good low-rank approximation, one would keep only the $D$ largest singular values at each decomposition step. This parameter is referred to as the bond dimension or rank of the tensor train.}
%
% Low-rank approximations in the QTT format limit the size of the virtual bonds to a maximum value of $D$, which is often referred to as the bond dimension or rank of the TT. 
(The terms bond dimension and rank will be used interchangeably.)
If the original dataset can be represented as a QTT of small bond dimension (ideally $D \sim \log(N)$) with only some small tolerable amount of error, the dataset is described as compressible and can be efficiently represented as a QTT. For concrete examples of common functions that can be efficiently represented in the QTT format, we refer the reader to~\citet[p. 56]{dolgov_tensor_2014}.  

% The indices $i_1, i_2, ... i_L$ encode the grid points in $x$. For example, if $d=2$, one possible encoding is a binary mapping, such that the $i^\text{th}$ grid point is indexed by its binary string $(i_1, i_2, ..., i_L)_2$. With this mapping, each tensor sequentially correspond to finer grid resolutions. 
% %
% Another possible encoding is a signed register, in which one takes the two's complement of the physical indices conditioned on the first index, such that $i \cong (i_1, i_1 + i_2, ..., i_1 + i_L)_2$. This mirror-like mapping may be more efficient for symmetric functions \citep{Ripoll2021quantuminspired}\edit{, and is used later in Section 5 on when representing Gaussian functions in velocity space}. 

% \textcolor{blue}{For the sake of the poor plasma physicists out there who would like to understand this, do you think it's possible to provide a worked out example (maybe as an appendix)? Also, sorry for being dumb, but why is this called a quantized tensor train? what's quantum about this?}

\edit{Linear operators acting on $h(x)$} can also be expressed in the QTT format, resulting in the QTT-Operator
\begin{align}
    \textsf{O}(x', x) & \cong \textsf{O}(o_1, o_2, ..., o_L; i_1, i_2, ..., i_L) \nonumber \\
    & = \sum_{\alpha_1=1}^{r_1} ... \sum_{\alpha_{L-1}=1}^{r_{L-1}} % \nonumber \\
    \textsf{M}^{(1)}_{\alpha_1} (o_1,i_1) \textsf{M}^{(2)}_{\alpha_1, \alpha_2}(o_2, i_2) \hdots % \nonumber \\ \hspace{3cm} 
    \textsf{M}^{(L)}_{\alpha_{L-1}} (o_L, i_L)
\end{align}
where each tensor $\textsf{M}^{(p)}$ now has two physical bonds ($o_p, i_p$) that are used to access the output $x'$ and input $x$ of the discretized operator $\textsf{O}$. % We label the bond dimension of QTT-Operators as D_W$. 
% The mapping used for the indices in QTT-states and QTT-operators must be consistent.

The QTT representation can be considered ``quantum-inspired'' since mapping classical data onto a quantum computer could be done in a similar fashion \citep{Ripoll2021quantuminspired, Lubasch2018multigrid, lubasch_variational_2020, jaksch2023variational}. In this case, each tensor would correspond to one qubit. \edit{The physical bond in the tensor represents the state of the qubit, while the virtual bonds capture the entanglement between the qubits}. \edit{The QTT-Operators are analogs to quantum gates or quantum circuits applied to the quantum computer.}

\begin{figure*}
    \centering
    \includegraphics[width=0.3\linewidth]{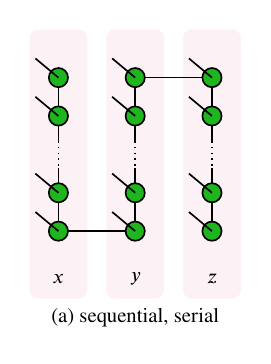}
    \includegraphics[width=0.32\linewidth]{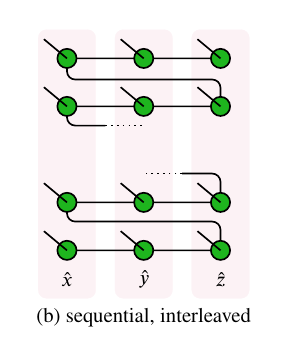}
    \includegraphics[width=0.3\linewidth]{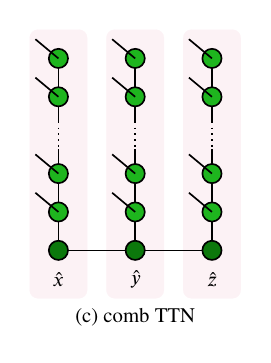}
    \caption{Tensor network diagrams for three-dimensional data. (a) Tensor train with tensors for different dimensions appended sequentially. (b) Tensor train with tensors for different dimensions combined in an interleaved fashion. (c) Comb-like tree tensor network with tensors for different dimensions on separate branches.}
    \label{fig:geometries}
\end{figure*}

\edit{More generally, for $K$-dimensional functions, each of the $K$ original dimensions is quantized. If the function is discretized using $d^L$ grid points along each of the original $K$ dimensions, then the quantization yields a $KL$-dimensional tensor of size $d$ along each new dimension. When decomposing this tensor into a QTN, several different ansatz\"{e} can be considered.}
% In the case of $K$-dimensional data sets with $K > 1$, one can consider several different QTN ansatz\"{e}.
% including those solving Navier-Stokes \citep{Gourianov2021, kiffner_tensor_2023, kornev_numerical_2023} or 1D1V Vlasov-Poission \cite{Ye2022}
% Recent works utilizing the QTN format thus far have only used the tensor train geometry. 
Even if one continues to use the tensor train geometry, as is done in most works, one can opt to append tensors corresponding to different problem dimensions sequentially (Fig.~\ref{fig:geometries}(a)) \citep{Ye2022, kiffner_tensor_2023, kornev_numerical_2023, peddinti_complete_2023} or in an interleaved fashion (Fig.~\ref{fig:geometries}(b)) \citep{Gourianov2021}. The caveat of sequential ordering is that interactions between different dimensions can be long-range. For example, any correlation between $x$ and $z$ are long-range and must be transmitted through all tensors corresponding to the $y$ grid points. With the interleaved geometry, in which tensors are ordered by grid scale, interactions between different dimensions are typically not as long-range. However, this ordering is less efficient for separable functions and operators. 
% , making it less ideal for PDEs like the Vlasov equation where most of the operators are separable or are sums of separable operators.

This work uses a comb-like tree tensor network (QTC) ansatz (also referred to as the QTT-Tucker format by \citet{Dolgov2013comb}). 
\edit{In this ansatz, the original $K$-dimensional function is first decomposed into a tensor train of $K$ tensors without quantization, so the physical bonds of each tensor correspond to the original problem dimensions, and their sizes are the number of grid points used along the corresponding dimensions. Then, the physical bond of each tensor is quantized, and each tensor is decomposed into a QTT.}
The resulting QTN is depicted in Fig.~\ref{fig:geometries}(c); the grid points for each dimension are represented by separate 1-D tensor trains (the branches) and each TT is connected to each other via another tensor train (the spine; darker green). Mathematically, this can be written like
\begin{align}
    \textsf{h}(x_{1,i}, & x_{2,i}, \dots, x_{K,i} ) \nonumber \\
    & \cong \textsf{h}(i^{(1)}_1, \dots, i^{(1)}_L, i^{(2)}_1, \dots, i^{(2)}_L, \dots, i^{(K)}_1, \dots, i^{(K)}_L) \nonumber \\ 
    % \hspace{0.5cm} 
    & =\sum_{\gamma_1=1}^{r_1} ... \sum_{\gamma_{K-1}=1}^{r_{K-1}} \textsf{B}^{(1)}_{\gamma_1} (i^{(1)}_1,...,i^{(1)}_L) \textsf{B}^{(2)}_{\gamma_1, \gamma_2}(i^{(2)}_1,..., i^{(2)}_L) % \hdots % \nonumber \\ & \hspace{3cm} 
    \hdots \textsf{B}^{(K)}_{\gamma_{L-1}} (i^{(K)}_1, ..., i^{(K)}) \, ,
\end{align}
where $\textsf{B}^{(k)}$ is a tensor train for the $k^\text{th}$ problem dimension
\begin{align}
    & \textsf{B}^{(k)}_{\gamma_k,\gamma_{k+1}}(i^{(k)}_1,...,i^{(k)}_L) = \nonumber \\
    & \hspace{0.5cm}
    \sum_{\beta_k} \tilde{\textsf{M}}^{(k)}_{\gamma_k, \gamma_{k+1},\beta_k}  % \nonumber \\ 
    \biggl( \sum_{\alpha_{(k,1)}}...\sum_{\alpha_{(k,L-1)}} 
    \textsf{M}^{(k,1)}_{\beta_{k},\alpha_{(k,1)}} (i^{(k)}_1) 
    % \nonumber \\ & \hspace{2cm}
    \textsf{M}^{(k,2)}_{\alpha_{(k,1)}, \alpha_{(k,2)}}(i^{(k)}_2) \hdots \textsf{M}^{(k,L)}_{\alpha_{(k,L-1)}} (i^{(k)}_L) \biggr) \, .
    \label{eq:branch}
\end{align}
To simplify notation, we assumed that all problems dimensions have the same number of grid points. However, this is not a necessary requirement.
\edit{In Eq.~\eqref{eq:branch}, tensor $\tilde{\textsf{M}}^{(k)}$ is the $k^\text{th}$ tensor in the spine of the comb.} Note that it does not have a physical index. In the low-rank approximation, the size of each dimension is capped at $S$. 
For simplicity, this work uses the same maximum bond dimension for all virtual bonds (i.e.,~$S=D$), though it may be advantageous to use different bond dimensions for the bonds in the branches ($\alpha_{(k,i)}$), along the spine ($\gamma_k$), and between the spine to the branches ($\beta_k$).

The motivation for this ansatz is to reduce the distance along the tensor network between different dimensions, while maintaining an efficient representation for product states. % (In other words, maintain an efficient representation of the differential operators).
Additionally, all algorithms for QTTs can be extended to the QTC geometry in a straightforward fashion, since the QTC geometry contains no loops and one can still define a center of orthogonality. The computational costs of relevant algorithms are summarized in Table~\ref{tab:costs} \edit{and described more in depth in the Supplemental Information (SI)}. The scaling of most algorithms are more expensive for the comb geometry because of the spine tensors, though this increased cost may be offset by increased representability, allowing one to use smaller bond dimension.

\begin{table*}
    \scriptsize
    \centering
    \begin{tabular}{l c c}
        \hline
         \textbf{Procedure} & \textbf{QTT} & \textbf{QTC-Spine}  \\
         \hline
         \textbf{Advection in $\textbf{x}$} & & \\
         % Naive & 
         %    $ \mathcal{O}(D^3 D_W^3 d^2) $ &
         %    $\mathcal{O}(D^3 D_W^3 d^2 + S^4 S_W^4 ) $ \\
         % Zip-up \cite{} & 
         %    $\mathcal{O}(D^3 D_W d^2) $ to $\mathcal{O}(D^3 D_W^3 d^2)$ & 
         %    $\mathcal{O}(D^3 D_w d^2 + S^4 S_W)$ to $\mathcal{O}(D^3 D_w^3 d^2 + S^4 S_W^4) $ \\
         \begin{tabular}{l}
             Matrix-vector multiplication 
             \\ with density matrix scheme % \cite{t}
         \end{tabular} & 
         \begin{tabular}{c}
                $\mathcal{O}(D^3 D_W^2 d + D^2 D_W^3 d^2) $ \\ %[-0.1cm]
                + ($Dd \times Dd$) eigenvalue problem
        \end{tabular}
        &
        \begin{tabular}{c}
            $\mathcal{O}(S^4 S_W)$ to $\mathcal{O}(S^4 S_W^4) $
        \end{tabular}
        \\
        \hline
        \textbf{Advection in $\textbf{v}$} &  & \\
        \begin{tabular}{l}
             Time-dependent variational \\ principle % [-0.1cm]
             with RK4
             % TDVP with RK4
         \end{tabular}
            & 
            $\mathcal{O}(2 D_W D^3 d + D_W^2 D^2 d^2)$ 
            &
            $\mathcal{O}(S^2 S_{W}^3 + S^4 S^2_{W}) $
            % \begin{tabular}{c}
            %     \shortstack[l]{$\mathcal{O}(D_W^2 D^2 d^2 + D^4 D_W d^2 + $ \\ \hspace{0.5cm} $S^2 S_{W}^3 + S^4 S_{W}^2 + S^6 S_{W})$}
            % \end{tabular}
        \\ %[0.5cm]
        % \begin{tabular}{l}
        %      Time-dependent variational \\ principle %[-0.1cm]
        %      with exact \\ diagonalization
        %      % \vspace{0.1cm}
        %      % TDVP with ED
        % \end{tabular}
        %     & 
        %     \begin{tabular}{c}
        %         $\mathcal{O}(D_W^2 D^2 d^2 + D^4 D_W d^2)$ \\
        %         $ + (D^2 d \times D^2 d) $ eigenvalue problem
        %     \end{tabular}
        %     &
        %     \begin{tabular}{c}
        %         $\mathcal{O}(S^2 S_{W}^3 + S^4 S_{W}^2 + S^6 S_{W})$ \\
        %         $ + (S^3 \times S^3) $ eigenvalue problem
        %     \end{tabular}
        %     % \begin{tabular}{c}
        %     %      \shortstack[l]{$\mathcal{O}(D_W^2 D^2 d^2 + D^4 D_W d^2 + D^6 d^3$ + \\ \hspace{0.5cm} $S^2 S_{W}^3 + S^4 S_{W}^2 + S^6 S_{W} + S^6 )$}
        %     % \end{tabular}
        %     \\
        \hline
        \textbf{Propagation of EM fields} & & \\
        % Compute $A_\text{eff}$ & 
        %     $\mathcal{O}(D_W^2 D^2 d^2 + D^4 D_W d^2)$ & 
        %     $\mathcal{O}(D_W^2 D^2 d^2 + D^4 D_W d^2 + S^2 S_{W}^3 + S^4 S_{W}^2 + S^6 S_{W})$ \\
        \begin{tabular}{l}
             Density matrix renormalization \\ group %[-0.1cm]
             with conjugate gradient \\ descent
             % DMRG with gradient descent
         \end{tabular}
            & 
            $\mathcal{O}(2 D_W D^3 d + D_W^2 D^2 d^2)$ & 
            \shortstack{not considered \\ in this work}
            % $\mathcal{O}(2 D_W D^3 d + D_W^2 D^2 d^2 + S^2 S_{W}^3 + 2 S^4 S^2_{W} + S^4 S_{W}) $
        \\
        \hline
    \end{tabular}
    \caption{Theoretical costs of the primary steps in the Vlasov-Maxwell solver for the tensor train (QTT) geometry and the spine in the comb geometry (QTC-Spine).  In the QTT geometry, all vectors have bond dimension $D$ while operators have bond dimension $D_W$. In the spine of the comb tensor network, all bonds are of size $S$ and $S_W$ for the vectors and operators, respectively. %In the comb geometry, vectors have bond dimension $D$ along the branches and bond dimension $S$ along the spine, while operators have bond dimension $D_W$ along the branches and bond dimension $S_W$ along the spine.
    Note that in practice, not all virtual bonds are necessarily of the same size, and the specified bond dimensions and the listed costs serve as upper bounds. 
    }
    \label{tab:costs}
\end{table*}

\section{Problem Overview}
The collisionless Vlasov equation is given by
\begin{equation}
    \frac{\partial f_s}{\partial t} + \textbf{v}_s \cdot \nabla_\textbf{x} f_s + \frac{1}{m_s} \textbf{F}_s \cdot \nabla_\textbf{v} f_s = 0,
    \label{eq:vlasov}
\end{equation}
where $f_s$ is the \edit{time-dependent} distribution function for particle species $s$ with coordinates in real space ($\textbf{x}$) and velocity space ($\textbf{v}$), and $m_s$ is the particle mass. %The subscript $s$ denotes the particle species (e.g., ions or electrons) that is being described. 
$\textbf{F}$ denotes the force that the particles are subject to. In the case of plasma dynamics, the force of interest is the Lorentz force,
\begin{equation}
    \textbf{F}_s = q_s \left( \textbf{v}_s \times \textbf{B} + \textbf{E} \right),
\end{equation}
where $q_s$ is the charge of particle species $s$, $\textbf{B}$ is the magnetic field, and $\textbf{E}$ is the electric field. Both fields have coordinates in real space ($\textbf{x}$) and vary over time.

% In the electrostatic case, one can obtain the electric field by solving for Poisson's equation,
% \begin{equation}
%     -\nabla^2 \Phi = \rho, \quad \vec{E} = -\nabla \cdot \Phi
% \end{equation}
% where $\Phi$ is the electrostatic potential and $\rho = \sum_s q_s \int \, d\textbf{v} f_s(\textbf{x},\textbf{v})$ is the charge density.
%
For an electromagnetic system, one propagates the electric and magnetic fields forward in time via Maxwell's equations,
\begin{align}
 \frac{\partial \textbf{B}}{\partial t} &= -\nabla \times \textbf{E}, \\
 \frac{\partial \textbf{E}}{\partial t} &= c^2 \nabla \times \textbf{B} - \frac{1}{\varepsilon_0} \textbf{J},
\end{align}
where $c$ is the speed of light, $\varepsilon_0$ is the vacuum permittivity, and $\textbf{J} = \sum_s q_s \int \, d\textbf{v} \, \textbf{v} f_s(\textbf{x},\textbf{v}\edit{,t})$ is the current density. The current density is the source of the nonlinearity, since the state of the plasma affects the electromagnetic force that is exerted on itself.

Eqs.~(3.1-3.4) are the equations we wish to solve in our Vlasov-Maxwell solver.

\section{Vlasov-Maxwell Solver}

% \subsection{Algorithm Overview}
We implement a semi-implicit finite-difference solver of the Vlasov-Maxwell equations utilizing the quantized tensor network (QTN) format. The distribution functions for each species and each component of the electromagnetic fields are represented as QTNs with bond sizes of at most $D$.
% with maximum bond dimension $D$. % (The actual size of the virtual bonds is allowed to be less than $D$.)
% In the QTT algorithm, one uses low-rank approximations of $f_s$, $\textbf{E}$, and $\textbf{B}$ in the QTT format. 
Likewise, the operators acting on these fields (e.g., the derivatives) are also represented in the QTN format (see Section I of the SI and~\citet{Ye2022} for details).

The distribution functions are updated via explicit time integration of the Boltzmann equation, with second-order operating splitting between the $\textbf{x}$ and $\textbf{v}$ advection steps. For advection in $\textbf{x}$, a split-step semi-Lagrangian scheme is used; the operator is split further into 1-D advection steps using a second-order scheme \citep{Durran2010numerical}. 
For advection in $\textbf{v}$, a new time evolution scheme specific to QTNs and based on the time-dependent variational principle (TDVP) \citep{Lasser2022various} is used. 
% \tofix{Although unnecessary, these operations are also split into 1-D advection steps using a second-order scheme.}
The electromagnetic fields are updated using a Crank-Nicolson time integration scheme, which is solved using an iterative scheme called density matrix renormalization group (DMRG) \citep{Dolgov2014alternating}. 
% we use either MacCormack time integration again with operator splitting into 1-D advection steps, or we use TDVP (Section \ref{}) without operator splitting.  

% When operator splitting is used for both advection in $\textbf{x}$ and $\textbf{v}$, we can order the advection like vx, vy, vz, x, y, z, z, y, x, vz, vy, vx

% In the following sections, we provide detail further how the solver is implemented using the QTN format.

% \section{Quantized tensor train algorithms}

\subsection{Ensuring positivity}
One problem with solving the Vlasov equation in the QTN format is that taking the low-rank approximation of $f$ generally does not preserve important properties, such as its norm and positivity. The norm can be corrected by rescaling the distribution function. 
It should be possible to develop methods that conserve momentum and energy (as is done in the non-quantized low-rank community \citep{Einkemmer2021conserve, Coughlin2023robust}) though these techniques are not implemented in this work. 
% It should also be possible to develop methods that conserve momentum and energy, though these techniques are not implemented in this work.
%
However, it is difficult to ensure positivity once in the QTN format.

A simple fix is to define $g$ such that $f=g^* g$, and instead evolve $g$ forward in time. Plugging this into Eq.~\eqref{eq:vlasov}, we obtain
\begin{equation}
    \frac{\partial g_s}{\partial t} + \textbf{v}_s \cdot \nabla_\textbf{x} g_s + \frac{1}{m_s} \textbf{F}_s \cdot \nabla_\textbf{v} g_s = 0
    \label{eq:vlasov_sq}
\end{equation}
and an equivalent equation for $g^*$. In the following calculations, we are actually solving Eq.~\eqref{eq:vlasov_sq} in order to ensure positivity in our calculations.
%
% However, note that depending on the collision operator used, one might not be able to use this trick in the presence of collisions.
% Note that one might not be able to use this trick in the presence of collisions. However, if one is only interested in adding collisions for numerical purposes, one can artificially relax $g$ to the square root of a Maxwellian using a Leonard-Bernstein-like collision operator
% \begin{equation}
%     C[g] = \beta \nabla_\textbf{v} \left( \textbf{v} g + \frac{1}{2} v_\text{th}^2 \nabla_\textbf{v} g \right)
% \end{equation}
% where $\beta$ is a parameter that determines the strength of the collisions. 

\subsection{Semi-Lagrangian time evolution}

With operator splitting, higher-dimensional advection problems are simplified to a series of 1-D advection problems of the form
\[ \frac{\partial \psi}{\partial t} + u \frac{\partial \psi}{\partial x} = 0 \]
where $\psi$ is the field of interest and $u$ is the advection coefficient that is constant along the $x$ axis. However, $u$ will in general be variable in all other dimensions of the problem.
In the semi-Lagrangian (SL) time evolution scheme, one approximates the field as a set of parcels, computes their exact trajectories, and then uses the updated parcels to estimate $\psi$ at the next time step.

Consider a uniform grid, with $x_j = j\Delta x$, and let $\phi$ denote a numerical approximation to field $\psi$; here, $\phi_j=\psi(x_j)$. Since there are no sources, $\phi$ at the next time step is given by \[ \phi(x_j, t^{n+1}) - \phi(\tilde{x}_j^n, t^n) = 0\]
where $\tilde{x}_j^n = x_j - u\Delta t$ is the departure point of the trajectory starting at time $t^n$ and arriving at $x_j$ at the next time step $t^{n+1}$.
In this work, the value of $\phi(\tilde{x}^n_j)$ is obtained using a cubic interpolation. For $u > 0$, this is given by \citep{Durran2010numerical}
\begin{align}
    \phi(\tilde{x}^n_j, t^n) = %  \nonumber \\ 
    & % \quad
    - \frac{\alpha (1-\alpha^2)}{6} \phi^n_{j-p-2} 
    + \frac{\alpha (1 + \alpha) (2-\alpha)}{2} \phi^n_{j-p-1} 
    \nonumber \\
    & % \quad
    + \frac{(1-\alpha^2)(2-\alpha)}{2} \phi^n_{j-p} 
    - \frac{\alpha (1-\alpha)(2-\alpha)}{6} \phi^n_{j-p+1}
    \label{eq:semilagrangian}
\end{align}
where $p$ is the integer part of $u \Delta t / \Delta x$, $\alpha=\frac{x_{j-p} - \tilde{x}_j^n}{\Delta x}$, and superscript $n$ denotes the time step. Thus, the next time step is obtained from $\phi^{n+1} = A_{SL} \phi^n$, where the matrix $A_{SL}$ encodes the above equation.

Obtaining the QTN-Operator decomposition of $A_{SL}$ can be computationally expensive. At the moment, we do not have an analytical form for the QTN-Operator, even for constant $u$, and any expression one does obtain would be difficult to generalize to higher-dimensional problems with variable $u$. One can compute the QTN-Operator numerically by performing a series of SVD operations in an iterative fashion \citep{Oseledets2010cross}. However, the cost of this naive but reliable procedure scales like $\mathcal{O}(N^2)$, \edit{where $N$ is the total number of grid points for the dimensions in which $u$ is not constant and the dimension along which the derivative is taken.} Interpolation algorithms such as TT-Cross \citep{Oseledets2010cross} often perform well for low-rank tensors, with computational costs that scale polynomially with rank. However, the resulting QTN is expected to require moderately large ranks, especially for larger time steps \citep{Kormann2015}. % In general, one cannot use aggressive low-rank approximations of operators as otherwise they may generate unphysical features.

Fortunately, while semi-Lagrangian time evolution in the QTN format does not appear to be advantageous for arbitrary problems, it can be relatively efficient for performing advection of the Vlasov equation along the spatial dimensions. For example, in the case of advection along the $x$ axis, \[ \frac{\partial}{\partial t} g(\textbf{x}, \textbf{v}) + v_x \frac{\partial}{\partial x} g(\textbf{x},\textbf{v}) = 0, \]
the operator $\textsf{A}_{SL}$ is effectively 2-dimensional, as it can be written in the form $\textsf{A}_{SL} = \tilde{\textsf{A}}_{x,v_x} \otimes \mathbb{I}_{y,z,v_y,v_z}$ \edit{(the subscripts denote the problem dimensions on which the operator acts)}. 
It is observed that the rank of the QTN representation of $\tilde{\textsf{A}}_{x,v_x}$ depends on the largest magnitude of integer $p$; so long as moderate time step and grid resolution is used the bond dimension remains manageable.
% It is observed that the rank of the QTN representation of $\tilde{A}_{x,v_x}$ increases with time step, but remains constant with increased resolution. 
Furthermore, if the time step is held constant, the QTN decomposition only needs to be performed once.

Once in the QTT format, the matrix-vector multiplication is performed using the density-matrix method \citep{mcculloch_from_2007}. %, tn_web}. %, which allows one to simultaneously perform the multiplication and QTT compression. 
Details of this algorithm for the comb geometry are given in Section III of the SI, and the computational costs are listed in Table~\ref{tab:costs}. 
{%\color{blue} 
Since the bond dimension of the operator $\textsf{A}_{SL}$ is expected to be modest for all time, its contribution to the scaling can be neglected, yielding costs that scale like $\mathcal{O}(D^3)$ for the QTT geometry and $\mathcal{O}(D^4)$ for the QTC geometry. For the Orszag-Tang vortex calculation, the observed scaling is closer to $\mathcal{O}(D^{2.5})$ (see Section VI.A.4 of the SI).}

\subsection{Time-dependent variational principle}
% Time-dependent variational principle (TDVP), time-dependent density renormalization group (td-DRMG). Critically, we can defeat the Courant-Friedrich-Levy stability condition using these time-stepping schemes. (Show Whistler waves with increasing resolution and insensitivity to time step for time-dependent methods.)

% (summarized from \citep{Lasser2022various})
% This algorithm is analogous to the time-dependent variational principle (TDVP) algorithm used in quantum simulation. (Time-dependent density matrix renormalization group is a related algorithm.)

% Thus far, the QTN time integration algorithms, either presented here (semi-Lagrangian scheme) or in prior works (RK4, Maccormack), are simply classical time integrators translated into the QTT format. These time integration schemes 

Thus far, prior works with QTNs utilize classical explicit time integrators (such as fourth-order Runge-Kutta (RK4) and MacCormack). Similar to the SL scheme discussed above, most explicit schemes only require matrix-vector multiplications and additions, and thus can easily be translated into the QTN format. These time integration schemes are also global schemes, since they involve manipulating the entire QTN state at once. 

% Most classical time integrators, including the semi-Lagrangian scheme presented above as well as those used in prior works, such as RK4 or the MacCormack algorithm, can readily be implemented in the QTN framework since the algorithms only rely on matrix-vector multiplications and additions. These algorithms as global time evolution schemes, since they involve manipulating the entire QTN state at once. 

% Prior works perform time integration using traditional classical methods such as RK4 or the MacCormack algorithm. Performing these algorithms in the QTN framework is straightforward, as they simply involve translating a series of matrix-vector multiplications and additions into a series of QTT-operator and QTT-vector multiplications and additions. These time integration schemes are global schemes, since they involve manipulating the entire QTT state at once.

We can instead consider a local time evolution scheme, in which individual tensors in the QTN are propagated forward in time in a sequential fashion. In quantum simulation, one such algorithm is the time-dependent variational principle (TDVP) scheme \citep{Haegeman_TDVP}. % and time-dependent density matrix renormalization group (td-DMRG) \cite{Schollwock_TDDMRG, Ronca_TDDMRG}. In this work, we focus on TDVP, though the two algorithms are similar in spirit.
This scheme is also closely related to the dynamical low-rank algorithm \citep{khoromskij2012efficient, Einkemmer2018}, which has been used for solving PDEs in the non-quantized tensor train community. % format (only utilizing low-rank tensor decompositions between but not within each problem dimension). 
% The dynamical low-rank algorithm, a numerical method for solving PDEs in the not quantized tensor train format (only utilizing low-rank tensor decompositions between but not within each problem dimension), uses an analogous time evolution scheme \cite{khoromskij2012efficient, Einkemmer2018}. 
More recently, the same underlying principle has been used in a quantum algorithm for solving the Schrodinger-Poisson equation \citep{cappelli2023vlasov}.

There are very fundamental differences between global and local time evolution.
In global time evolution, the state is first propagated in time and then projected onto the manifold for QTT of rank $D$. In contrast, in local schemes, the dynamics themselves are approximated by projection onto that manifold. Because the state remains in the low-rank manifold and does not require compression of a high-rank object, local methods are cheaper than global methods.
% The projector can sometimes incorrectly confine one to a smaller subspace due to symmetry; we found it imperative to use a global time evolution scheme such as RK4 for the first time step.
% Unfortunately, this means that local methods like TDVP (both 1-site and 2-site) are prone to ``getting stuck," yielding incorrect dynamics \citep{yang2020time}. 

TDVP utilizes the Dirac-Frenkel variational principle, which states that the dynamics of a unitary system at any given point in time lies in the tangent space of the current state \citep{Frenkel1934wave, Lasser2022various}. By treating each tensor \edit{in the QTN} as a variational parameter, one can obtain the effective dynamics for each tensor (given by anti-Hermitian operator $\textsf{A}_\text{eff}^{(i)}$ for the $i^\text{th}$ tensor) and sequentially evolve them forward in time. 
More specifically, if the unitary dynamics acting on state $\psi$ is described by anti-Hermitian operator $\textsf{A}$, the equation of motion for the $i^\text{th}$ tensor \edit{in the QTN} $\textsf{M}_i$ is
\begin{align}
    \frac{\partial}{\partial t} \textsf{M}_i & = \textsf{A}_\text{eff}^{(i)} \textsf{M}_i\, ,
    \label{eq:tdvp_site}
\end{align}
with
\begin{align}
    \textsf{A}_\text{eff}^{(i)} = \partial_{\textsf{M}_i} \, \partial_{\textsf{M}_i^*} \langle \psi(\textsf{M}_1, ..., \textsf{M}_L) \, | \, \textsf{A} \, | \, \psi(\textsf{M}_1, ... \textsf{M}_L) \rangle \, ,
    \label{eq:tdvp_Aeff}
\end{align}
assuming that the QTN is canonicalized such that the orthogonality center is located at the $i^\text{th}$ tensor.
As a result, the original dynamics is decomposed into roughly $2KL$ equations of motion of size $\mathcal{O}(D^2)$. (The factor of two arises from the algorithm when sweeping across all tensors.) 
\tofix{For a second-order scheme, one updates all the tensors sequentially in a sweeping fashion with increasing $i$ and then decreasing $i$, using half of the original time step at each update.}
We refer the reader to the Section IV of the SI and 
\citet{Paeckal_2019_time} for a more in-depth explanation and the algorithmic details. 

In this work, Eq.~\eqref{eq:tdvp_site} is solved using RK4, so the computational cost scales like $\mathcal{O}(D^3 D_W d + D^2 D_W^2 d^2)$, where $D_W$ is the bond dimension of the QTT-O representation for time evolution operator $A$. 
{%\color{blue}
For the QTC geometry, treatment of the tensors along the spine formally scales like $\mathcal{O}(S^2 S_W^3 + S^4 S_W^2)$, where $S$ and $S_W$ are the bond dimensions of the spine tensors in the QTC representations of state $\psi$ and operator $A$.}
% {\color{blue} For the QTC geometry, the bulk of the cost lies in the treatment of the tensors along the spine, formally scaling like $\mathcal{O}(S^2 S_W^3 + S^4 S_W^2)$, where $S$ and $S_W$ are the bond dimensions of the spine tensors in the QTT representations of $x$ and $A$. In the case of 2D3V calculations of the Vlasov equation, the $S_W$ is small for most bonds, so the scaling is expected to be closer to $\mathcal{O}(S^4)$}
%
% \textcolor{blue}
{In the case of the Vlasov equation with two spatial dimensions, $S_W$ can be neglected, but $D_W \sim D$ due to the nonlinear term. Setting $S=D$, the overall expected scaling is $\mathcal{O}(D^4)$,
which is what is observed for the Orszag-Tang vortex calculation (see Section VI.A.4 of the SI). 
% $\mathcal{O}(D^{2.78})$ scaling is observed (see Section V.A.3 of the SI).
% The wall-time of this step for one of the test problems is plotted with respect to $D$ in Section V.A.3 of the SI, and a $\mathcal{O}(D^{2.78})$ scaling is observed.
% The observed scaling with respect to $D$ are reported in Section V.A.3 of the SI.
}

The time evolution scheme presented above is the single-site variant of TDVP. By construction, the time-evolved state will always have the same bond dimension as the initial state. As a result, the algorithm is cheaper than global methods since it avoids needing to compress the state from an enlarged bond dimension. However, this feature can be problematic when the time evolved state requires larger bond dimension, particularly when the initial bond dimension is very small. To counter this, one would typically use the slightly more expensive two-site variant in which two neighboring tensors are updated simultaneously so the QTN bond dimension can be adjusted as needed. 
However, in our solver, single-site TDVP appears to be sufficient because it is used in conjunction with global time evolution schemes that already increase QTN rank. For both variants, we found it imperative to use a global time evolution scheme such as RK4 for the first time step. 
% We also note that there are single-site and two-site variants for tangent-space methods. The primary difference is that with the single-site variant, the time-evolved state will always have the same bond dimension as the initial state. This often can be an issue when the initial bond dimension is small. In contrast, the two-site variant allows one to alter the rank of the QTN. In this work, because TDVP is used in conjunction with global time evolution schemes that do allow for increases in QTN rank, the slightly cheaper single-site variant is used. We also found it imperative to use a global time evolution scheme such as RK4 for the first time step. % In the naive 1-site variant used here,  % However, when used in conjunction with the SL advection time integrator, which does allow for the bond dimension to increase at each time step, this is not a serious issue.

%%%
Particularly interesting is that even when Eq.~\eqref{eq:tdvp_site} is solved approximately using RK4, we observe that the local methods allow one to use larger time steps than what global methods are limited to (e.g., by the Courant-Friedrich-Levy (CFL) constraint \citep{cfl1967partial}). In the following example, it appears that the time steps of the local algorithm can be larger by about a factor of two. % \tofix{\sout{at least a factor of four} about a factor of two}.
This is perhaps possible because global information is incorporated into the effective operator for each individual tensor. The ability to go beyond CFL is crucial for QTN finite difference solvers, as it allows one to use higher spatial resolution (for improved accuracy) without requiring one to significantly increase temporal resolution.

% However, we observe that local methods allow one to use larger time steps than what global methods are limited to (e.g., from the Courant-Friedrich-Levy (CFL) constraint \cite{cfl1967partial}). We believe this is possible because global information is incorporated when determining the effective dynamics for each individual tensor. The ability to overcome CFL is crucial for QTT finite difference solvers, as it allows one to use higher spatial resolution for a given temporal resolution.

\subsubsection{Advection with time varying electric field}

We demonstrate TDVP time evolution with the advection of a 0D2V uniform electron Maxwellian distribution in the presence of constant background magnetic field $B_{0,z} \hat{z}$, and uniform but time-varying electric field $E_x(t) = E_0 \cos(\omega t) \hat{x}$. The resulting dynamics is the drifting of the Maxwellian with velocities
% \begin{align}
%     v_x(t) = -\omega_x(t) - v_{x,0} \cos(t) - v_{y,0} \sin(t) \\
%     v_y(t) = -\omega_y(t) + v_{x,0} \sin(t) - v_{y,0} \cos(t)
% \end{align}
\edit{
\begin{align}
    v_x(t) = u_x(t) + v_{x,0} \cos(\omega_c t) + v_{y,0} \sin(\omega_c t), \\
    v_y(t) = u_y(t) - v_{x,0} \sin(\omega_c t) + v_{y,0} \cos(\omega_c t),
\end{align}
where $\omega_c = q B_0 /m$ and $B_0>0$, and
\begin{align}
    u_x(t) &= \frac{E_0}{B_0} \frac{\omega_c^2}{\omega_c^2 -\omega^2} \left(\sin(\omega_c t) - \frac{\omega}{\omega_c} \sin(\omega t) \right), \\
    u_y(t) &= \frac{E_0}{B_0} \frac{\omega_c^2}{\omega_c^2-\omega^2} \left(\cos(\omega_c t) - \cos(\omega t) \right),
\end{align}
for $\omega \neq 1$.  The drift velocity in the $z$ direction remains constant and therefore is not considered. In the following calculations, we use units such that $B_0=1$ and $q/m=-1$.
}
% and 
% \begin{align}
%     \omega_x(t) &= \frac{E_0}{2} \left( t \cos(t) + \sin(t) \right) \\
%     \omega_y(t) &= - \frac{E_0}{2} t \sin(t)
% \end{align}
% for $\omega = 1$.

We perform time evolution using a second-order semi-Lagrangian \edit{(SL)} scheme, fourth-order Runge-Kutta (RK4), and TDVP. The calculations use $2^L$ grid points along the $v_x$ and $v_y$ axes (with bounds $[-12 v_{th}, 12 v_{th})$). The test electron distribution is initialized with drift velocities $v_{x,0}=v_{y,0}=0$ and thermal velocity of 1, and is evolved until a time of 100 using time step $\Delta t$. 
\tofix{Because split-step schemes are used, the results are shown with respect to the smallest time-step used in the time-evolution scheme ($\Delta t_\text{step}$) for a more fair comparison.}
The calculations were performed using a sequential layout with data along both axes encoded via a binary mapping. Both second-order and fourth-order centered finite difference stencils are considered.

\begin{figure}
    \centering
    \includegraphics[width=0.53\linewidth]{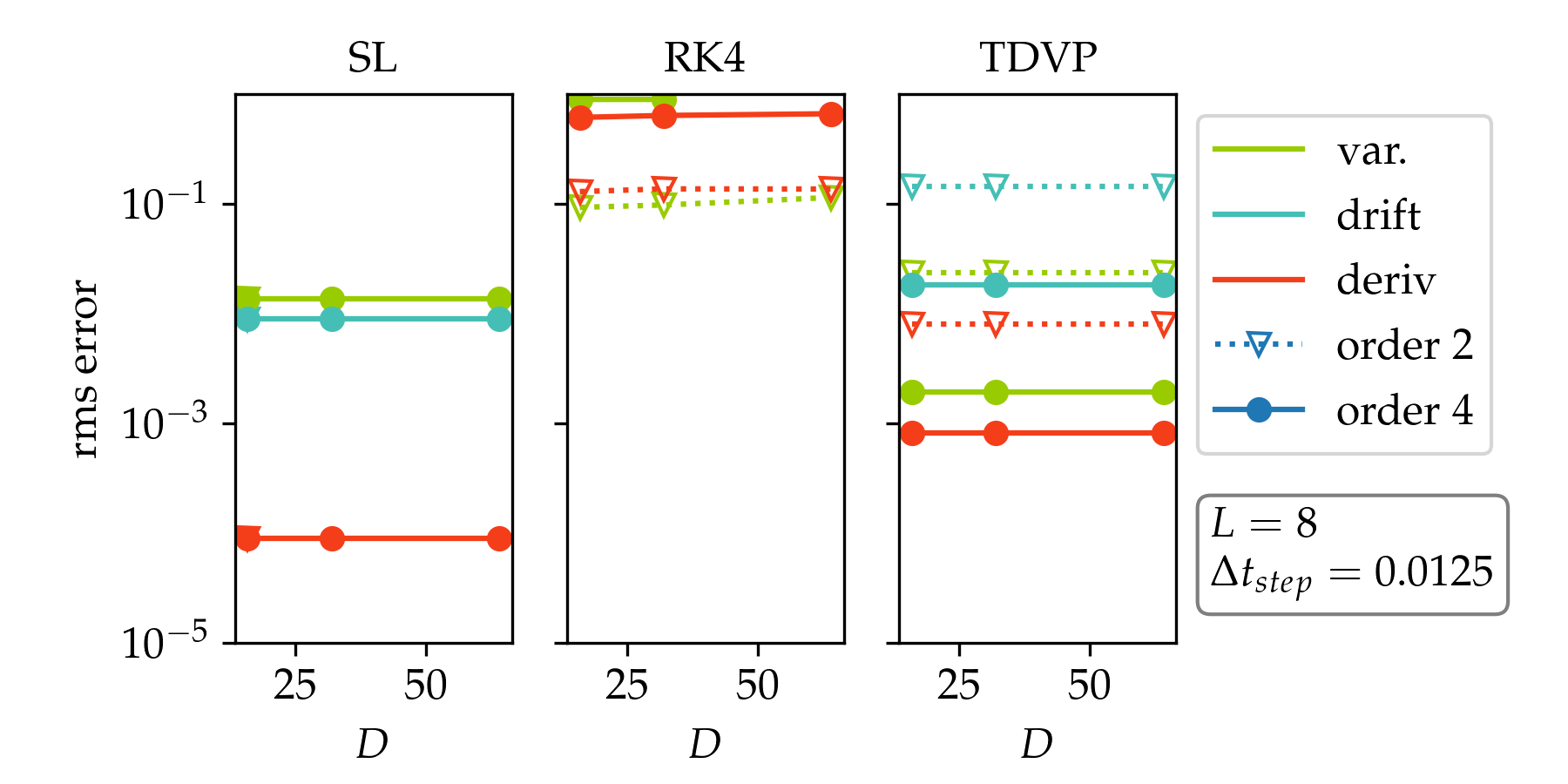}
    \includegraphics[width=0.53\linewidth]{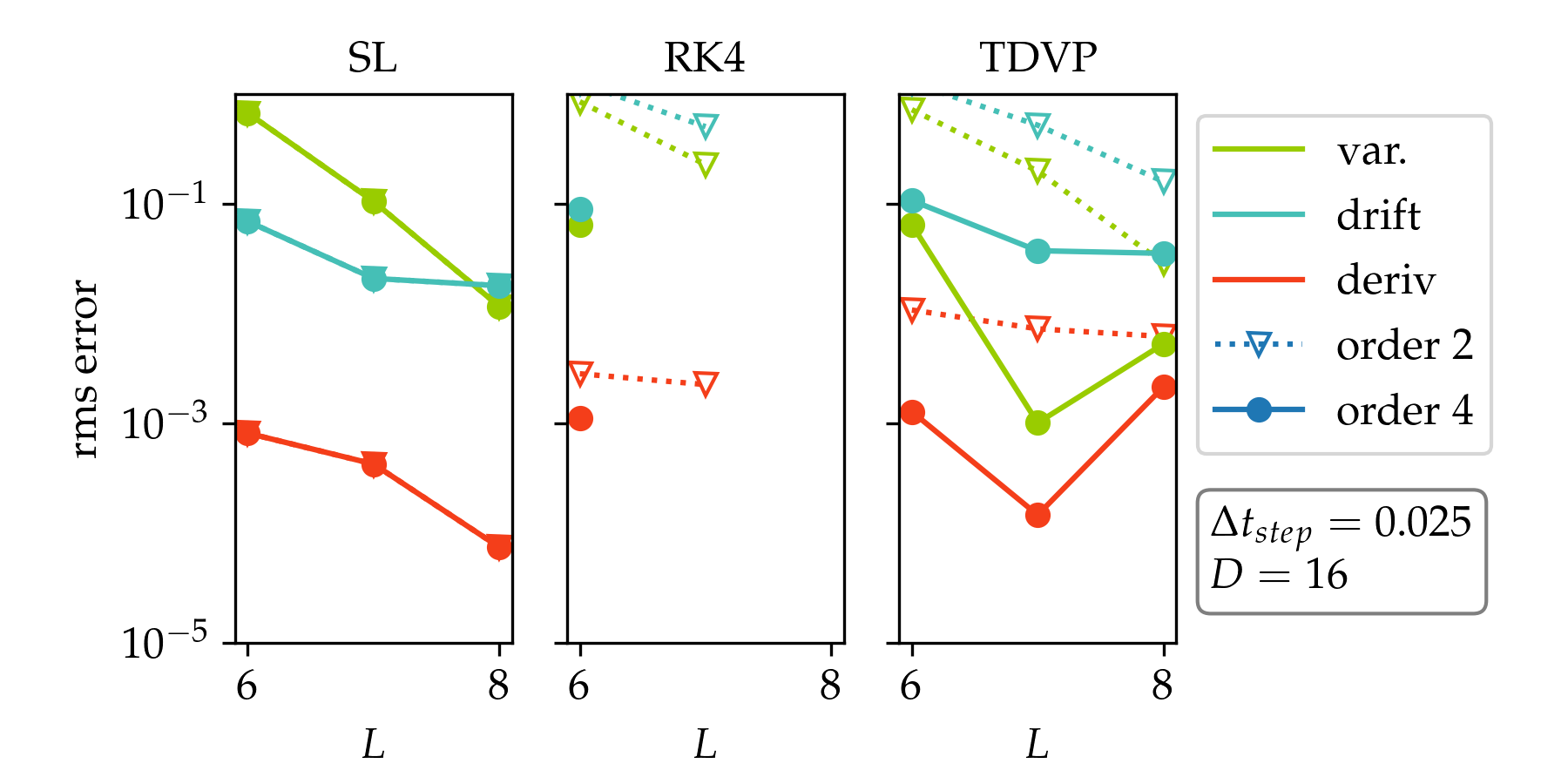}
    \includegraphics[width=0.53\linewidth]{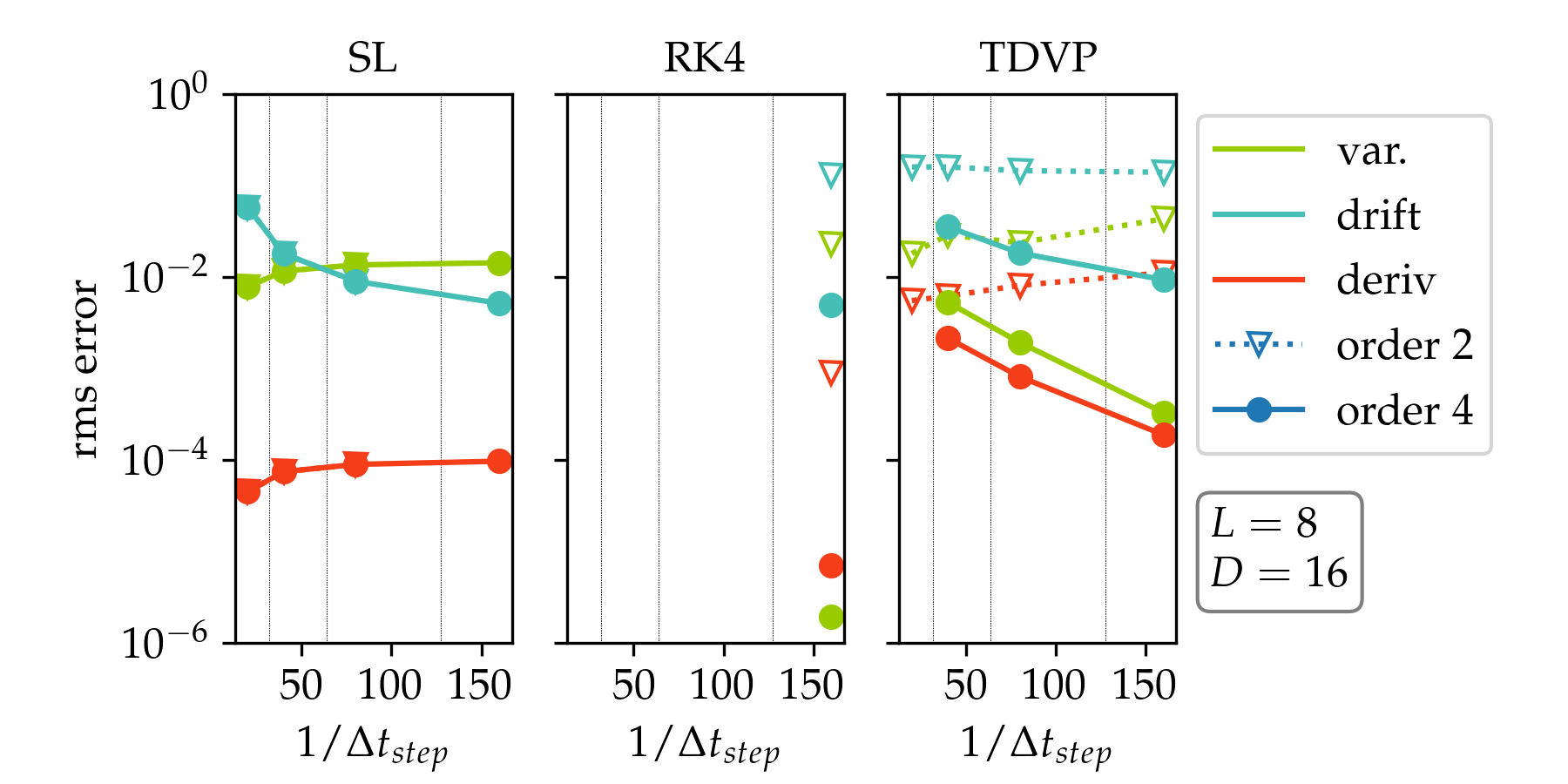}
    \caption{Rms errors in the dynamics for advection with time-varying electric field. Test parameters are set to be $\omega=0.4567$, $E_0=0.9$, and $B_{0,z}=1$. Calculations are performed with bond dimension $D$, grid resolution $2^L$ along each dimension, and time step $\Delta t$. \tofix{The finest interval of each time step is $\Delta t_\text{step}$}. (top) Results for varied $D$ and fixed $L=8$ and \tofix{$\Delta t_\text{step} = 0.0125$}. (middle) Results for varied $L$ and fixed $D=16$ and \tofix{$\Delta t_\text{step}=0.05$}. (bottom) Results for varied \tofix{$\Delta t_\text{step}$} and fixed $L=8$ and $D=16$. \tofix{The dashed black lines depict multiples of the CFL limit.} The different plots along each row are labeled with the time-evolution scheme used: semi-Lagrangian (SL), fourth-order Runge-Kutta (RK4), and single-site TDVP. Errors with respect to {analytical results for} the variance (i.e.,~thermal velocity squared, green), drift velocity (blue), and smoothness of the Maxwellian distribution (Eq.~\eqref{eq:deriv_err}; red) are plotted in different colors. The two symbols denote calculations performed with a second-order or fourth-order centered finite-difference stencil. By construction, the stencil plays no role in SL time evolution. Missing data is a result of numerical instabilities.}
    \label{fig:tdvp}
\end{figure}

At each time step, the distribution function $|g(v_x,v_y)|^2$ obtained after time evolution is fit to a Maxwellian $\mathcal{M}$ and the errors in the variance ($\epsilon_{\sigma}$) and drift velocity ($\epsilon_{v}$) are computed as the norm of the differences of the fit with respect to the theoretical values along the two dimensions. The ``error in the smoothness'' is quantified as
\begin{equation}
    \epsilon_\nabla = \frac{1}{2^L} \left \lVert \left( \frac{d}{dv_x} + \frac{d}{dv_y} \right) \left( |g(v_x,v_y; t)|^2 - \mathcal{M}_\text{fit} \right) \right \rVert,
    \label{eq:deriv_err}
\end{equation}
where the derivatives are computed with a first-order finite difference stencil. The total error of the simulation is the root-mean-square error, 
$\lVert \epsilon(t) \rVert / \sqrt{N_t}$, where $N_t$ is the total number of time steps.
% \begin{equation}
%     \epsilon_\nabla = \lVert \nabla \left( |g(v_x,v_y; t)|^2 - \mathcal{M}_\text{fit} \right) \rVert
% \end{equation}

The results are shown in Fig.~\ref{fig:tdvp}. Plots in the top row show that a bond dimension of $D=16$ is \edit{sufficiently large} for these calculations, since the error is not reduced with increased bond dimension. In the single-site TDVP calculations, the bond dimension throughout the simulation is fixed at the bond dimension of the initial state, which is about $D=10$. Plots in the middle row show how error scales with increasing grid resolution, for fixed bond dimension $D=16$ and time step $\Delta t_\text{step}=0.05$. 
This time step is too large for RK4 with resolutions greater than \tofix{$L=7$}, leading to numerical instabilities.
In the SL case, increasing resolution most notably improves the error in the variance (temperature) of the Maxwellian, as well of the smoothness of the resulting distribution function. For TDVP, increasing resolution also reduces errors of the drift velocity, but the distribution function exhibits more numerical noise (larger $\epsilon_\nabla$) at higher grid resolutions. The bottom row compares errors for calculations with $D=16$, $L=8$ for different time steps. With SL time evolution, error in the drift velocity decreases with reduced time step, but the error in variance and smoothness increases. This suggests that the error arises from the projection of the Lagrangian dynamics onto the grid at each time step. With TDVP, for the calculation with the fourth-order finite-difference stencil, performance improves with reduced time step, as is expected since the sequential iterative updates have an associated commutation error. However, with the second-order stencil, % the error appears to be dominated by lack of grid resolution since 
reducing time step does not reduce error.

In summary, as a local time evolution scheme, in practice, TDVP is computationally cheaper than both SL and RK4 time evolution. The above results show that both SL time evolution and TDVP allow one to take larger time steps than with \tofix{explicit} grid-based methods like RK4. However, note that when TDVP is performed with RK4 as the local time integrator (as is done here), there still exists a maximum time-step above which time evolution becomes unstable.
%The 1-site variant appears to allow a larger time-step than the 2-site variant. 
In general, the error in the thermal velocity is smaller with TDVP than SL time evolution (with which the Maxwellian tends to broaden). However, the distribution function from SL tends to be smoother than the results from TDVP. This may become problematic for long-time time integration using TDVP.

\subsection{Implicit solver for Maxwell's equations}

The electric and magnetic fields are propagated forward in time implicitly using Crank-Nicolson. Given the fields at time step $n$, the updated fields at the next time step are obtained by solving 
% To update Maxwell's equations, we use the Crank-Nicolson time evolution integrator, in which the fields at the $(n+1)^\text{th}$ time step is obtained by solving
\begin{align}
    &
    \begin{bmatrix}
        \frac{1}{\Delta t} \mathbb{I} & -\frac{c^2}{2} \nabla \times \\[0.2cm]
        \nabla \times & \frac{1}{\Delta t} \mathbb{I}
    \end{bmatrix}
    \begin{bmatrix}
        \textbf{E}^{n+1} \\[0.2cm] \textbf{B}^{n+1}
    \end{bmatrix}
    = %\nonumber \\
    %& \hspace{0.5cm}
    \begin{bmatrix}
        \frac{1}{\Delta t} \mathbb{I} & \frac{c^2}{2} \nabla \times \\[0.2cm]
        -\nabla \times & \frac{1}{\Delta t} \mathbb{I}
    \end{bmatrix}
    \begin{bmatrix}
        \textbf{E}^{n} \\[0.2cm] \textbf{B}^{n}
    \end{bmatrix}
    + 
    \begin{bmatrix}
        c^2 \nabla \times \textbf{B}_0 - \frac{1}{\varepsilon_0} \textbf{J}^{n} \\[0.2cm] 
        -\nabla \times \textbf{E}_0
    \end{bmatrix}
    % +
    % \begin{bmatrix}
    %     \frac{1}{\varepsilon_0} \vec{J}^{n} \\ 0
    % \end{bmatrix}
    \label{eq:em_update}
\end{align}
where $\textbf{E}_0$ and $\textbf{B}_0$ are time-invariant background contributions to the electric and magnetic fields, and $\textbf{J}^n$ is the current density measured from the distribution functions at the $n^\text{th}$ time step.

Linear equations of the form $Ax=b$ can be solved approximately in the QTT format using the density matrix renormalization group (DMRG) algorithm \citep{Oseledets2012solution, Dolgov2014alternating, Schollwock_MPS, Gourianov2021}, in which each tensor (or pairs of neighboring tensors) are optimized sequentially to minimize the cost function $ x^\dagger A x - 2 x^\dagger b$. (Details are provided in the Section V of the SI.) If the local optimizations are performed using conjugate gradient descent (or conjugate gradients squared \citep{sonneveld_cgs_1989} for non-Hermitian $A$), then the algorithm primarily consists of tensor contractions and the cost of the algorithm scales like $\mathcal{O}((D_x^2 D_b + D_b D_x^2) D_A^2)$, where $D_x$ and $D_b$ are the bond dimensions of QTT-States $x$ and $b$, and $D_A$ is the bond dimension of QTT-Operator $A$. For the operator in Eq.~\eqref{eq:em_update}, $D_A$ is moderate and independent of system size, since $A$ can be built directly from differential operators and the calculations are performed on a uniform Cartesian grid.

If the QTT bond dimension is sufficiently large and the time-step is small enough such that matrix $A$ is well-conditioned, then DMRG converges quickly to the expected solution. In practice, DMRG often converges to some value larger than the tolerated error threshold (of $10^{-6}$). In this case, we split a single time step into at most 4 smaller equally sized time steps; if DMRG shows better convergence at each step, then we continue to solve Maxwell's equations with these reduced time steps. Otherwise, we simply accept that the QTT bond dimension is likely not large enough and continue with the simulation as is. Certainly, a more advanced implicit time evolution scheme should be considered in the future.

Note that the DMRG algorithm can be extended to more general QTN geometries. However, in this work we are only solving Maxwell's equation in two dimensions, and the QTC with two branches is equivalent to a QTT. As such, we do not need to modify the original algorithm.

\section{Results}
In the following section, we evaluate the performance of the proposed QTN methods on two common test problems. Our code is built off the \edit{Python} tensor network software package quimb \citep{gray2018quimb}, \edit{though one could use any of the many other tensor network software packages available.}

\subsection{Orszag-Tang vortex}
\begin{figure*}
    \vspace{0.2cm}
    \centering
    {
    \scriptsize 
    \begin{tabular}{c c c}
        \hspace{0.5cm} (a) $D = 32$, $t=21 \Omega_{c,p}^{-1}$ \hspace{0.5cm}
        & \hspace{0.6cm} (b)  $D=64$, $t=21 \Omega_{c,p}^{-1}$ 
        \hspace{0.2cm} 
        & \hspace{0.3cm} (c) $D_f=32, \, D_{EM}=128$, $t=9 \Omega_{c,p}^{-1}$
    \end{tabular}
    }
    \\[0.05cm]
    \subfloat{
        \includegraphics[width=0.34\linewidth,trim={0.5cm 0cm 1.25cm 1.35cm}, clip]
            {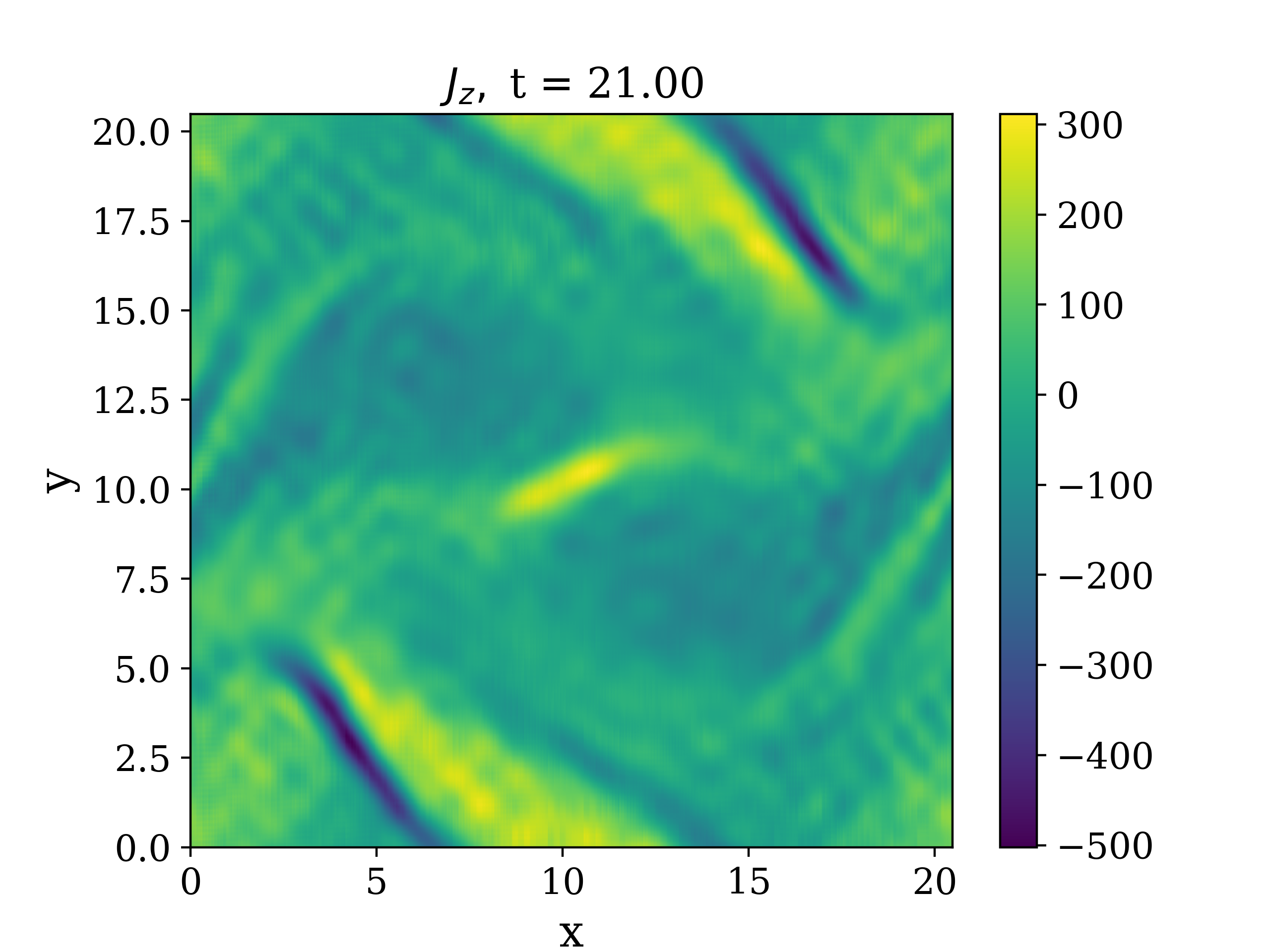}
            % {figs/orszag/D32_dt0.001_t21.png}
        \hspace{0.1cm}
        \includegraphics[width=0.30\linewidth,trim={2.25cm 0cm 1.25cm 1.35cm}, clip] 
            {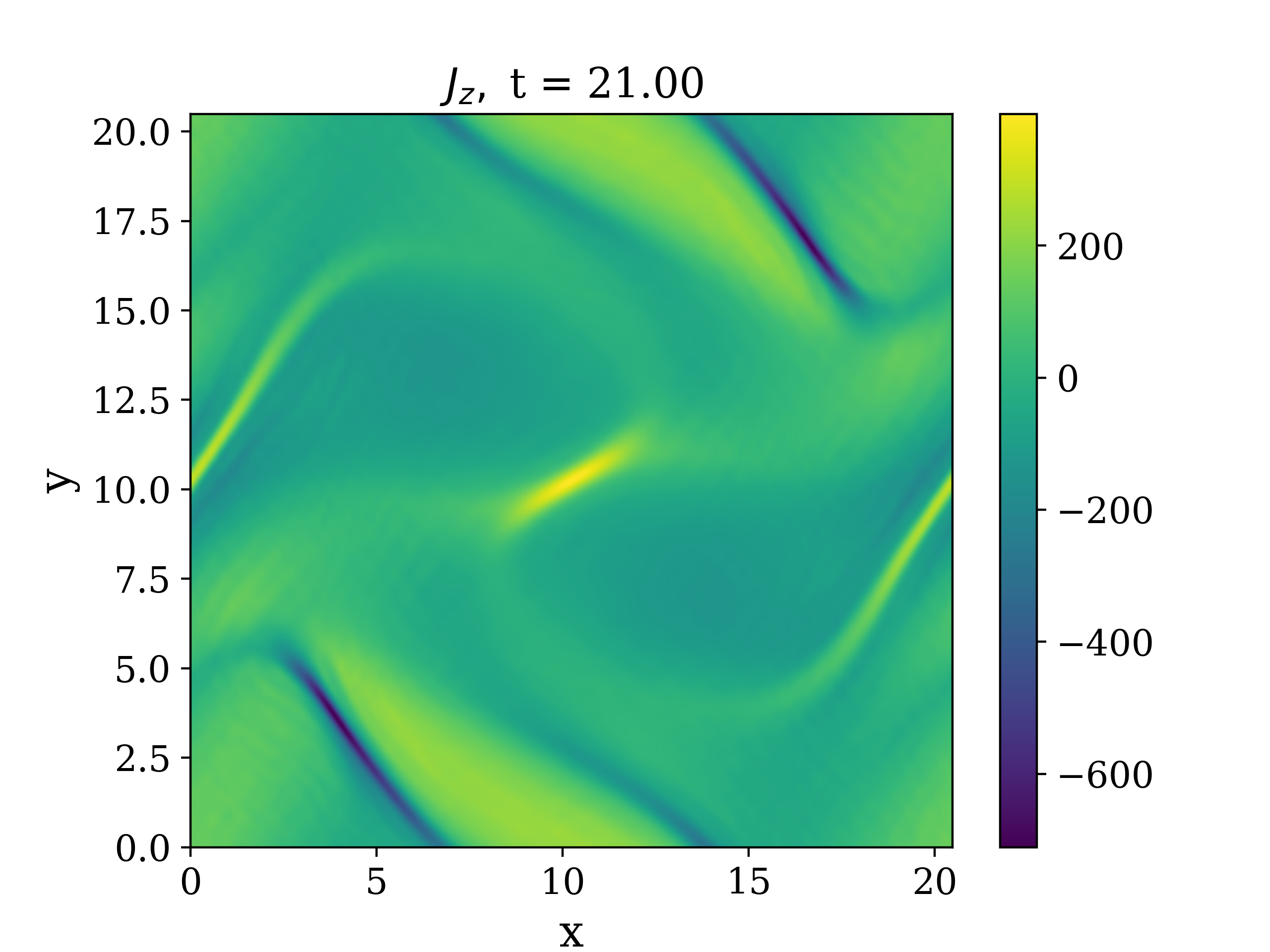}
            % {figs/orszag/D64_dt0.001_t21.png}
        \hspace{0.1cm}
        \includegraphics[width=0.30\linewidth,trim={2.25cm 0cm 1.25cm 1.35cm}, clip] 
            {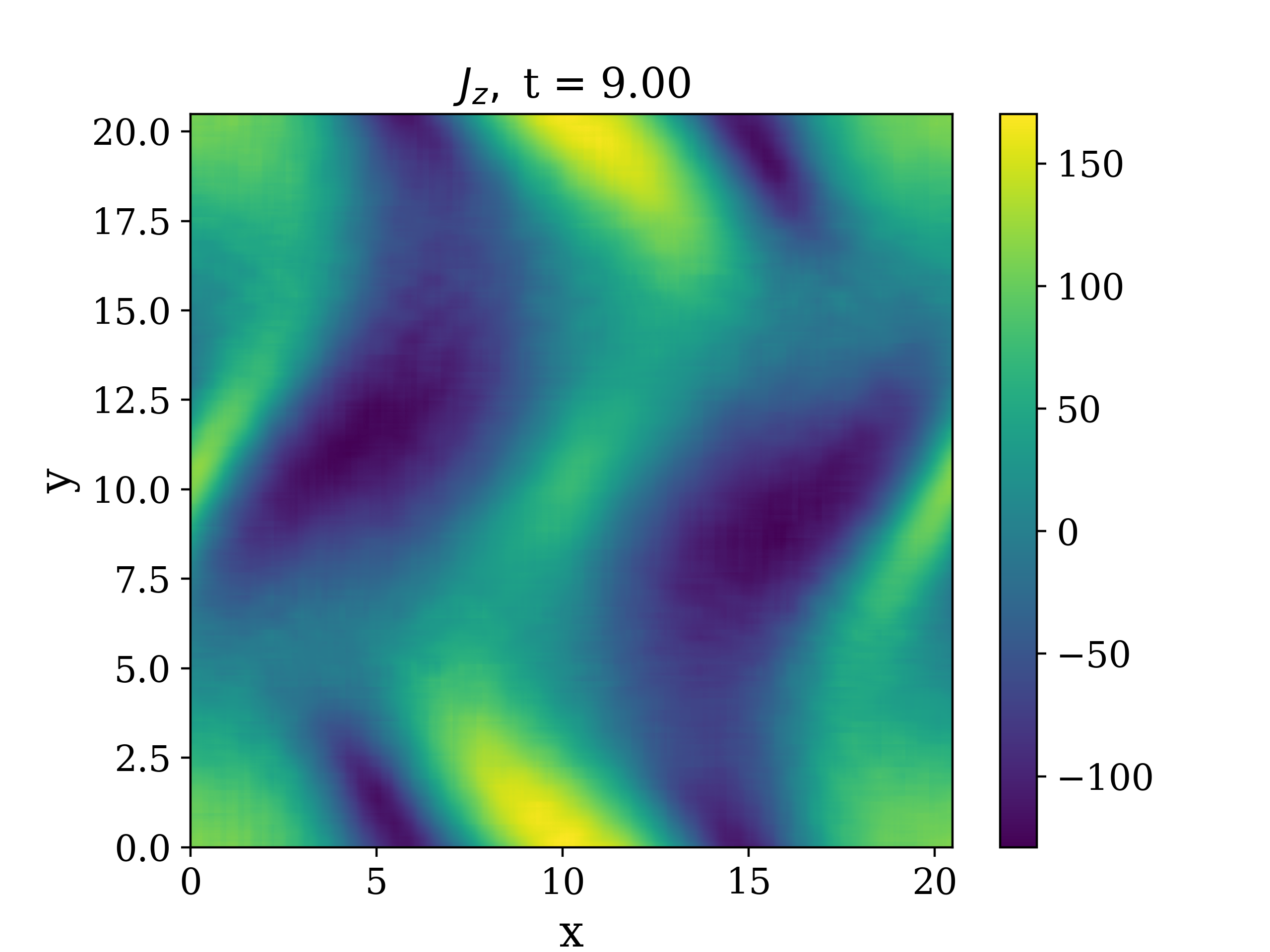}
            % {figs/orszag/D32_DE128_dt0.001_t09.png}
    }
    \caption{Plots of the out-of-plane current % and the energy spectrum
    for the Orszag-Tang vortex at the specified time for (a) $D=32$, (b) $D=64$, and (c) bond dimension $D_f=32$ for the distribution functions and bond dimension $D_{EM}=128$ for the electromagnetic fields. Calculations were performed with a time step of $0.001 \Omega_{c,p}^{-1}.$
    }
    \label{fig:orszag-tang}
\end{figure*}

\edit{The Orszag-Tang vortex \citep{orszag_tang_1979} is a well-documented nonlinear test problem that captures the onset and development of turbulence.}
We consider a (fully nonlinear and kinetic) 2D3V calculation with protons and electrons.
The initial conditions are \citep{Juno2018discontinuous}
\begin{align}
    \textbf{B} &= -B_0 \sin \left( k_y y \right) \hat{x} + B_0 \sin \left( 2k_x x \right) \hat{y} + B_g \hat{z}, \\
    \textbf{u}_p &= -u_{0,p} \sin \left( k_y y \right) \hat{x} + u_{0,p} \sin \left( k_x x \right) \hat{y}, \\
    \textbf{J} &= e n_0 (\textbf{u}_p - \textbf{u}_e) = \frac{1}{\mu_0} \nabla \times \textbf{B}, \\
    \textbf{E} &= -\textbf{u}_e \times \textbf{B},
\end{align}
where $k_x=k_y=2\pi/L_\text{box}$, with $L_\text{box}$ being the simulation domain size in real space. $B_g$ is the out-of-plane guide field. The amplitudes of the perturbations are set to $B_0 = 0.2B_g$ and $u_{0,p} = 0.2 v_{A,p}$, where $v_{A,p} = B_g / \sqrt{\mu_0 n_0 m_p}$ is the Alfv\'{e}n velocity of the protons.
The initial velocities ($\textbf{u}_s$) are included as initial drift velocities of the Maxwellian distribution, 
\begin{align}
    % f_s(\textbf{v}) = \left( \frac{1}{2\pi v_{th,s}^2 } \right)^{3/2} \exp \left( -\frac{(\textbf{v}-\textbf{u}_s)^2}{2 v_{th,s}^2} \right)
    f_s(\textbf{v}) = \left( 2\pi v_{th,s}^2  \right)^{-3/2} \exp \left( -\frac{(\textbf{v}-\textbf{u}_s)^2}{2 v_{th,s}^2} \right),
\end{align}
where $v_{th,s}^2 = k_B T_s / m_s $ is the thermal velocity of species $s$, with $k_B$ the Boltzmann constant, $m_s$ the mass, and $T_s$ the temperature.

The plasma is initialized with density $n_0=1$, mass ratio $m_p/m_e = 25$, temperature ratio of $T_p/T_e = 1$, proton plasma beta $\beta_p = 2 v^2_{th,p}/v^2_{A,p} = 0.08 $ and speed of light $c = 50 v_{A,p}$. 

The simulation is performed with $L_\text{box}= 20.48 d_p$, where $d_p = \sqrt{ m_p / \mu_0 n_0 e^2}$ is the proton skin depth, with $e$ being the elemental charge. The limits of the grid in velocity space for species $s$ are $\pm 7v_{th,s}$. Periodic boundary conditions are used in all dimensions, including the velocity dimensions, so that the time evolution operator is unitary and TDVP can be used. 
The calculations are performed with $2^9$ grid points along each spatial axis, and $2^6$ grid points along each axis in velocity space. With this grid resolution, the grid-spacing in real space is 0.04$d_p$, so the electron skin depth is well-resolved. We utilize the QTC ansatz with tensors corresponding to each dimension on its own branch, ordered as $(x, y, v_x, v_y, v_z)$, and the data is encoded with binary mapping such that fine grid tensors are attached to the spine for all dimensions. At each time step, the proton and electron distribution functions, as well as each component of the electric and magnetic fields, are compressed to bond dimension $D$.

Fig.~\ref{fig:orszag-tang} shows the out-of-plane current density % and the $x$-component of the magnetic field
at time $21 \Omega_{c,p}^{-1}$ obtained from simulations with bond dimension (a) $D=32$ and (b)  $D=64$.
% , and (c) $L_v=7$ and bond dimension $D=64$. In the last calculation, a fourth-order centered finite-difference stencil is used instead of a second-order stencil.  
Qualitatively, both calculations exhibit the expected dynamics. 
However, the features in the $D=32$ calculation appear less sharp than in the $D=64$ case, and there is significantly more numerical noise. Unfortunately, the numerical noise will continue to grow for both calculations. 
Visual inspection suggests that the noise arises from the Maxwell equation solver in the electric field (see Section VI.A.1 of the SI for these plots). However, when increasing the bond dimension of the QTTs representing the electric and magnetic fields to $128$, the noise actually worsens and one is limited to even shorter simulation times (Fig.~\ref{fig:orszag-tang}(c)). \edit{It thus appears that} the low-rank approximation provides some noise mitigation; with less compression, the noise grows unchecked.
%
% This suggests that the low-rank approximation actually provides some amount of noise mitigation. 
% In the presence of significant numerical noise, the fields will likely require larger bond dimension than the true solution. 
% Over time, as fine grid-scale noise continues to grow, one might expect that the QTT will eventually only accurately capture features on an effective grid determined by $D$, i.e.,~the grid resolution for which the specified $D$ would yield the exact result. 

\edit{
Since the $D=32$ calculations yield broader features than the $D=64$ case, one might be tempted to say that the QTT is only able to capture features at an ``effective grid resolution'' determined by $D$, or the grid resolution for which the specified $D$ would yield no compression. (For the Orszag-Tang vortex calculation, since the EM fields are 2-D, $D=32$ would yield no compression for a grid with 32 points along each dimension. This grid would have a grid-spacing of 0.64 $d_p$. For $D=64$, the grid would have 64 points along each dimension and grid-spacing 0.32 $d_p$. These length scales also appear as special scales in the spectra; see Section V1.A.1 of the SI.)
However, we stress that one should not interpret $D$ as an effective resolution. The observed correlation here is likely caused by numerical noise. Random and unstructured functions are not compressible in the QTT format. As such, numerical noise will likely reduce the compressibility of the original function. When compressing a noisy function, one typically will not obtain the original function. 
Instead, some noise will persist and continue to accumulate over time.
}

% Random and unstructured functions are not compressible in the QTT format. As such, numerical noise will likely reduce the compressibility of the original function. Unfortunately, it is unlikely one will obtain the original function when making low-rank approximation on a noisy function. Instead, it is more likely that we obtain low-rank noise that exhibits some structure. 
% Given the observation that the features in the $D=32$ calculation are broader than in the $D=64$ case, it appears that after the noise accumulates over time, the QTT is only able to capture features at an ``effective grid resolution" determined by $D$, i.e.,~the grid resolution for which the specified $D$ would yield no compression.  
%
% For example, in these calculations, $D=32$ for the electromagnetic fields (which are 2D) would correspond to a grid resolution of $0.64 d_p$, while $D=64$ would correspond to a grid resolution of $0.32 d_p$. % This would explain why the features in the $D=32$ calculation are broader than in the $D=64$ case. 
% The same observations can be made using the spectra of the electromagnetic fields (see Section V.A of the SI). 
% These length scales also appear as special scales in the spectra (see Section V.A of the SI).

\subsection{Reconnection problem}

\begin{figure*}
    \subfloat[$D=32$]{
        \includegraphics[width=0.33\linewidth,trim={0.75cm 0cm 1.0cm 0cm}, clip] 
            {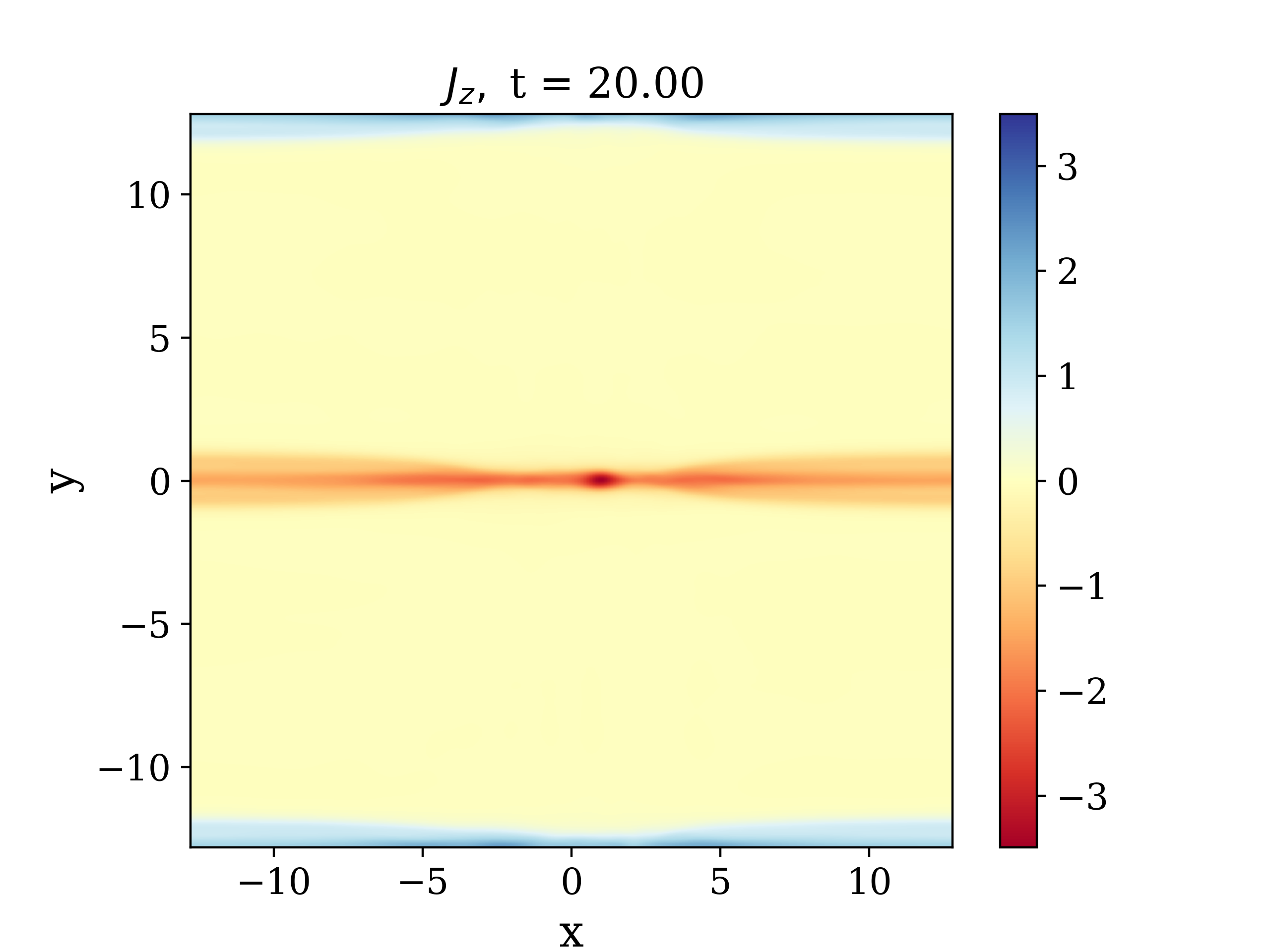}
            % {figs/reconn/D32_Jz20.png}
        \includegraphics[width=0.33\linewidth,trim={0.75cm 0cm 1.0cm 0cm}, clip] 
            {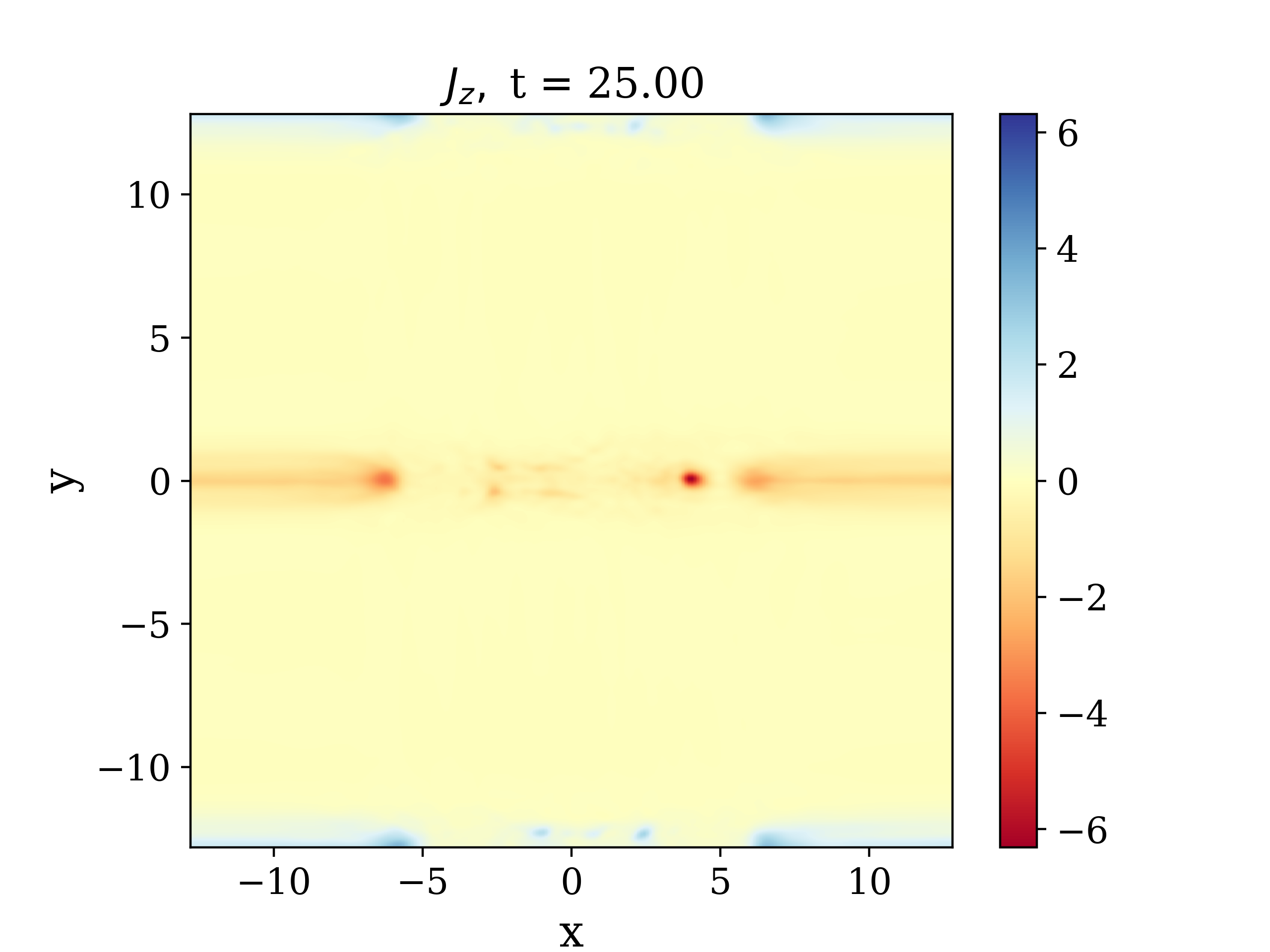}
            % {figs/reconn/D32_Jz25.png}
        \includegraphics[width=0.33\linewidth,trim={0.75cm 0cm 1.0cm 0cm}, clip] 
            {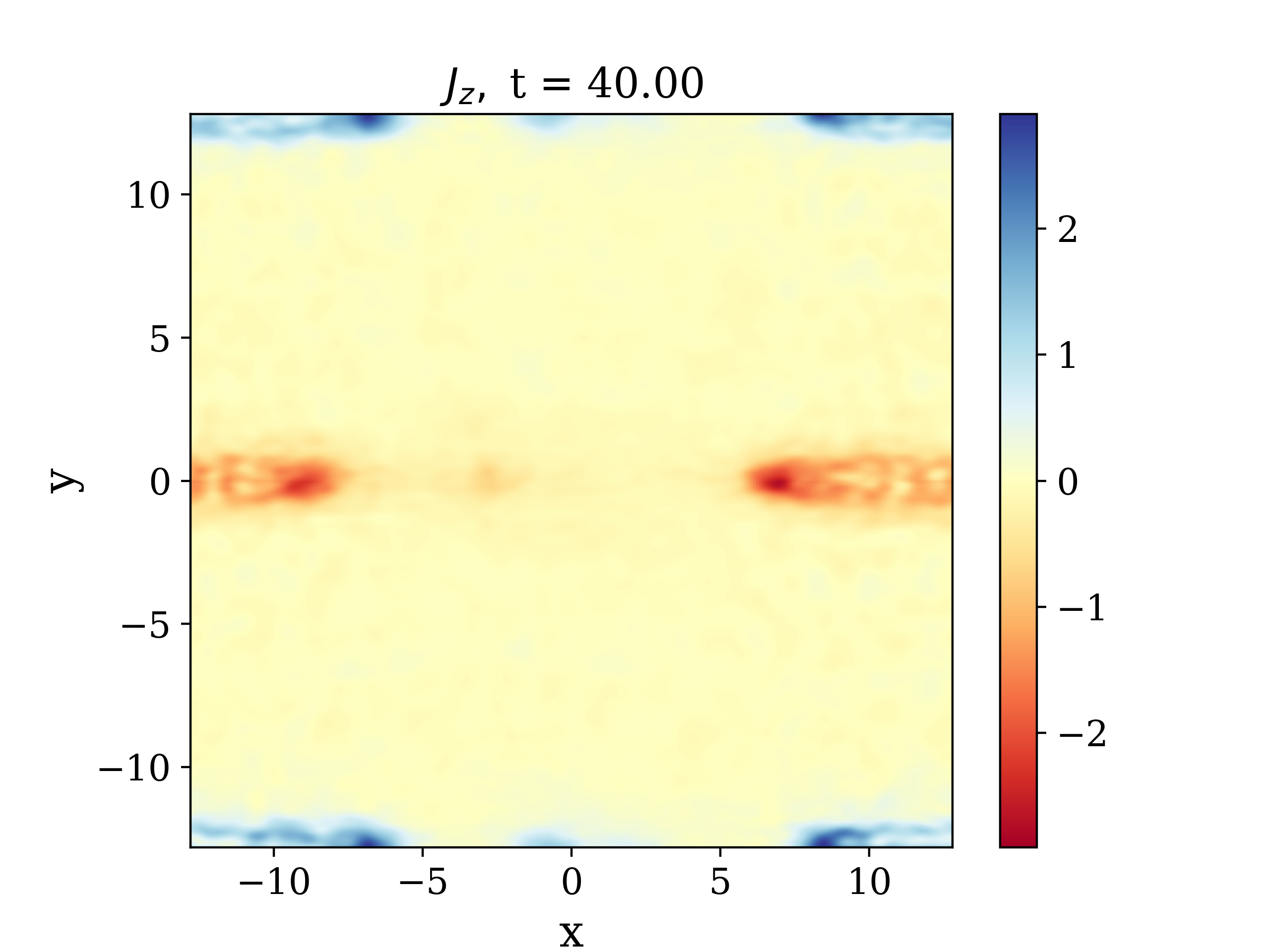}
            % {figs/reconn/D32_Jz40.png}
    }
    \\[-0.05cm]
    \subfloat[$D=64$]{
        \includegraphics[width=0.33\linewidth,trim={0.75cm 0cm 1.0cm 0cm}, clip] 
            {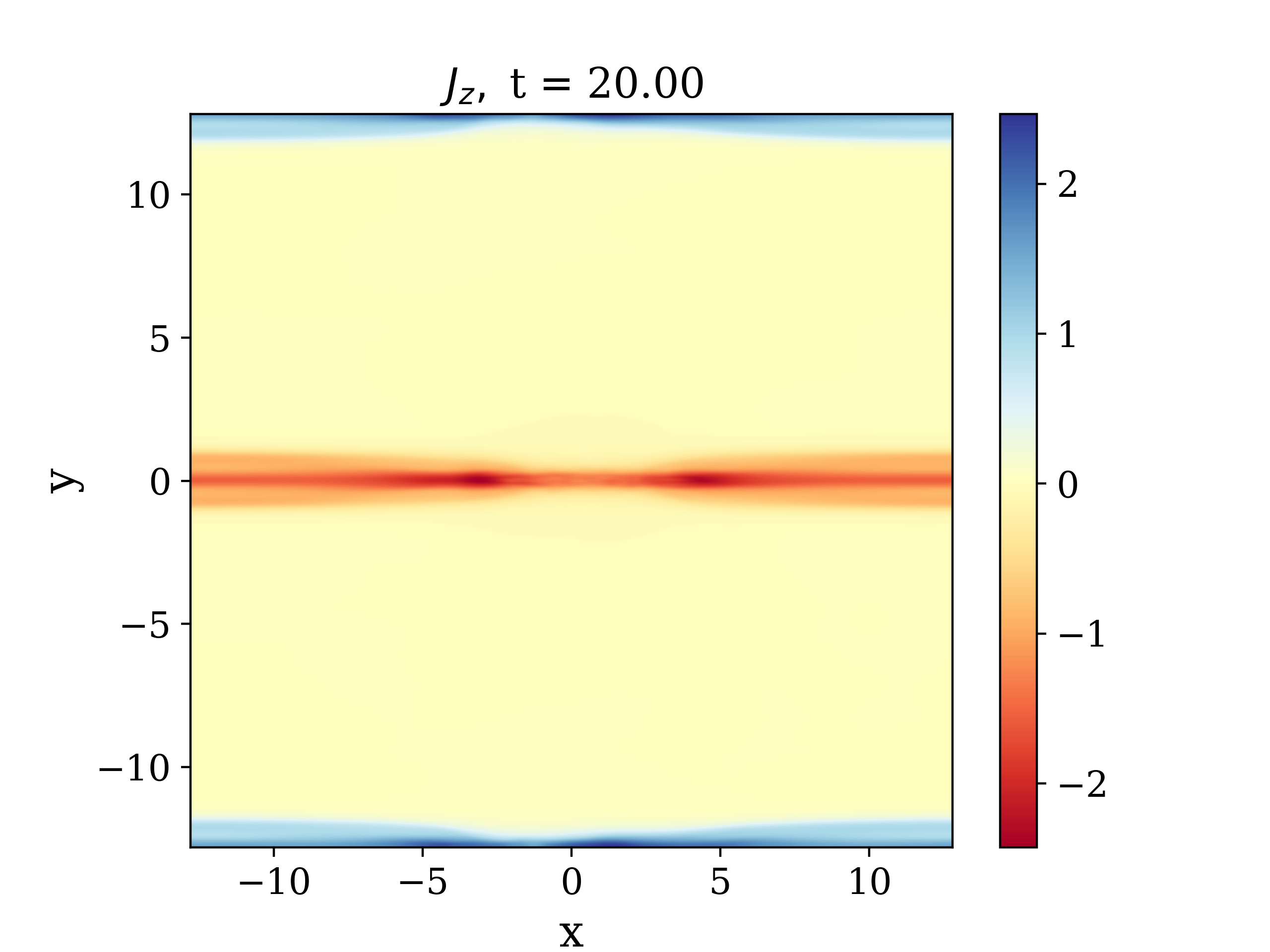}
            % {figs/reconn/D64_Jz20.png}
        \includegraphics[width=0.33\linewidth,trim={0.75cm 0cm 1.0cm 0cm}, clip] 
            {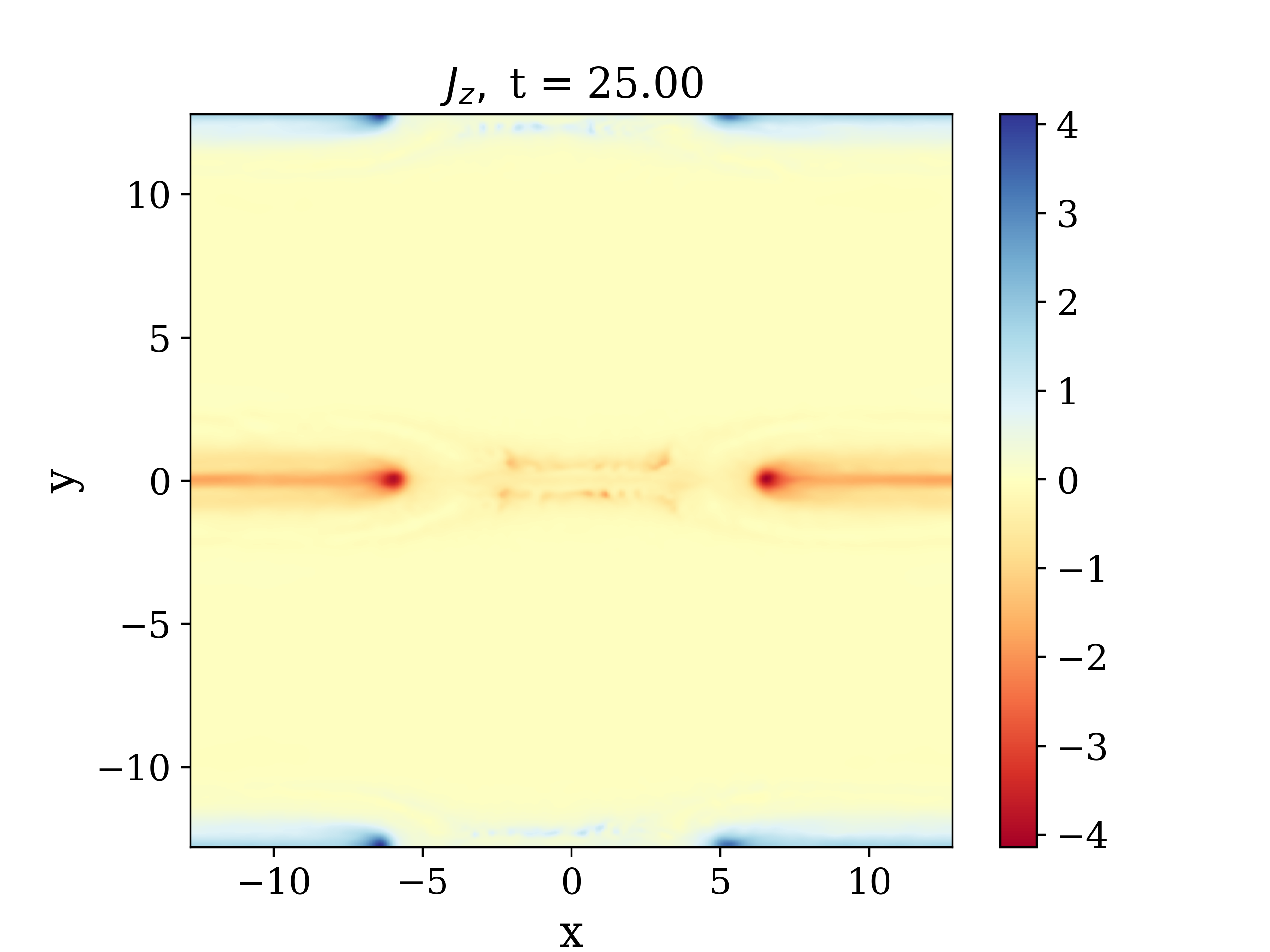}
            % {figs/reconn/D64_Jz25.png}
        \includegraphics[width=0.33\linewidth,trim={0.75cm 0cm 1.0cm 0cm}, clip] 
            {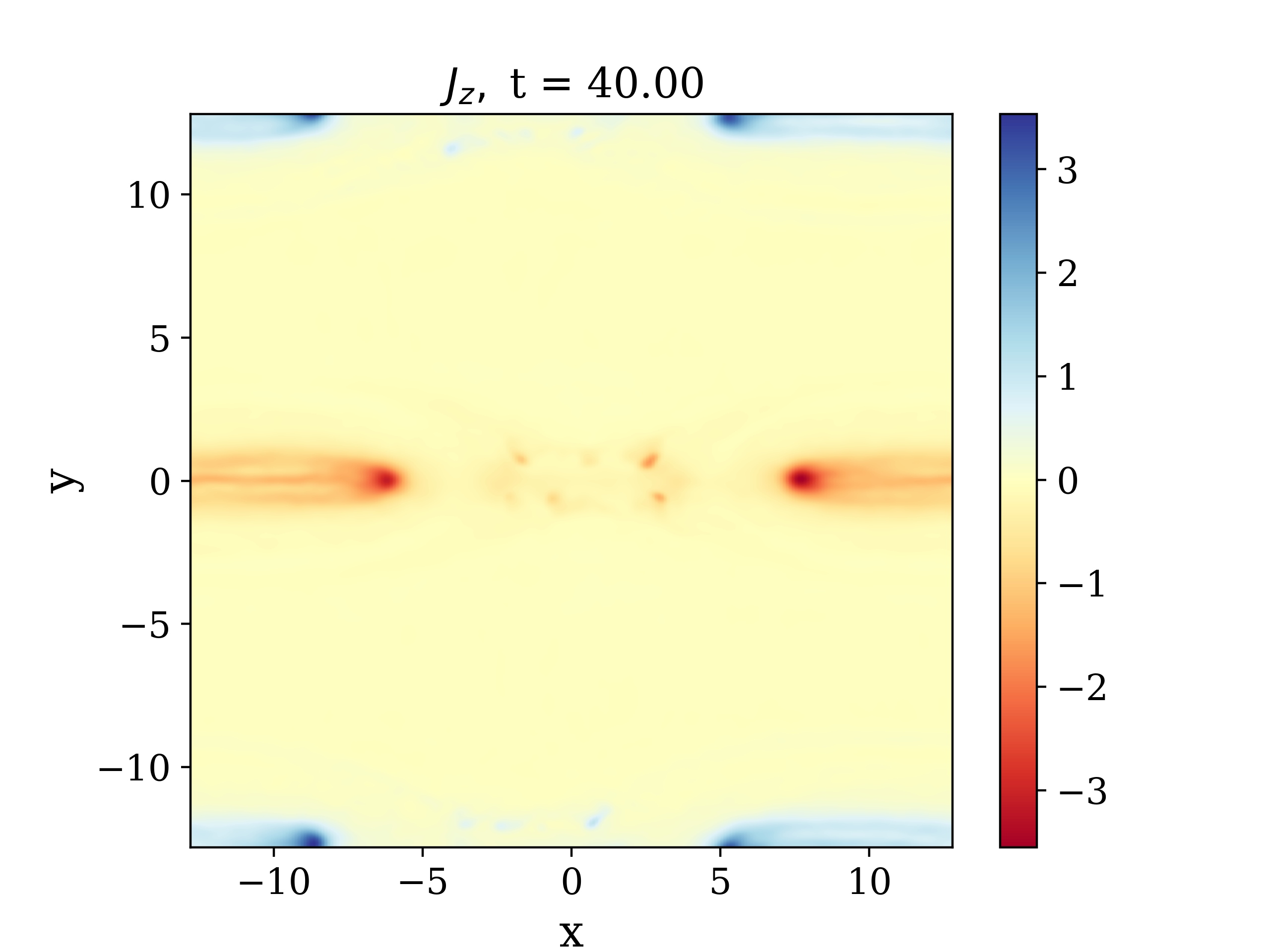}
            % {figs/reconn/D64_Jz40.png}
    }
    \caption{Out-of-plane current density for the \edit{GEM reconnection problem}. Calculations were performed with with bond dimension (a) $D=32$ and (b) $D=64$, and time step $\Delta t=0.005 \Omega_{c,p}^{-1}$. Time is in units of  $\Omega_{c,p}^{-1}.$}
    \label{fig:reconnection_jz}
\end{figure*}

The next test we consider is the 2D3V \edit{Geospace Environmental Modeling (GEM) challenge \citep{birn_geospace_2001}: a magnetic reconnection set-up that has become a classical code-benchmark problem}. %\edit{which simulates island formation in a slightly perturbed Harris sheet.
%Incorporating kinetic effects is critical for correctly computing the magnetic reconnection flux.}
As in the Orszag-Tang vortex test problem, we adapt the initial conditions to fully kinetic simulations \citep{schmitz_kinetic_2006, Rahn2022parallel6D}, with additional modifications so that periodic boundary conditions can be used.

In the traditional set-up, the initial configuration for each species $s$ is 
\begin{align}
    f_s = f_{0,s} + f_{b,s},
\end{align}
where 
\begin{align}
    f_{b,s} = n_b (2\pi v_{th,s}^2)^{-3/2} \exp(-\textbf{v}^2/2 v_{th,s}^2)    
\end{align}
is a uniform background distribution and
$f_{0,s}$ is the Harris sheet equilibrium \citep{harris1962plasma}. Given the magnetic field
\begin{align}
    \textbf{B}(y) = B_0 \tanh \left( \frac{y}{\lambda}\right) \hat{x},
\end{align}
where $\lambda$ is the thickness of the sheet, the proton/electron distribution is
\begin{align}
    f_{0,s} = \frac{n_0(y)}{(2 \pi v_{th,s}^2)^{3/2}} \exp \left( -\frac{ v_x^2 + v_y^2 + (v_z - v_{0,s})^2 } {2 {v^2_{th,s}}} \right), 
\end{align}
where 
\begin{align}
    n_0(y) = n_0 \sech^2 \left(\frac{y}{\lambda}\right).
\end{align}
The drift velocities are constrained by the equilibrium conditions,
\begin{align}
    & \frac{B_0^2}{2\mu_0} = n_0 (T_p + T_e), \\
    & \lambda = \frac{2}{e B_0} \frac{T_p + T_e}{v_{0,p} - v_{0,e}}, \\
    & \frac{v_{0,e}}{v_{0,p}} = - \frac{T_e}{T_p}.
\end{align}
% In addition to the Harris sheet equilibrium, there is also a uniform background distribution for both species, with density $n_b=0.2n_0$ and zero drift velocity. The temperatures are the same as above. 
%The temperatures are as before but there is no drift velocity.

The initial perturbation in the magnetic field is 
\begin{align}
    B_x &= - \frac{A_0 k_y}{2} \cos (k_x x) \, \sin (k_y y/2) \, , \\
    B_y &= A_0 k_x \sin(k_x x)\, \cos (k_y y/2) \, .
\end{align}

To allow for periodic boundary conditions, the simulation domain (suppose previously $\left[-L_{\text{box},x}/2, L_{\text{box},x}/2 \right)$ and $\left[ -L_{\text{box},y}/2, L_{\text{box},y}/2 \right)$) is doubled in $y$, and an inverted Harris sheet placed at the simulation boundary:
\begin{align}
    B_x(y) = B_0 \tanh \left( \frac{y}{\lambda}\right) 
    % \nonumber \\
    - B_0 \tanh \left( \frac{y+L_{\text{box},y}}{\lambda}\right) 
    % \nonumber \\
    - B_0 \tanh \left( \frac{y-L_{\text{box},y}}{\lambda}\right) \,,
\end{align}
and
\begin{align}
    n(y) = n_0 \sech((y+L_{\text{box},y})/\lambda^2) 
    % \nonumber \\
    + n_0 \sech((y-L_{\text{box},y})/\lambda^2) \,.
\end{align}
The perturbation in the magnetic field remains the same.

Simulations are performed with $L_{\text{box},x}=25.6 d_p$, $L_{\text{box},y}=12.8 d_p$, and a Harris sheet thickness of $\lambda = 0.5 d_p$. The plasma is defined to have a mass ratio of $m_p/m_e = 25$, temperature ratio of $T_p/T_e = 5$, proton plasma beta $\beta_p = 5/6$, and electron Alfv\'{e}n velocity $v_{A,e}=0.25 c$. The strength of the perturbation is $A_0 = 0.1 B_0 d_p$. The background density is set to $n_b=0.2 n_0$. The limits of the grid in velocity space are % $\pm 5 v_{A,p}$ 
$\pm 7.75 v_{th,p}$
for the protons. 
For the electrons, the limits are % $\pm 10 v_{A,p}$ 
$\pm 7 v_{th,e}$ in the $v_x$ and $v_y$ directions, and % $\pm 20 v_{A,p}$ 
$\pm 14 v_{th,e}$ in the $v_z$ direction. The resolution of the calculations are $2^8 \times 2^9 \times 2^7 \times 2^7 \times 2^8$ along the $x,\, y,\ v_x,\, v_y,$ and $ v_z$ axes respectively. We again utilize the QTC ansatz with tensors corresponding to each dimension on its own branch. In contrast to above, the data is encoded with the mirror (two's complement) mapping for all dimensions because it showed marginally better results. Additionally, the bond dimension for the electric and magnetic fields are set to 128 so that the DMRG solver for Maxwell's equations might converge more quickly.

\begin{figure}
    \centering
    \subfloat{
        \includegraphics[width=0.5\linewidth,trim={0cm 0cm 0cm 0cm}]
            {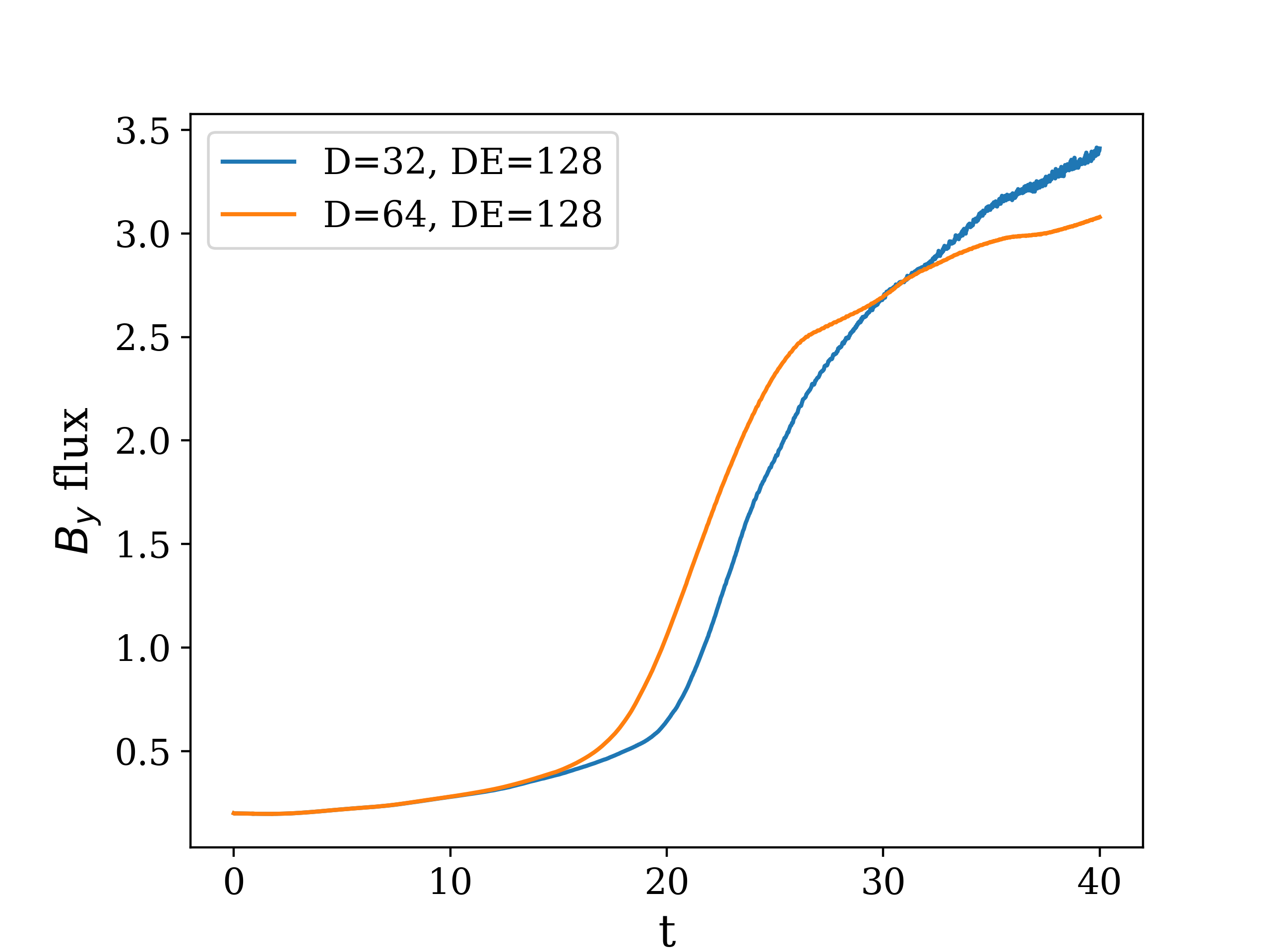}
            % {figs/reconn/plot_Byflux_L89-778_te3164-223_COMBMM_dt0.005.png}
    }
    % \\[-0.05cm]
    % \subfloat[$D=64$]{
    %     \includegraphics[width=0.33\linewidth,trim={0cm 0cm 0cm 0cm}]
    %         {figs/plot_Byflux__L89-778_te3164-223_COMBMM_D64.png}
    % }
    \caption{Reconnected magnetic flux computed with time step $\Delta t=0.005 \Omega_{c,p}^{-1}$ and bond dimension $D=32$ (blue) and $D=64$ (orange). Time is in units of $\Omega_{c,p}^{-1}$.}
    \label{fig:reconnection_flux}
\end{figure}

Fig.~\ref{fig:reconnection_jz} shows the current density and Fig.~\ref{fig:reconnection_flux} plots the reconnected magnetic flux $\Psi= \int_0^{L_{\text{box},x}/2} B_y(x,y=0) \, dx$ for calculations with bond dimensions $D=32$ and $D=64$. Additional results are presented in Section VI.B of the SI.
% In contrast to above, Maxwell's equations are solved implicitly with bond dimension 128 so that DRMG would converge more quickly. 
% In general, if the time evolution is numerically stable, 
Again, for both bond dimensions considered here, we roughly obtain the expected reconnected flux over time \citep{schmitz_kinetic_2006}. However, the dynamics appear sensitive to numerical noise, as evidenced by unexpected secondary island formation in the Harris sheet for the $D=32$ case.   % At longer times, the simulations exhibit numerical noise, which is likely caused by both the reduced bond dimension and the relatively large time step.

\section{Discussion}

The above results are a promising first demonstration that high-dimensional simulations of the Vlasov-Maxwell equations can be performed efficiently with the QTC ansatz. Our calculations use $N=2^{36}$ total grid points for the Orszag-Tang vortex, and $N=2^{39}$ total grid points for the GEM reconnection problem. However, we are able to observe the expected phenomenological results satisfactorily with a maximum bond dimension of just $D=64$. Despite the computational overhead in the QTC algorithm, ignoring constant factors, the $\mathcal{O}(D^4)$ cost per time step is a fraction of the $\mathcal{O}(N)$ scaling for traditional methods, and the simulations can be run on a single compute node. 
% (Tighter bounds can be obtained by distinguishing between the virtual bonds along the spine and along the branches). 
Furthermore, by using the local time evolution scheme TDVP, one can use a larger time step (observed to be larger by about a factor of 4) than allowed in global time evolution schemes.
% one critically avoids the linear increase in number of time steps with respect to grid resolution. 
In addition to computational cost, memory storage costs are also reduced, from $\mathcal{O}(N)$ to $\mathcal{O}(D^3)$. 
\edit{Extension to full 3D3V calculations is straightforward, limited only by the current implementation of our code for the QTC geometry. % (we currently do not have a 3-D DMRG solver for Maxwell's equations in the QTC geometry).
}

\edit{
One can measure the efficiency of the QTN ansatz by measuring the von Neumann or bipartite entanglement entropy (EE) at each bond in the QTN. For simplicity, consider a QTT of length $L$. The first $j$ tensors form the left partition, and the remaining tensors form the right partition. One can always partition the QTT via singular value decomposition, giving rise to
\begin{align}
    \textsf{T}(i_1, ... i_j, i_{j+1}, ... i_L) = \textsf{U}(i_1, ... i_{j}) \Sigma \textsf{V}^\dagger(i_{j+1}, ..., i_{L}),
\end{align}
where $U$ and $V$ are unitary matrices representing the left and right partitions, respectively, and $\Sigma$ is a diagonal matrix containing the singular values $\sigma$.
The EE is then given by
\begin{equation}
    EE^{(j)} = - \sum_i \tilde{\sigma}_i^2 \log_2 (\tilde{\sigma}_i^2),
\end{equation}
where $\tilde{\sigma}$ are the singular values normalized by $\sum_i \sigma_i^2 = 1$. The minimum EE of zero occurs when there is only one non-zero singular value. Maximum entanglement occurs when all of the singular values are of equal weight, which would be $\log_2(D)$ for a bond of size $D$. A smaller EE suggests that one can typically use a smaller bond dimension for the QTN. However, note that while the EE describes the amount of information transferred along the virtual bond, it does not provide information about the error that would arise from truncating the singular values.
}

\begin{figure}
    \subfloat{
        \includegraphics[width=0.50\linewidth,trim={0cm 0cm 0cm 0cm}]
            {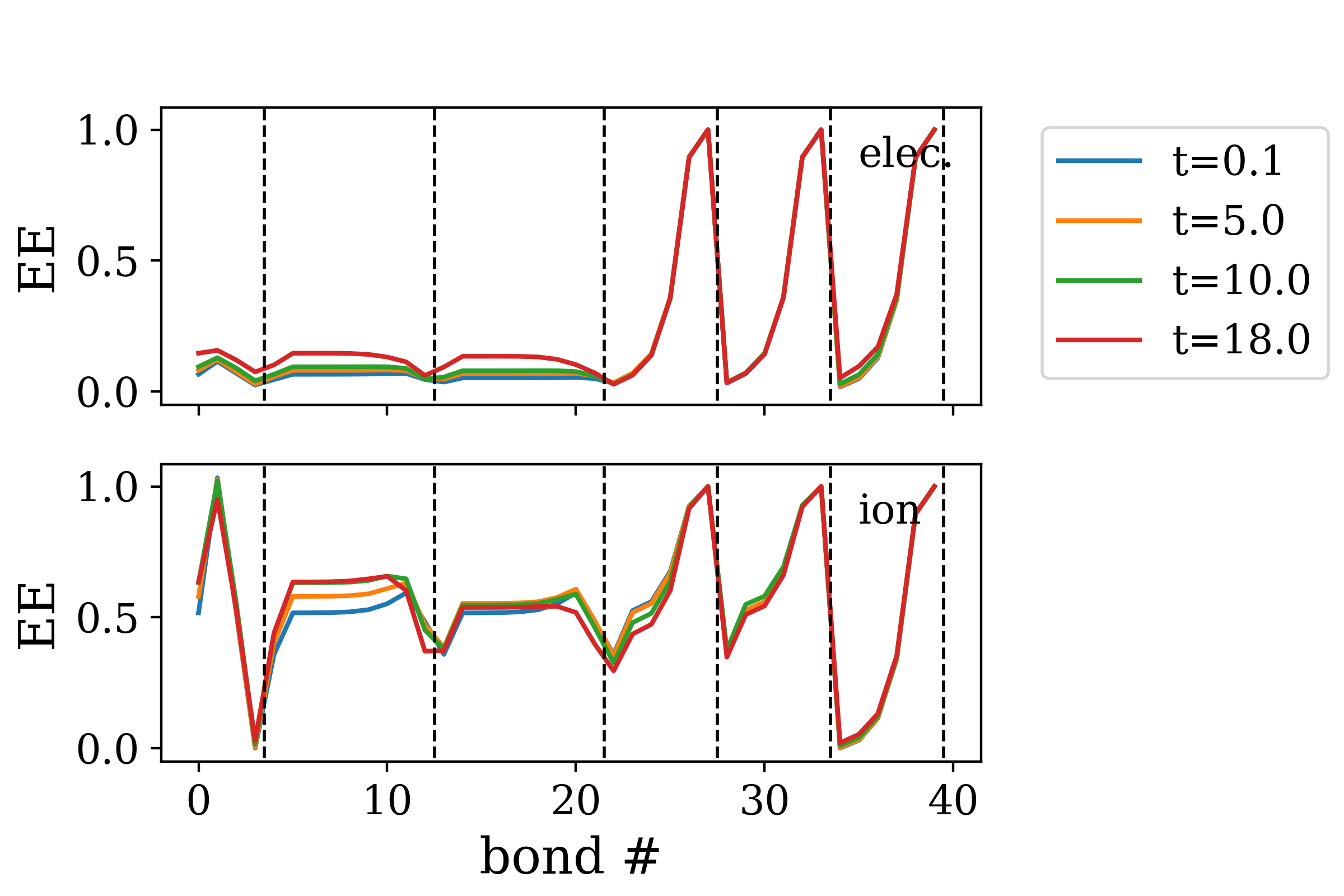}
            % {figs/orszag/EE_f_t_D64_dt0.001.png}
    }
    \subfloat{
        \includegraphics[width=0.50\linewidth,trim={-0.25cm 0cm 0cm 0cm}] 
            {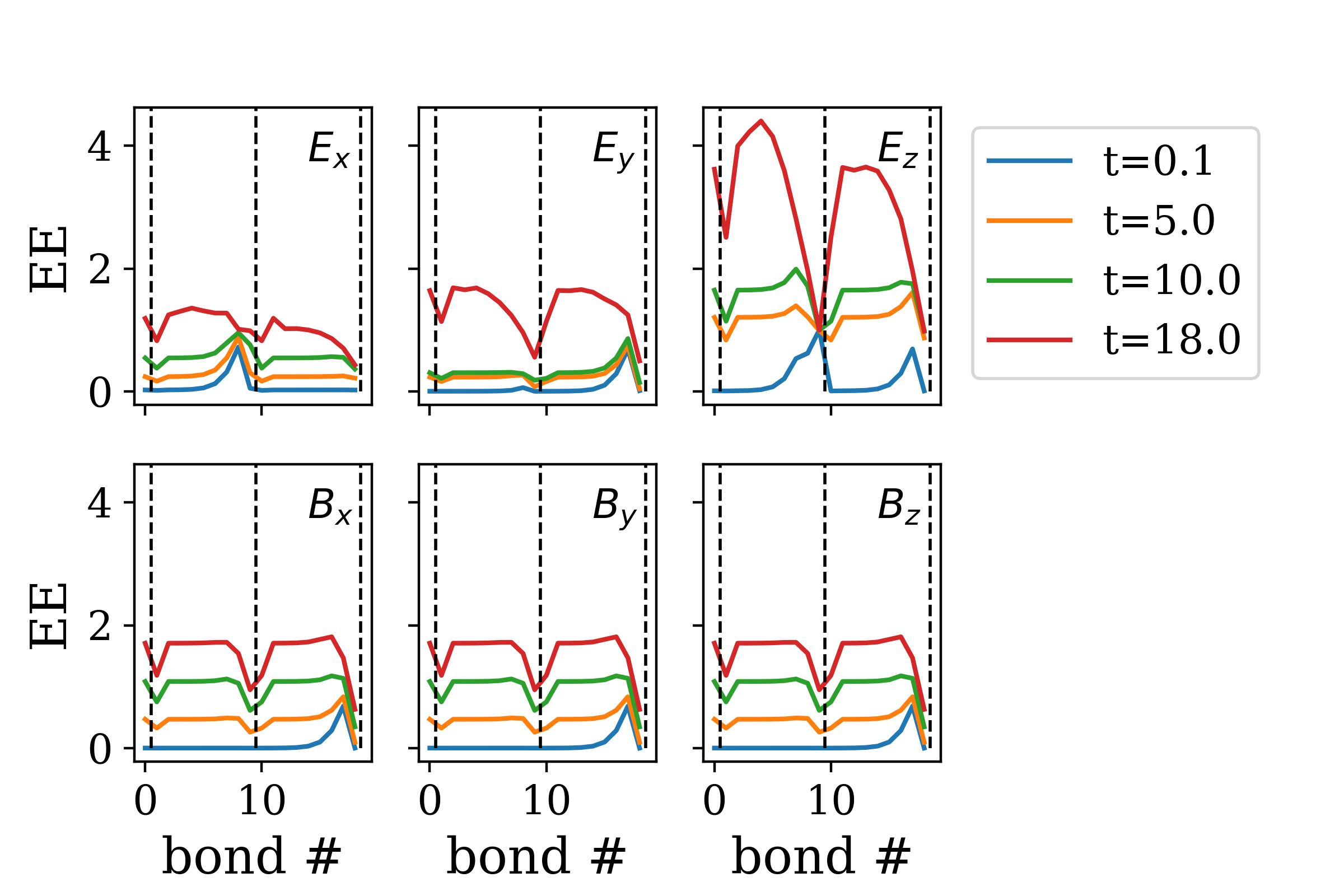}
            % {figs/orszag/EE_EB_t_D64_dt0.001.png}
    }
    \caption{Entanglement entropy at each bond for the Orszag-Tang problem. Bond number is ordered by bonds in the spine, and then bonds in each branch.}
    \label{fig:EE_ot}
\end{figure}

\begin{figure}
    \subfloat{
        \includegraphics[width=0.50\linewidth,trim={0cm 0cm 0cm 0cm}]
            % {figs/reconn/EE_T_A.png}
            {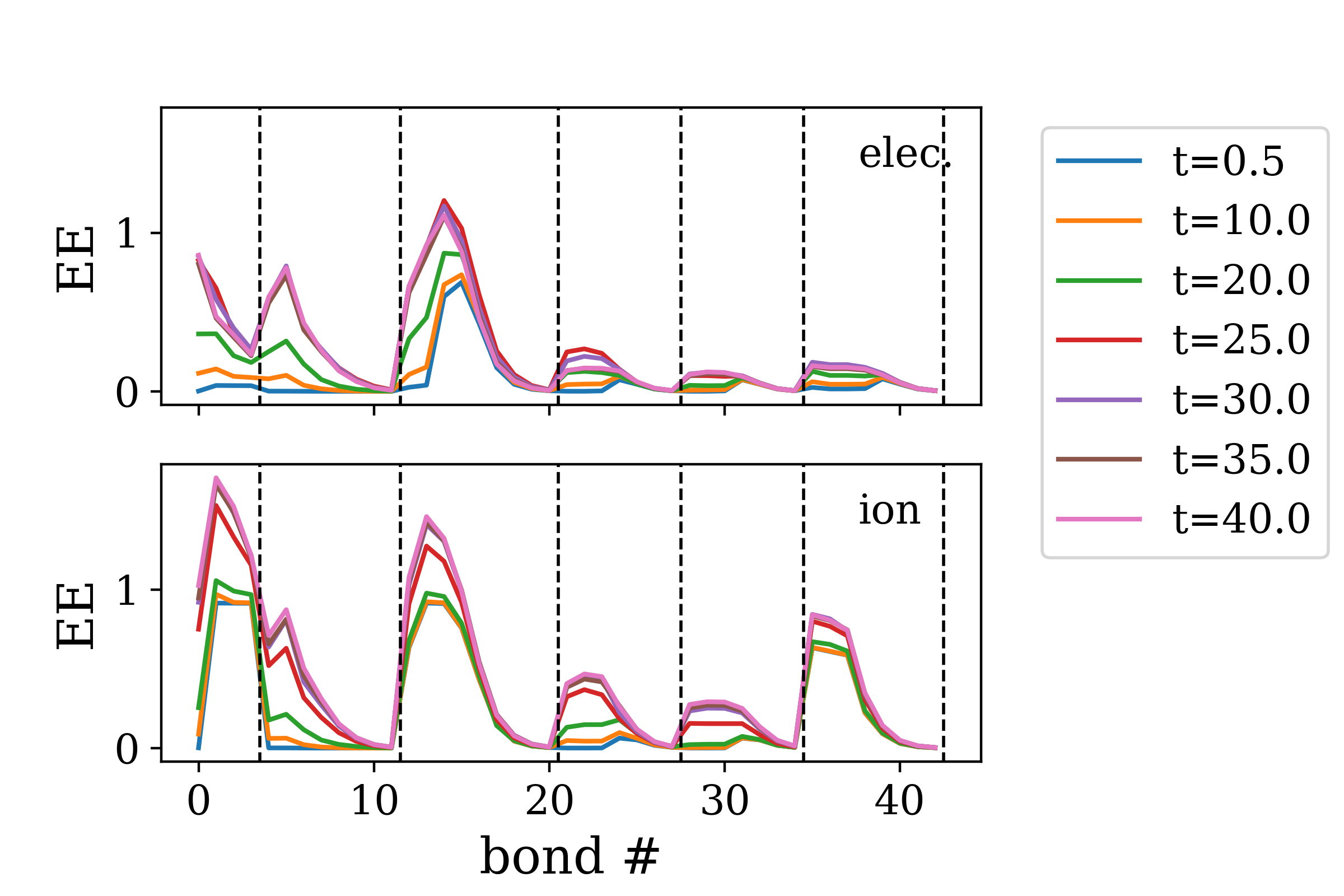}
    } 
    \subfloat{
        \includegraphics[width=0.50\linewidth,trim={-0.25cm 0cm 0cm 0cm}] 
            % {figs/reconn/EE_EB_1.png}
            {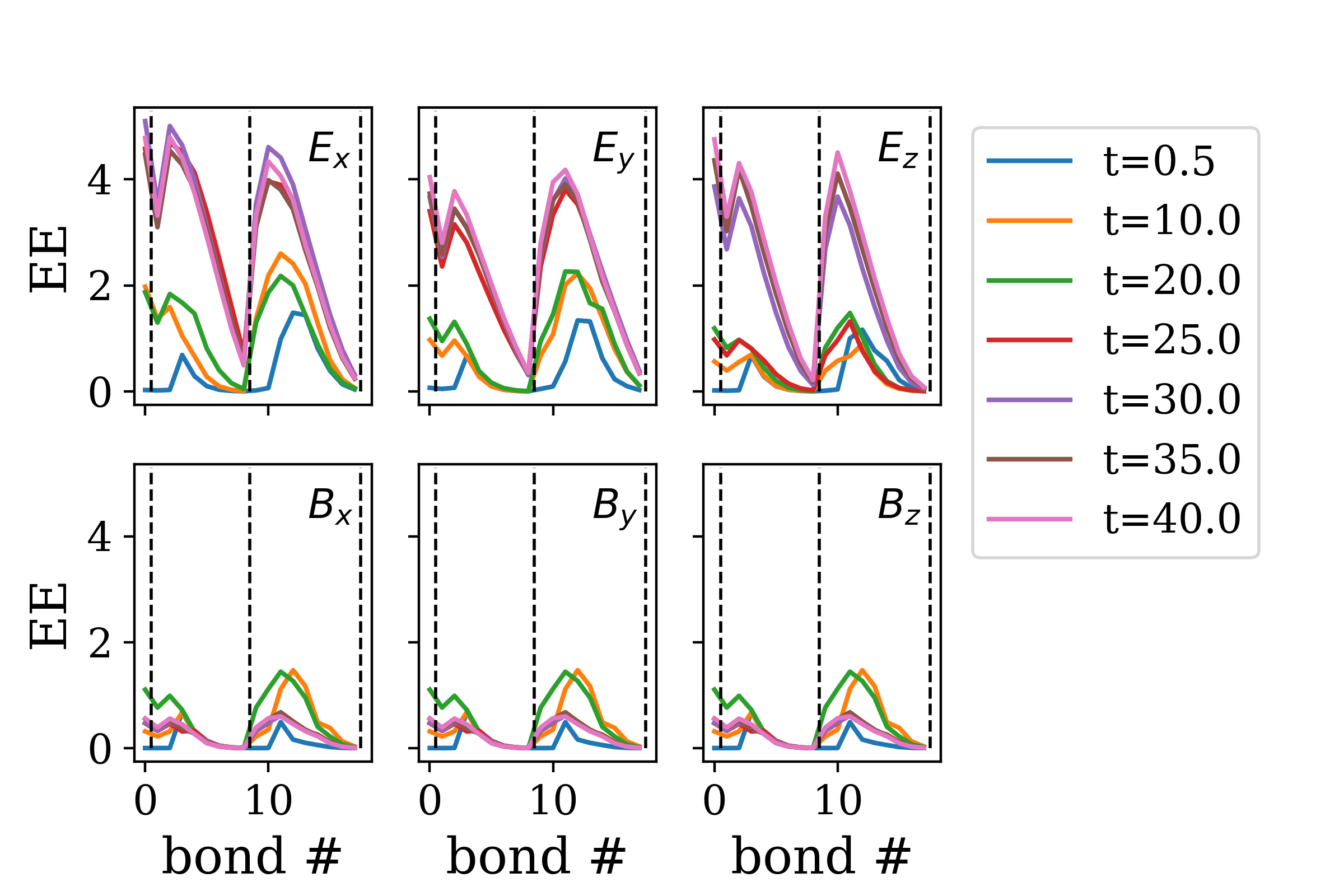}
    }
    \caption{Entanglement entropy at each bond for the reconnection problem. Bond number is ordered by bonds in the spine, and then bonds in each branch.}
    \label{fig:EE_reconn}
\end{figure}

% The von Neumann or bipartite entanglement entropy (EE) is one measure of efficiency of the QTC representation. If considering the $j^\text{th}$ bond in the tensor network, it is given by
% \begin{equation}
%     EE^{(j)} = - \sum_i \tilde{\sigma}_i^2 \log_2 (\tilde{\sigma}_i^2)
% \end{equation}
% where $\tilde{\sigma}$ are the singular values at bond $j$ when the tensor network's center of center of orthogonality is located at bond $j$. (The singular values are normalized such that $\sum_i \tilde{\sigma}_i^2 = 1$.)
%
We compute the EE at each bond in the QTC for the ion distribution, electron distribution, electric fields, and magnetic fields. Results for the Orszag-Tang vortex are shown in Fig.~\ref{fig:EE_ot}, and results for the reconnection problem are shown in Fig.~\ref{fig:EE_reconn}. Interestingly, in both cases, the EE of the ion and electron distributions remain relatively small, and the EE of the electric and magnetic fields tend to grow. Some of that is likely due to noise, as suggested by the sudden increase in EE and the visible increase of noise in the fields themselves (see the Section VI.A.2 and VI.B.2 SI for these figures). 
Most of the noise appears to be injected through the electric field. This suggests that the issue is likely not due to insufficient bond dimension but rather numerical noise injected through various stages of the solver. This may be because our current QTT format solvers (i.e.,~TDVP and DMRG) are prone to introducing noise, but it also may be due to the simulation parameters used---the time step may have been too large for the Crank-Nicolson scheme, and the grid resolution used still is not fine enough to resolve the electron Debye length.
% Since most of the noise seems to be injected through the electric field, this suggests that our solver for Maxwell's equations needs to be improved, or that the time step used for the Crank-Nicolson scheme may be too large.
% Fortunately, it should be possible to remedy both types of issues in the future.
\edit{
We intend to further investigate and characterize potential sources of numerical noise so that these issues can be remedied in the future. 
}

\edit{Fortunately, it should be possible to improve upon our basic implementation to obtain better results. For example,} as with most local optimization schemes, the DMRG solver used for the implicit time evolution of Maxwell's equations is prone to slow convergence and only finding local minima, particularly when the bond dimension is small. Our current implementation is very simple; using a preconditioner and a better initial guess will likely improve performance. One might also benefit from adding additional constraints, such as one to preferentially find smooth results, to help improve convergence. Whether or not these constraints can be also be used to improve numerical stability of the implicit solver as well is also worth more in-depth investigation.  % Improving this solver, and extending it beyond the QTT geometry, would be beneficial. Perhaps one could also develop a local time evolution scheme for Maxwell's equations. \textcolor{red}{(maybe remove?)}
Alternatively, one could include a filter or artificial dissipation that removes grid-scale noise. Preliminary results of the Orszag-Tang vortex calculation performed with an un-centered Crank-Nicolson scheme are shown in Section VI.B.3 of the SI.

In addition to improving our solver for Maxwell's equations, we also need to improve our understanding of the TDVP time evolution scheme.
In particular, while single-site TDVP time evolution is used in this work with promising results, local tangent-space time evolution schemes need be examined further, since they have their limitations. Most notably, TDVP (both 1-site and 2-site) is known to sometimes yield incorrect dynamics \citep{yang2020time}, even as one reduces the time step. Understanding the lack of convergence with respect to time step and grid resolution must be addressed in order to trust results for problems we do not know the solution to. 
Additionally, when using TDVP with RK4, there still is a time-step constraint, but we currently do not have an estimate for it. We also need to extend TDVP to non-Hermitian systems to account for non-periodic boundary conditions, so that a greater variety of problems can be considered. \edit{This should be straightforward, as any non-Hermitian system can be embedded in a Hermitian system that is twice the original system size.} 
It may be possible to improve our basic implementation to better ensure conservation laws. For example, for simulation of closed quantum systems, the TDVP algorithm naturally conserves system energy and system norm, because the Hamiltonian describes both system energy and dynamics. Though this is not true in the case of the Vlasov equation, there may be an alternative formulation that allows one to take advantage of these conservation properties. There also may be ways to impose additional conservation properties, as is done in the dynamical low-rank literature \citep{einkemmer_quasi-conservative_2019, einkemmer_low-rank_2020, Einkemmer2021conserve}.

% There are also many practical improvements to be made. 
As of now, our code runs serially on a single CPU node, with the exception of the internal threading performed by Python's numpy package. As such, while the amount of resources required to perform and store these calculations is much less than conventional methods, the overall run-time is likely longer than that of an optimized solver. % Finding ways to trade savings in computational resources for overall run-time
% 
%While our code could likely be optimized, 
\edit{
The most expensive steps in the calculation are the advection in velocity space, and the DMRG solver of the EM fields when many iterations are required. As mentioned earlier in this section, the DMRG solver can be improved by using a preconditioner to achieve faster convergence. For the advection step, since the bulk of the cost arises during tensor contraction, the runtime could be improved by distributing those calculations.
}
However, a more fundamental problem is that the QTN algorithms discussed here are inherently serial, thus making parallel calculations unattractive. The QTC geometry may be more amenable for parallelization, with algorithms modified to allow one to work with each branch independently. However, computations on the spine generally have worse scaling than the branches, and would face the same challenges for parallelization as QTTs. \edit{While parallelized algorithms for tensor trains do exist \citep{secular2020parallel}, they must be carefully investigated and tested, as they may introduce additional errors and are prone to numerical instabilities.} The parallelized algorithms likely will also be less efficient than serial algorithms, requiring repetitive calculations. For example, instead of computing the necessary information for just one branch and then building the information for the remaining branches as one proceeds through the algorithm, one may need to compute the required information for each branch all at once. However, given the low computational costs of the serial algorithm for modest bond dimension, increasing the cost linearly with problem dimension for parallelization is affordable.

%%%% compare to others
One might argue that the observed efficiency and compressibility of the QTN representation is due to the inefficiency of the grid-based representation, and one could instead use a spectral or finite element basis. 
% Also, while these simulations yield reasonable results, not all physical length scales, such as the electron Debye length, are resolved. One would need to increase resolution even further to accurately capture those phenomena.
\citet{Juno2018discontinuous} reports a discontinuous Galerkin finite element calculation of the Orszag-Tang vortex with similar parameters. They use a $40^2\times 8^3$ grid in 2D3V and a second-order polynomial basis, resulting in $N_p = 112$ degrees of freedom per grid cell. Thus, the distribution function is represented using $N N_p$  ($9 \times 10^7$) degrees of freedom (d.o.f.), which is greater than the approximate upper bound of $5D^3$ for the QTC ansatz ($1.3 \times 10^6$ d.o.f. for $D=64$). % (The run-time cost is expected to scale like $\mathcal{O}(N N_p^2)$ per time step.) 
By this rough comparison, the QTC ansatz indeed provides a smaller but approximate representation of the distribution function. For reference, newer finite element simulations on a $112^2 \times 16^3$ grid up to a time of $75\Omega_{c,p}^{-1}$ used 10k node-hours \citep{Juno2018discontinuous, Hakim2020conservative, Hakim2020alias}. 

In comparison to non-quantized low-rank methods, the proposed QTN method exhibits a much more modest reduction in computational resources for the problems discussed above. One advantage of the QTN methods is the ability to use time-steps \tofix{somewhat} larger than the CFL limit because of the local time evolution scheme, without having to perform semi-Lagrangian time evolution. Furthermore, QTNs are more likely to demonstrate a bigger advantage for problems requiring orders of magnitude more grid points along each dimension, as will be necessary for more realistic problems (e.g., with the proper mass ratio).
A more thorough investigation on the efficiency of the QTN format for different simulation parameters needs to be performed. Additionally, testing the proposed algorithm for more complex problems, including those in different coordinate systems and  \edit{ especially those that are currently not tractable using conventional methods}, is certainly worth pursuing. 
The QTN format is also not limited to a real-space, grid-based interpretation. Since the convolution operation can be performed efficiently within the QTN format \citep{kazeev2013multilevel}, one can also implement an efficient QTN solver with spectral bases. 
\edit{Spectral representations are often more compact than grid-based representations, so the degree to which QTNs can further compress these representations should be measured.}
Existing work also discusses finite-element solvers in the QTN format \citep{Kazeev2016qttfem, Kazeev2018quantizedfem, Kazeev2020quantizedfem, kornev_numerical_2023}. Performing a careful comparison to understand the benefits and caveats of using different 
% \edit{(particularly, more compact)} 
representations would be of practical interest and may improve our understanding of QTNs more generally.

%%%
There are many immediate and practical uses for QTN-based solvers. 
Because the low-rank approximations in the QTN format are solely numerical, one is able to solve the Vlasov equation at reduced cost without relying on assumptions that would limit the range of problems that can be studied. As such, one could perform a rough parameter sweep using the proposed solver before identifying a specific regime in which more accurate calculations are needed. The QTN solvers can also help with generating approximate but physically accurate training data for machine learning methods.

%%%%

Lastly, we also hope that the QTN representation can act as another tool for understanding turbulence. Drawing from ideas in quantum information theory, one can better understand information flow between different length scales. Additionally, by connecting QTNs to other concepts like wavelets \citep{oseledets_algebraic_2011} and renormalization group, one may be able to build a consistent picture between different interpretations.

% \textcolor{blue}{Would it be worth (and would you be comfortable with) speculating a little bit here about how the fact that QTTs work for the Vlasov equation suggests that they may be an efficient data representation for the training of neural networks?}

% In practice, for more complex simulations, even if the chosen QTN ansatz is an (at least semi-)efficient representation of the PDE in question, larger bond dimensions than those used here may be needed. Thus, it may also be worth performing the basic linear algebra operations in a distributed fashion. 

% This could be mitigated by requiring the spine to have a smaller bond dimension than the branches. 
% Judging from the not quantized tensor-train literature, ranks are ... (can't find any on Vlasov-Maxwell? Allman-Rahn only has LRD in v-space.)

% Better comb geometry algorithms?

\section{Conclusion}
Understanding the nonlinear behavior of collisionless plasmas requires first-principles descriptions such as provided by the coupled Vlasov-Maxwell system of equations. 
However, solving the Vlasov equation with traditional numerical algorithms is extremely \edit{resource-intensive}, so an alternative approach that offers significantly reduced computational cost but nonetheless captures the important physical effects is very much needed.
To address this problem, in this work we pioneer a grid-based Vlasov-Maxwell solver utilizing a quantum-inspired approach: the quantized tensor network (QTN) format.
Moreover, we use semi-Lagrangian time integration and a QTN time evolution scheme, both of which allow for timesteps \tofix{somewhat} larger than the CFL limit. % \tofix{\sout{, thus overcoming a major issue plaguing high-resolution calculations}}.
The cost of our algorithm at worst scales roughly like $\mathcal{O}(D^4)$, where $D$ is the rank or bond dimension of the QTN, and determines the accuracy of the calculation. In the test cases considered, we find that QTNs of at most $D=64$ are sufficient for capturing the expected dynamics with reasonable accuracy, even though the simulation is performed on grids of around $2^{36}$ grid points in total. 
%
%% some blurb about QI and plasma physics
{% \color{blue}
This success suggests that the quantum-inspired representation can compress the Vlasov equation 
without significant loss of relevant physical effects,
% without losing any kinetic effects, 
and thus may be a useful tool for interpreting and and exploiting the multi-scale properties of plasma dynamics. % for example, potentially leading to quantum-inspired reduced-order models.
More practically, the low-rank calculation offers
}
%
% this corresponds to 
approximately a $10^5$ reduction in the number of degrees of freedom compared to the corresponding traditional finite difference representation, and a $10^2$ reduction with respect to a comparable finite element calculation. 
Due to the significant reduction in degrees of freedom and larger time-steps, these simulations require significantly fewer resources than conventional numerical methods. % Critically, despite these much greater efficiencies, 
Critically, despite achieving much greater efficiencies,
the key physical processes of the test problems investigated are retained.
Therefore, we expect that QTN-based algorithms will enable Vlasov simulations that would have remained out of reach with conventional numerical methods, and thus accelerate the pace of discovery in plasma physics and associated fields.

\section{Code Availability}
{Code can be made available upon request.}

\section{Supplementary Information}
{Supplementary material is available at \href{https://doi.org/10.1017/jfm.2019...}{https://doi.org/10.1017/jfm.2019...}}

\section{Declaration of Interest}
{The authors report no conflict of interest.}

\section{Funding}
{The authors were supported by awards DE-SC0020264 and DE-SC0022012 from the Department of Energy. 
}

\section{Acknowledgments}
{EY thanks Jimmy Juno for discussions regarding the cost of the discrete-Galerkin finite element Vlasov solver in Gkeyll.}

\section{Author ORCIDs}
{E. Ye, https://orcid.org/0000-0001-9694-568X; N. F. Loureiro, https://orcid.org/0000-0001-9755-6563}.

\bibliographystyle{jpp}
\bibliography{main} % , references}

\newpage


\section*{Supplementary material}

# Quantized Tensor Networks for the Vlasov-Maxwell Equations: Supplementary Information

## I. EXAMPLES OF QUANTIZED TENSOR TRAINS

As stated in the main text, the quantized tensor train format is

$$h_i = h(x_i) \cong h(i_1, i_2, ..., i_L) = \sum_{\alpha_1=1}^{r_1} ... \sum_{\alpha_{L-1}=1}^{r_{L-1}} \mathsf{M}_{\alpha_1}^{(1)}(i_1) \mathsf{M}_{\alpha_1,\alpha_2}^{(2)}(i_2) ... \mathsf{M}_{\alpha_{L-1}}^{(L)}(i_L), \tag{1}$$

and the quantized tensor train operator (QTT-Operator) format is

$$\mathsf{O}(x_i', x_i) \cong \mathsf{O}((o_1, o_2, ..., o_L), (i_1, i_2, ..., i_L)) = \sum_{\alpha_1=1}^{r_1} ... \sum_{\alpha_{L-1}=1}^{r_{L-1}} \mathsf{O}_{\alpha_1}^{(1)}(o_1, i_1) \mathsf{O}_{\alpha_1,\alpha_2}^{(2)}(o_2.i_2) ... \mathsf{O}_{\alpha_{L-1}}^{(L)}(o_L, i_L). \tag{2}$$

### A. QTT Operators for the Vlasov Equation

In this section, we explicitly write the QTT representations of the operators used in the Vlasov equation.

#### 1. QTT-Operator for first derivative

The QTT-Operator for the first derivative along one axis can be written explicitly, assuming uniform grid points and a finite difference scheme. With periodic boundary conditions and a centered second-order stencil, it is

$$\frac{\partial}{\partial x} \approx \frac{1}{\Delta x} \begin{bmatrix} \mathbb{I} & S^- + S^+ & S^+ + S^- \end{bmatrix} \begin{bmatrix} \mathbb{I} & S^- & S^+ \\ 0 & S^+ & 0 \\ 0 & 0 & S^- \end{bmatrix} \begin{bmatrix} \mathbb{I} & S^- & S^+ \\ 0 & S^+ & 0 \\ 0 & 0 & S^- \end{bmatrix} ... \begin{bmatrix} \mathbb{I} & S^- & S^+ \\ 0 & S^+ & 0 \\ 0 & 0 & S^- \end{bmatrix} \begin{bmatrix} 0.5(S^- - S^+) \\ 0.5S^+ \\ -0.5S^- \end{bmatrix}, \tag{3}$$

where

$$\mathbb{I} = \begin{bmatrix} 1 & 0 \\ 0 & 1 \end{bmatrix}, \qquad S^+ = \begin{bmatrix} 0 & 0 \\ 1 & 0 \end{bmatrix}, \qquad S^- = \begin{bmatrix} 0 & 1 \\ 0 & 0 \end{bmatrix}.$$

In this notation, the matrices from left to right are the tensors $O^{(1)}$ to $O^{(L)}$ in Eq. (2). For the $p^{\text{th}}$ tensor, the dimensions of the blocks are labeled by $o_p$ and $i_p$, while the dimensions of the explicitly written matrices are the virtual bonds $\alpha_{p-1}$ and $\alpha_p$.

#### 2. Application of QTT-Operators to QTTs

**Example 1.** As a brief aside, consider a system with 8 grid points and suppose our vector of interest is all ones. The corresponding QTT is rank 1, and given by

$$h = [1, 1, 1, 1, 1, 1, 1, 1] \cong \mathbf{1} \otimes \mathbf{1} \otimes \mathbf{1} = \begin{bmatrix} \mathbf{1} \end{bmatrix} \begin{bmatrix} \mathbf{1} \end{bmatrix} \begin{bmatrix} \mathbf{1} \end{bmatrix} \qquad \text{where } \mathbf{1} = \begin{bmatrix} 1 \\ 1 \end{bmatrix}.$$

Applying the finite difference operator in Eq. (3) to this QTT, we obtain the following QTT:

$$\frac{\partial}{\partial x} h \approx \frac{1}{\Delta x} \begin{bmatrix} \mathbb{I}\mathbf{1} & (S^+ + S^-)\mathbf{1} & (S^+ + S^-)\mathbf{1} \end{bmatrix} \begin{bmatrix} \mathbb{I}\mathbf{1} & S^-\mathbf{1} & S^+\mathbf{1} \\ 0 & S^+\mathbf{1} & 0 \\ 0 & 0 & S^-\mathbf{1} \end{bmatrix} \begin{bmatrix} 0.5(S^- - S^+)\mathbf{1} \\ 0.5S^+\mathbf{1} \\ -0.5S^-\mathbf{1} \end{bmatrix}. \tag{4}$$

By combining blocks that are the same within each matrix, the QTT can be simplified to

$$\frac{\partial}{\partial x}h \approx 0.5 \begin{bmatrix} \mathbf{1} & \mathbf{1} & \mathbf{1} \end{bmatrix} \begin{bmatrix} \mathbf{1} & \begin{bmatrix} 1 \\ 0 \\ 0 \\ 1 \end{bmatrix} & \begin{bmatrix} 0 \\ 1 \\ 0 \\ 0 \end{bmatrix} \\ 0 & \begin{bmatrix} 0 \\ 1 \end{bmatrix} & 0 \\ 0 & 0 & \begin{bmatrix} 1 \\ 0 \end{bmatrix} \end{bmatrix} \begin{bmatrix} \begin{bmatrix} 1 \\ -1 \end{bmatrix} & \begin{bmatrix} 0 \\ 1 \end{bmatrix} & \begin{bmatrix} -1 \\ 0 \end{bmatrix} \end{bmatrix} = 0.5 \begin{bmatrix} \mathbf{1} \end{bmatrix} \begin{bmatrix} \mathbf{1} & \mathbf{1} & \mathbf{1} \end{bmatrix} \begin{bmatrix} \begin{bmatrix} 1 \\ -1 \\ 0 \\ 1 \\ -1 \\ 0 \end{bmatrix} \end{bmatrix} = 0.5 \begin{bmatrix} \mathbf{1} \end{bmatrix} \begin{bmatrix} \mathbf{1} \end{bmatrix} \begin{bmatrix} 0 \end{bmatrix} = 0 , \quad (5)$$

which is the expected result.

**Example 2.** Now consider the vector $h = [-4, -3, -2, -1, 0, 1, 2, 3]$. There exists an analytical expression for QTTs representing the linear function $av + b$ on a uniform grid. Let the grid be defined along the interval $[-d/2, d/2]$, such that the $i^{\text{th}}$ grid point is $v_i = -d/2 + di/N$. For a grid of $N = 2^L$ grid points ($\vec{v}$), the QTT of length $L$ is

$$a\vec{v} + b = \begin{bmatrix} \mathbf{1} & 2^{-1}ad\vec{n} + (-ad/2 + b)\mathbf{1} \end{bmatrix} \begin{bmatrix} \mathbf{1} & 2^{-2}ad\vec{n} \\ 0 & \mathbf{1} \end{bmatrix} \cdots \begin{bmatrix} \mathbf{1} & 2^{-(L-1)}ad\vec{n} \\ 0 & \mathbf{1} \end{bmatrix} \begin{bmatrix} 2^{-L}ad\vec{n} \\ \mathbf{1} \end{bmatrix} \quad \text{where} \quad \vec{n} = \begin{bmatrix} 0 \\ 1 \end{bmatrix} . \quad (6)$$

Thus, taking $L = 3$ and plugging in $a = 1$, $b = 0$, and $d = 8$ into Eq. (6), we arrive at

$$h = [-4, -3, -2, -1, 0, 1, 2, 3] \cong \begin{bmatrix} \begin{bmatrix} 1 \\ 1 \end{bmatrix} & \begin{bmatrix} -4 \\ 0 \end{bmatrix} \end{bmatrix} \begin{bmatrix} \begin{bmatrix} 1 \\ 1 \end{bmatrix} & \begin{bmatrix} 0 \\ 2 \\ 1 \\ 1 \end{bmatrix} \\ 0 & \begin{bmatrix} 0 \\ 2 \\ 1 \\ 1 \end{bmatrix} \end{bmatrix} \begin{bmatrix} \begin{bmatrix} 0 \\ 1 \\ 1 \\ 1 \end{bmatrix} \end{bmatrix} .$$

After applying the finite difference operator in Eq. (3), we obtain the QTT (rank 6, with 3 tensor cores)

$$\frac{\partial}{\partial x}h \approx \frac{1}{\Delta x} \begin{bmatrix} \begin{bmatrix} \mathbb{I} \begin{bmatrix} 1 \\ 1 \end{bmatrix} & (S^+ + S^-) \begin{bmatrix} 1 \\ 1 \end{bmatrix} \end{bmatrix} & \begin{bmatrix} \mathbb{I} \begin{bmatrix} -4 \\ 0 \end{bmatrix} & (S^+ + S^-) \begin{bmatrix} -4 \\ 0 \end{bmatrix} & (S^+ + S^-) \begin{bmatrix} -4 \\ 0 \end{bmatrix} \end{bmatrix} \end{bmatrix} \times$$

$$\begin{bmatrix} \begin{bmatrix} \mathbb{I} \begin{bmatrix} 1 \\ 1 \end{bmatrix} & S^- \begin{bmatrix} 1 \\ 1 \end{bmatrix} & S^+ \begin{bmatrix} 1 \\ 1 \end{bmatrix} \\ 0 & S^+ \begin{bmatrix} 1 \\ 1 \end{bmatrix} & 0 \\ 0 & 0 & S^- \begin{bmatrix} 1 \\ 1 \end{bmatrix} \end{bmatrix} & \begin{bmatrix} \mathbb{I} \begin{bmatrix} 0 \\ 2 \end{bmatrix} & S^- \begin{bmatrix} 0 \\ 2 \end{bmatrix} & S^+ \begin{bmatrix} 0 \\ 2 \end{bmatrix} \\ 0 & S^+ \begin{bmatrix} 0 \\ 2 \end{bmatrix} & 0 \\ 0 & 0 & S^- \begin{bmatrix} 0 \\ 2 \end{bmatrix} \end{bmatrix} \\ & \begin{bmatrix} \mathbb{I} \begin{bmatrix} 1 \\ 1 \end{bmatrix} & S^- \begin{bmatrix} 1 \\ 1 \end{bmatrix} & S^+ \begin{bmatrix} 1 \\ 1 \end{bmatrix} \\ 0 & S^+ \begin{bmatrix} 1 \\ 1 \end{bmatrix} & 0 \\ 0 & 0 & S^- \begin{bmatrix} 1 \\ 1 \end{bmatrix} \end{bmatrix} \\ 0 & \end{bmatrix} \times \begin{bmatrix} 0.5(S^- - S^+) \begin{bmatrix} 0 \\ 1 \end{bmatrix} \\ 0.5S^+ \begin{bmatrix} 0 \\ 1 \end{bmatrix} \\ -0.5S^- \begin{bmatrix} 0 \\ 1 \end{bmatrix} \\ 0.5(S^- - S^+) \begin{bmatrix} 1 \\ 1 \end{bmatrix} \\ 0.5S^+ \begin{bmatrix} 1 \\ 1 \end{bmatrix} \\ -0.5S^- \begin{bmatrix} 1 \\ 1 \end{bmatrix} \end{bmatrix}$$

$$= 0.5 \begin{bmatrix} \begin{bmatrix} \begin{bmatrix} 1 \\ 1 \end{bmatrix} \end{bmatrix} & \begin{bmatrix} \begin{bmatrix} -4 \\ 0 \end{bmatrix} & \begin{bmatrix} 0 \\ -4 \end{bmatrix} \end{bmatrix} \end{bmatrix} \begin{bmatrix} \begin{bmatrix} \begin{bmatrix} 1 \\ 1 \end{bmatrix} \\ 0 \end{bmatrix} & \begin{bmatrix} \begin{bmatrix} 0 \\ 2 \\ 1 \\ 1 \end{bmatrix} & \begin{bmatrix} 2 \\ 0 \\ 1 \\ 0 \end{bmatrix} & \begin{bmatrix} 2 \\ 0 \\ 0 \\ 1 \end{bmatrix} \\ \begin{bmatrix} 0 \\ 2 \\ 1 \\ 0 \end{bmatrix} & \begin{bmatrix} 1 \\ 0 \\ 1 \\ 0 \end{bmatrix} \end{bmatrix} \end{bmatrix} \begin{bmatrix} \begin{bmatrix} \begin{bmatrix} 0 \\ 1 \\ -1 \\ 0 \\ 1 \\ -1 \\ 0 \end{bmatrix} \end{bmatrix} \end{bmatrix}$$

where we use the same simplification techniques as before. Now, removing the zero terms and using the fact that

$\begin{bmatrix} 1 \\ 1 \end{bmatrix} = \begin{bmatrix} 0 \\ 1 \end{bmatrix} + \begin{bmatrix} 1 \\ 0 \end{bmatrix}$ and $\begin{bmatrix} 1 \\ -1 \end{bmatrix} = \begin{bmatrix} 0 \\ 1 \end{bmatrix} + \begin{bmatrix} 1 \\ 0 \end{bmatrix}$, we can further simplify the QTT and obtain the expected result:

$$
\begin{aligned}
\ldots &= 0.5 \left[ \left[ \begin{bmatrix} 1 \\ 0 \end{bmatrix} \begin{bmatrix} 0 \\ 1 \end{bmatrix} \right] \left[ \begin{bmatrix} 1 \\ 0 \end{bmatrix} \begin{bmatrix} 0 \\ 1 \end{bmatrix} \right] \right] \left[ \begin{bmatrix} \begin{bmatrix} 0 \\ 2 \\ 0 \\ 2 \end{bmatrix} & \begin{bmatrix} 0 \\ 2 \\ 0 \\ 2 \end{bmatrix} & \begin{bmatrix} 2 \\ 0 \\ 2 \\ 0 \end{bmatrix} & \begin{bmatrix} 2 \\ 0 \\ 2 \\ 0 \end{bmatrix} \end{bmatrix} \\
&\qquad \begin{bmatrix} \begin{bmatrix} -4 \\ -4 \end{bmatrix} & \begin{bmatrix} -4 \\ -4 \end{bmatrix} & \begin{matrix} -4 \\ 0 \\ 0 \\ -4 \end{matrix} & \begin{matrix} 0 \\ -4 \\ -4 \\ 0 \end{matrix} \end{bmatrix} \end{bmatrix} \left[ \begin{bmatrix} \begin{bmatrix} 1 \\ 0 \end{bmatrix} \\ \begin{bmatrix} 0 \\ -1 \end{bmatrix} \\ \begin{bmatrix} 0 \\ 1 \end{bmatrix} \\ \begin{bmatrix} -1 \\ 0 \end{bmatrix} \end{bmatrix} \right] \\
&= 0.5 \left[ \begin{bmatrix} 1 \\ 0 \end{bmatrix} \begin{bmatrix} 0 \\ 1 \end{bmatrix} \right] \begin{bmatrix} \begin{bmatrix} 2 \\ 2 \end{bmatrix} & \begin{bmatrix} 6 \\ -2 \end{bmatrix} \\ \begin{bmatrix} 2 \\ -6 \end{bmatrix} & \begin{bmatrix} -2 \\ -2 \end{bmatrix} \end{bmatrix} \begin{bmatrix} \begin{bmatrix} 0 \\ 1 \end{bmatrix} \\ \begin{bmatrix} -1 \\ 0 \end{bmatrix} \end{bmatrix} = [-3, 1, 1, 1, 1, 1, 1, -3]
\end{aligned}
\tag{7}
$$

### 3. Element-wise Multiplication

The Vlasov equation also requires performing elemental multiplication of generic functions $h_1(x)h_2(x)$. This computation is performed by writing $h_1(x)$ as a diagonal matrix and $h_2(x)$ as a vector, and then performing matrix-vector multiplication.

For the spatial advection term, we require diagonalizing $h_1(v) = v$. Analogous to Eq. (6), a matrix with $a\vec{v} + b$ along its diagonal (assuming a uniform grid along the interval $[-d/2, d/2)$) can be written as an QTT-Operator of bond dimension 2. For a a grid with $N = 2^L$ grid points, we obtain the QTT-Operator of length $L$,

$$
\mathrm{diag}(a\vec{v}+b) = \begin{bmatrix} \mathbb{I} & 2^{-1}ad\hat{n} + (-ad/2+b)\,\mathbb{I} \end{bmatrix} \begin{bmatrix} \mathbb{I} & 2^{-2}ad\hat{n} \\ 0 & \mathbb{I} \end{bmatrix} \cdots \begin{bmatrix} \mathbb{I} & 2^{-(L-1)}ad\hat{n} \\ 0 & \mathbb{I} \end{bmatrix} \begin{bmatrix} 2^{-L}ad\hat{n} \\ \mathbb{I} \end{bmatrix} \quad \text{where} \quad \hat{n} = \begin{bmatrix} 0 & 0 \\ 0 & 1 \end{bmatrix}. \tag{8}
$$

For the velocity advection term, we require performing elemental multiplication of the distribution function $f$ with components of the Lorentz force $F$. The force must be computed at each time step so we do not have an analytical form. But, since $F$ is already represented in the QTT format, the tensors of the QTT-Operator representing $\mathrm{diag}(F)$ are simply

$$
O^{(p)}_{\alpha_{p-1}, \alpha_p}(o_p, i_p) = \sum_{x_p \in \{0,1\}} M^{(p)}_{\alpha_{p-1}, \alpha_p}(x_p) \delta_{x_p, i_p, o_p}, \tag{9}
$$

where $M^{(p)}$ is the $p^{\text{th}}$ tensor in the original QTT, and $\delta_{ijk}$ as the multi-index Kronecker delta function.

### 4. Multi-dimensional operators

While the operators in the above discussion act only on one dimension, they can still be represented in a multi-dimensional space. To do this, one simply pads the original QTT with a tensor train of identity matrices for the remaining dimensions. For example, for a 2-D system with dimensions along $x$ and $v$, if a sequential ordering of the tensors is used, the QTT representation of operator $O: f(x,v) \mapsto \frac{\partial}{\partial x} f(x,v)$ would look like

$$
\frac{\partial}{\partial x} \otimes \mathbb{I}_v = \underbrace{\begin{bmatrix} \mathbb{I} & S^+ + S^- & S^+ + S^- \end{bmatrix} \begin{bmatrix} \mathbb{I} & S^- & S^+ \\ 0 & S^+ & 0 \\ 0 & 0 & S^- \end{bmatrix} \begin{bmatrix} \mathbb{I} & S^- & S^+ \\ 0 & S^+ & 0 \\ 0 & 0 & S^- \end{bmatrix} \cdots \begin{bmatrix} \mathbb{I} & S^- & S^+ \\ 0 & S^+ & 0 \\ 0 & 0 & S^- \end{bmatrix} \begin{bmatrix} 0.5(S^- - S^+) \\ 0.5 S^+ \\ -0.5 S^- \end{bmatrix}}_{x\text{-axis tensors}} \underbrace{\begin{bmatrix} \mathbb{I} \end{bmatrix} \ldots \begin{bmatrix} \mathbb{I} \end{bmatrix}}_{y\text{-axis tensors}}, \tag{10}
$$

where $\mathbb{I}_v$ is the identity operator acting on dimension $v$. (The multidimensional QTC is constructed similarly, except one must now also define the spine tensors to be of size $1 \times 1 \times 1$ and value 1.)

Similarly, the representation of operator $O: f(x,v) \mapsto vf(x,v)$ is

$$
\mathbb{I}_x \otimes \mathrm{diag}(v) = \underbrace{\begin{bmatrix} \mathbb{I} \end{bmatrix} \ldots \begin{bmatrix} \mathbb{I} \end{bmatrix}}_{x\text{-axis tensors}} \underbrace{\begin{bmatrix} \mathbb{I} & v_{\max}(\hat{n} - \mathbb{I}) \end{bmatrix} \begin{bmatrix} \mathbb{I} & 2^{-1}v_{\max}\hat{n} \\ 0 & \mathbb{I} \end{bmatrix} \cdots \begin{bmatrix} 2^{-L+1}v_{\max}\hat{n} \\ \mathbb{I} \end{bmatrix}}_{y\text{-axis tensors}}, \tag{11}
$$

where we take the grid points along the $v$-axis to be uniformly spaced along the interval $[-v_{\max}, v_{\max}]$.

These 1-D operators can be combined via the tensor train analog of matrix multiplication. For example, to obtain operator $O: f(x, v) \mapsto v \frac{\partial}{\partial x} f(x, v)$, one would apply the QTT-operator defined in Eq. (10) to that defined in Eq. (11) (or vice versa; the operators commute in this case), yielding

$$
\frac{\partial}{\partial x} \otimes \mathrm{diag}(v) = \underbrace{\begin{bmatrix} \mathbb{I} & S^+ + S^- & S^+ + S^- \end{bmatrix} \begin{bmatrix} \mathbb{I} & S^- & S^+ \\ 0 & S^+ & 0 \\ 0 & 0 & S^- \end{bmatrix} \begin{bmatrix} \mathbb{I} & S^- & S^+ \\ 0 & S^+ & 0 \\ 0 & 0 & S^- \end{bmatrix} \ldots \begin{bmatrix} \mathbb{I} & S^- & S^+ \\ 0 & S^+ & 0 \\ 0 & 0 & S^- \end{bmatrix} \begin{bmatrix} 0.5(S^- - S^+) \\ 0.5 S^+ \\ -0.5 S^- \end{bmatrix}}_{x\text{-axis tensors}}
$$

$$
\otimes \underbrace{\begin{bmatrix} \mathbb{I} & v_{\max}(\hat{n} - \mathbb{I}) \end{bmatrix} \begin{bmatrix} \mathbb{I} & 2^{-1} v_{\max} \hat{n} \\ 0 & \mathbb{I} \end{bmatrix} \ldots \begin{bmatrix} 2^{-L+1} v_{\max} \hat{n} \\ \mathbb{I} \end{bmatrix}}_{y\text{-axis tensors}}. \quad (12)
$$

## II. RELEVANT BASIC CONCEPTS IN TENSOR NETWORKS

In this section, we will introduce key tensor network concepts relevant for this work. This is not meant to be a thorough introduction; we instead refer the reader to other works [1–3]. Throughout the supplementary information, we will use the same notation introduced in the main paper to describe discretized functions represented as quantized tensor trains (QTTs)

$$
h_i = h(x_i) \cong h(i_1, i_2, ..., i_L) = \sum_{\alpha_1=1}^{r_1} ... \sum_{\alpha_{L-1}=1}^{r_{L-1}} H^{(1)}_{\alpha_1}(i_1) H^{(2)}_{\alpha_1, \alpha_2}(i_2) \ldots H^{(L)}_{\alpha_{L-1}}(i_L), \quad (13)
$$

or as quantized tree tensor networks with a comb geometry (QTCs) (see Fig. 1(c) of the main text)

$$
\begin{aligned}
f_{i^{(1)}, i^{(2)}, ..., i^{(K)}} &= f(x_{1, i^{(1)}}, x_{2, i^{(2)}}, \ldots, x_{K, i^{(K)}}) \\
&\cong f(i_1^{(1)}, \ldots, i_L^{(1)}, i_1^{(2)}, \ldots, i_L^{(2)}, \ldots, i_1^{(K)}, \ldots, i_L^{(K)}) = \\
&= \sum_{\gamma_1=1}^{r_1} ... \sum_{\gamma_{K-1}=1}^{r_{K-1}} B^{(1)}_{\gamma_1}(i_1^{(1)}, ..., i_L^{(1)}) B^{(2)}_{\gamma_1, \gamma_2}(i_1^{(2)}, ..., i_L^{(2)}) \ldots B^{(K)}_{\gamma_{L-1}}(i_1^{(K)}, ..., i^{(K)}), \quad (14)
\end{aligned}
$$

where $B^{(k)}$ is a QTT for the $k^{\text{th}}$ dimension

$$
B^{(k)}_{\gamma_k, \gamma_{k+1}}(i_1^{(k)}, ..., i_L^{(k)}) = \sum_{\beta_k} S^{(k)}_{\gamma_k, \gamma_{k+1}, \beta_k} \left( \sum_{\alpha_{(k,1)}} ... \sum_{\alpha_{(k,L-1)}} \tilde{M}^{(k;1)}_{\beta_k, \alpha_{(k,1)}}(i_1^{(k)}) M^{(k;2)}_{\alpha_{(k,1)}, \alpha_{(k,2)}}(i_2^{(k)}) \ldots M^{(k;L)}_{\alpha_{(k,L-1)}}(i_L^{(k)}) \right). \quad (15)
$$

The respective quantized tensor networks (QTNs) for operators (QTT-Os and QTC-Os) have the same form, but with two physical indices (e.g., $o$, $i$) at each tensor instead of just one (denoted by $i$ in the above equations).

### A. Tensor Network Differentiation

The working principle behind most local tensor network algorithms involve updating a single tensor or a pair of neighboring tensors in a sequential fashion. For optimization problems (e.g. finding extreme eigenvalues and solving linear equations), this involves minimizing the cost function (a scalar value) assuming all but one tensor (or pair of tensors) remain fixed. This requires taking the derivative of the cost function with respect to the complex conjugate of that tensor.

Consider the expectation value $\langle x|A|x \rangle$,

$$
\langle x|A|x \rangle = \quad \begin{matrix} \langle x| \\ A \\ |x\rangle \end{matrix} \quad .
$$

Let $X^{(i)}$ denote the $i^{\text{th}}$ tensor in the QTT for $x$. Because the tensor network depends linearly on $X^{(i)*}$, (treating $X^{(i)}$ and its complex conjugate as independent), taking the derivative with respect to $X^{(i)*}$ amounts to removing that tensor from the tensor network. For example, for $i = 3$,

$$\frac{\partial}{\partial X^{(3)*}} \langle x|A|x\rangle \quad = \quad \begin{array}{l} \frac{\partial}{\partial X^{(3)*}} \langle x| \\ A \\ |x\rangle \end{array}.$$

An equivalent representation is

$$\left(\frac{\partial}{\partial X^{(3)*}} \frac{\partial}{\partial X^{(3)}} \langle x|A|x\rangle\right) X^{(3)} \quad = \quad A_{\text{eff}} X^{(3)} \quad = \quad \begin{array}{l} \frac{\partial}{\partial X^{(3)*}} \langle x| \\ A \\ \frac{\partial}{\partial X^{(3)}} |x\rangle \; X^{(3)} \end{array}.$$

In the figure, $X^{(3)}$ is highlighted in orange. The expression in the parentheses can be considered an effective operator for the reduced problem;

$$A_{\text{eff}} \quad = \quad \qquad \qquad .$$

Also note that though the figures depict the tensor networks for the tensor train geometry, the same concept holds for the comb geometry. If the portions of the tensor network in the dashed boxes are precomputed (see the following section), then the cost of computing $A_{\text{eff}}$ scales like $\mathcal{O}(D^2 D_W^2 d^2 + D^4 D_W d^2)$, where $D$ and $D_W$ are the bond dimensions of $x$ and $A$, respectively. The cost of computing $A_{\text{eff}} X$ is actually cheaper, scaling like $\mathcal{O}(2D^3 D_W d + D^2 D_W^2 d^2)$.

### B. Environment tensors

Computing expectation values and overlaps require performing a series of tensor contractons. While the most efficient contract pattern may vary depending on the size of each bond, building environment tensors is often most practical since they are used in multiple steps of most algorithms.

#### 1. Tensor train geometry

Let us consider tensor trains of length $L$. In this case, the left or right environments for $\langle b|A|x\rangle$ are (see Fig 1(a) and (b))

$$(E_L^{(p)})_{\alpha_p^b, \alpha_p^A, \alpha_p^x} = \sum_{\alpha_{p-1}^b, \alpha_{p-1}^A, \alpha_{p-1}^x} \sum_{o_p, i_p} B_{\alpha_{p-1}^b, \alpha_p^b}^{(p)*}(o_p) A_{\alpha_{p-1}^A, \alpha_p^A}^{(p)}(o_p, i_p) X_{\alpha_{p-1}^x, \alpha_p^x}^{(p)}(i_p)(E_L^{(p-1)})_{\alpha_{p-1}^b, \alpha_{p-1}^A, \alpha_{p-1}^x}, \quad (16)$$

$$(E_R^{(p)})_{\alpha_{p-1}^b, \alpha_{p-1}^A, \alpha_{p-1}^x} = \sum_{\alpha_p^b, \alpha_p^A, \alpha_p^x} \sum_{o_p, i_p} B_{\alpha_{p-1}^b, \alpha_p^b}^{(p)*}(o_p) A_{\alpha_{p-1}^A, \alpha_p^A}^{(p)}(o_p, i_p) X_{\alpha_{p-1}^x, \alpha_p^x}^{(p)}(i_p)(E_R^{(p+1)})_{\alpha_p^b, \alpha_p^A, \alpha_p^x}, \quad (17)$$

where $p$ denotes the site of the tensor in the tensor train and $B$, $A$ and $X$ denote tensors in the QTTs for $b$, $A$, and $x$, respectively. In essence, the left/right environments at site $p$ are the left/right half of the tensor network for $\langle b|A|x\rangle$, including the tensors at site $p$. If the bond dimensions of $x$, $b$, and $A$ are $D_x$, $D_b$, and $D_A$, respectively, and they all have physical bonds of size $d$, then the cost of building the environment scales like $\mathcal{O}((D_x^2 D_b + D_b^2 D_x)D_A d + D_x D_b D_A^2 d^2)$. Similarly, the left and right environments for $\langle b|x\rangle$ are (see Fig 1(c) and (d))

$$(F_L^{(p)})_{\alpha_p^b, \alpha_p^x} = \sum_{\alpha_{p-1}^b, \alpha_{p-1}^x} \sum_{i_p} B_{\alpha_{p-1}^b, \alpha_p^b}^{(p)*}(i_p) X_{\alpha_{p-1}^x, \alpha_p^x}^{(p)}(i_p)(F_L^{(p-1)})_{\alpha_{p-1}^b, \alpha_{p-1}^x} \quad (18)$$

$$(F_R^{(p)})_{\alpha_{p-1}^b, \alpha_{p-1}^x} = \sum_{\alpha_p^b, \alpha_p^x} \sum_{i_p} B_{\alpha_{p-1}^b, \alpha_p^b}^{(p)*}(i_p) X_{\alpha_{p-1}^x, \alpha_p^x}^{(p)}(i_p)(F_R^{(p+1)})_{\alpha_p^b, \alpha_p^x} \quad (19)$$

The cost of the tensor contraction scales like $\mathcal{O}(D_x^2 D_b d + D_b^2 D_x d)$.

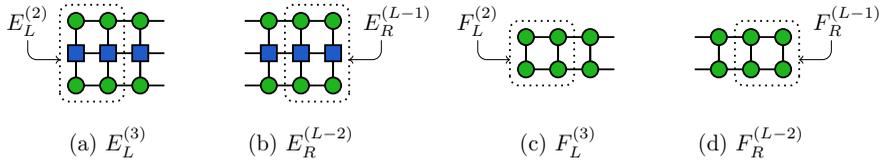

(a) $E_L^{(3)}$     (b) $E_R^{(L-2)}$     (c) $F_L^{(3)}$     (d) $F_R^{(L-2)}$

FIG. 1. Tensor network diagrams for left environments ($E_L^{(3)}$, $F_L^{(3)}$) and right environments ($E_R^{(L-2)}$, $F_R^{(L-2)}$), where $L$ is the length of the QTT. Building environments is typically done recursively, with portions of the TN in the dotted boxes computed in the previous iteration. Tensors in QTT-vectors $(x, b)$ are in green, while tensors in the QTT-operator $(A)$ are in blue.

Note that the environments are defined recursively. This is because in addition to computing the next the environment tensor, they can also be used to compute the effective operators introduced in the previous section. At the ends of the tensor train, for $p = 0$ or $p = L + 1$, the environment tensor is equal to one (and the indices of the environment tensor are dummy indices of size one).

### 2. Comb geometry

For a comb tensor network with $K$ branches, one first computes the environments starting from the ends of the branch as described above, obtaining the right environments $\{E_{R,\text{branch}}^{(k)}\}$. The primary difference lies in contracting the tensor along the spine.

Again, for concreteness, let us consider the expectation value $\langle b|A|x\rangle$. In this case the $k^{\text{th}}$ branch environment tensor has three indices, $\beta_k^b, \beta_k^A, \beta_k^x$, where $\beta$ denotes the bond connecting the spine tensor ($\bar{M}^{(k)}$ in Eq. (15)) to the first tensor of the branch tensor train ($M^{(k;1)}$).

For the first branch, the environment tensor including the spine is (see Fig. 2(a))

$$\left(E_{L,\text{spine}}^{(1)}\right)_{\gamma_1^b, \gamma_1^A, \gamma_1^x} = \sum_{\beta_1^b, \beta_1^A, \beta_1^x} \left(E_{R,\text{branch}}^{(1)}\right)_{\beta_1^b, \beta_1^A, \beta_1^x} X_{\beta_1^x, \gamma_1^x}^{(1)} A_{\beta_1^A, \gamma_1^A}^{(1)} B_{\beta_1^b, \gamma_1^b}^{(1)*}. \tag{20}$$

Again, $B$, $A$, and $X$ correspond to tensors in the QTCs for $b$, $A$, and $x$, respectively. For subsequent branches ($k > 1$), the environment tensor is (see Fig. 2(b))

$$\left(E_{L,\text{spine}}^{(k)}\right)_{\gamma_k^b, \gamma_k^A, \gamma_k^x} = \sum_{\substack{\beta_k^b, \beta_1^A, \beta_1^x \\ \gamma_k^b, \gamma_k^A, \gamma_k^b}} \left(E_{L,\text{spine}}^{(k-1)}\right)_{\gamma_{k-1}^b, \gamma_{k-1}^A, \gamma_{k-1}^x} \left(E_{R,\text{branch}}^{(k)}\right)_{\beta_k^b, \beta_k^A, \beta_k^x} X_{\beta_k^x, \gamma_k^x}^{(k)} A_{\beta_k^A, \gamma_k^A}^{(k)} B_{\beta_k^b, \gamma_k^b}^{(k)*}. \tag{21}$$

The optimal order for tensor contraction will greatly depend on the size of the bonds. For most PDEs (in which operators are sums of differentials), the sizes of bonds $\gamma^A$ and $\beta^A$ are of order 1. In this case, using a tensor contraction order of $E_{L,\text{spine}}$, $A$, $X$, $E_{R,\text{branch}}$, and finally $B$, the computational cost scales like $\mathcal{O}(S_b^3 S_x + S_x^3 S_b)$, where $S_x$ and $S_b$ are the bond dimension of the spine (both $\gamma$ and $\beta$) for QTC-vectors $x$ and $b$. The right spine environment ($E_{R,\text{spine}}^{(k)}$) would be computed analogously starting from the last tensor in the spine.

At times, one would like to compute the left environment of the spine (starting the from the side connected to the branch. In this case, one has to utilize the spine environment tensors (see Fig. 2(c)):

$$\left(E_{L,\text{branch}}^{(k;0)}\right)_{\beta_k^b, \beta_k^A, \beta_k^x} = \sum_{\substack{\gamma_{k-1}^b, \gamma_{k-1}^A, \gamma_{k-1}^x \\ \gamma_k^b, \gamma_k^A, \gamma_k^x}} \left(E_{L,\text{spine}}^{(k-1)}\right)_{\gamma_{k-1}^b, \gamma_{k-1}^A, \gamma_{k-1}^x} \left(E_{R,\text{spine}}^{(k+1)}\right)_{\gamma_k^b, \gamma_k^A, \gamma_k^x} B_{\gamma_{k-1}^b, \gamma_k^b, \beta_k^b}^{(k)*} A_{\gamma_{k-1}^A, \gamma_k^A, \beta_k^A}^{(k)} X_{\gamma_{k-1}^x, \gamma_k^x, \beta_k^x}^{(k)}. \tag{22}$$

## III. DENSITY-MATRIX ALGORITHM FOR MATRIX-VECTOR MULTIPLICATION

The naive algorithm for performing matrix-vector multiplication involves contracting the $p^{\text{th}}$ tensor in the QTN and QTN-O together for all sites $p$ in the tensor network, and then compressing the newly obtained QTN [1, 6]. The

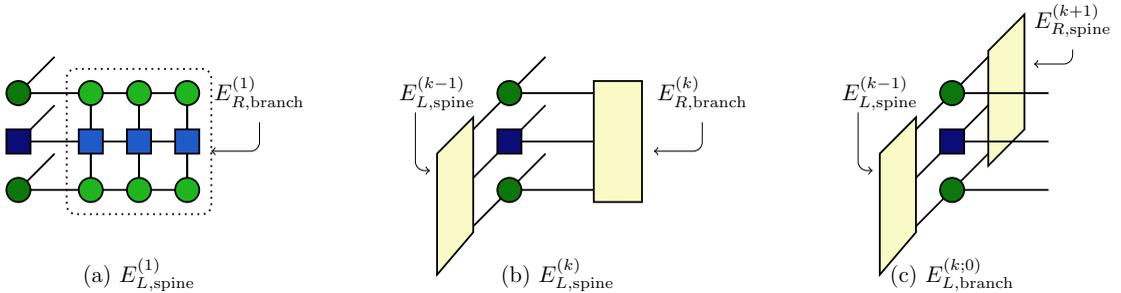

(a) $E_{L,\text{spine}}^{(1)}$  (b) $E_{L,\text{spine}}^{(k)}$  (c) $E_{L,\text{branch}}^{(k;0)}$

FIG. 2. Tensor network diagrams depicting Eqs. (20) –(22). Tensors in QTC-vectors $(x, b)$ are in green, while tensors in the QTC-operator $(A)$ are in blue. Darker shades denote tensors in the spine.

| Procedure | QTT | QTC-spine |
|---|---|---|
| **Operator-vector multiplication** | | |
| Naive [4] | $\mathcal{O}(D^3 D_W^3 d^2)$ | $\mathcal{O}(S^4 S_W^4)$ |
| Zip-up [5] | $\mathcal{O}(D^3 D_W d^2)$ to $\mathcal{O}(D^3 D_W^3 d^2)$ | $\mathcal{O}(S^4 S_W)$ to $\mathcal{O}(S^4 S_W^4)$ |
| Density matrix [2] | $\mathcal{O}(D^3 D_W^2 d + D^2 D_W^3 d^2)$ $+ (Dd \times Dd)$ eigenvalue problem | $\mathcal{O}(S^5 S_W^2)$ $+ (S^2 \times S^2)$ eigenvalue problem |
| **Local update schemes** | | |
| Build environment tensor (with $n$ operators) | $\mathcal{O}(D^3 (D_w)^n d + n D^2 D_W^2 d^2)$ | $\mathcal{O}(S^4 S_W^n + n S^2 S_W^5)$ |
| Compute $A_{\text{eff}}$ | $\mathcal{O}(D_W^2 D^2 d^2 + D^4 D_W d^2)$ | $\mathcal{O}(S^2 S_W^3 + S^4 S_W^2 + S^6 S_W)$ |
| Compute $A_{\text{eff}} x$ | $\mathcal{O}(2 D_W D^3 d + D_W^2 D^2 d^2)$ | $\mathcal{O}(S^2 S_W^3 + 2 S^4 S_W^2 + S^4 S_W)$ |

TABLE I. Theoretical costs of common procedures for the quantized tensor train (QTT) and the spine of the comb-like tree tensor network (QTC-spine). In the QTT geometry, vectors have bond dimension $D$ while operators have bond dimension $D_W$. In the spine of the comb geometry, vectors and operators have bond dimension $S$, $S_W$ along the spine, respectively. The costs are obtained assuming $S_W$ and $D_W$ are smaller than $S$ and $D$. If this is not the case, other tensor contraction orderings may be optimal.

compression (in particular, the canonicalization, which is like a preconditioning step) is expensive; for the tensor train geometry, the computational cost scales like $\mathcal{O}(D^3 D_W^3 d)$, where $D$ is the bond dimension of the QTT and $D_W$ is the bond dimension of the QTT-O.

Instead, one could consider algorithms that canonicalize the resulting QTT as the tensors are being contracted. One such algorithm is the zip-up algorithm, whose cost scales like $\mathcal{O}(D D_W \tilde{D}^2 d^2)$, where $\tilde{D}$ is an intermediate bond dimension that is problem dependent, ranging in value from $D$ to $D D_W$ [5]. In this work, we mainly used the density-matrix algorithm [2, 7], detailed below. Note that after these algorithms are used, the QTNs are in canonical form. We observe better performance after compressing again using SVD, sweeping in the opposite direction. This calculation is no longer as costly, since the bond dimension has already been reduced.

### A. Basic Algorithm for QTTs

In the density matrix algorithm, instead of canonicalizing the QTT and compressing using SVD, it involves computing the partial density matrix of $\langle x | A^\dagger A | x \rangle$ at each site, and computing its eigenvalues (which one truncates in the low-rank approximation). The basic algorithm for QTTs is written in Alg. 1. However, we recommend referring to Institute [2] for diagrammatic representations of the tensor contractions.

---

**Algorithm 1** Density Matrix Compression

---

**Input:** QTT $x$, QTT-O $A$ of length $L$
**Output:** QTT $y$, the low-rank approximation of $Ax$, and of bond dimension $D$
Notes: sums are performed over all repeated indices

**Build left environments**
If $E_L^{(0)}$ is not given, set to one
Build left environments $\{E_L^{(p)}\}$ for $\langle x|A^\dagger A|x\rangle$ starting from $E_L^{(0)}$

**Compute compressed tensors**
**procedure** COMPRESS REDUCED DENSITY MATRIX (END)
    Compute density matrix $\rho_{o'_L,o_L}^{(L)} \leftarrow \sum E_{\alpha'_{L-1},\beta'_{L-1},\beta_{L-1},\alpha_{L-1}}^{(L-1)} X_{\alpha'_{L-1}}^{(L)*}(i'_L) A_{\beta'_{L-1}}^{(L)*}(i'_L,o'_L) A_{\beta_{L-1}}^{(L)}(o_L,i_L) X_{\alpha_{L-1}}^{(L)}(i_L)$

    Solve eigenvalue problem $U_{o'_L,\eta_L}^{(L)},\lambda_{\eta_L}^{(L)} \leftarrow \mathrm{EIG}(\rho_{o'_L,o_L}^{(L)})$, keeping only the $D$ largest eigenvalues

    Define $M_{\alpha_{L-1}}^{(L)}(i_L) \leftarrow U_{o'_L,\eta_L}$, REINDEX$(\eta_L \rightarrow \alpha_{L-1}, o'_L \rightarrow i_L)$

    Compute right (compressed) environment $C_{\alpha_{L-1},\beta_{L-1},\eta_L}^{(L)} \leftarrow \sum X_{\alpha_{L-1}}^{(L)}(i_L) O_{\beta_{L-1}}^{(L)}(o_L,i_L) U_{o_L,\eta_L}^{*(L)}$

**for** site $p = L-1, ..., 2$ **do**
    **procedure** COMPRESS REDUCED DENSITY MATRIX
        Compute density matrix $\rho_{o'_p,n'_{p+1},o_p,\eta_{p+1}}^{(p)} \leftarrow \sum E_{\alpha'_{p-1},\beta'_{p-1},\beta_{p-1},\alpha_{p-1}}^{(p-1)} X_{\alpha'_{p-1},\alpha'_p}^{(p)*}(i'_p) A_{\beta'_{p-1},\beta'_p}^{(p)*}(i'_p,o'_p)$

$$A_{\beta_{p-1},\beta'_p}^{(p)}(o_p,i_p) X_{\alpha_{p-1},\alpha_p}^{(p)}(i_p) C_{\alpha_o,\beta_p,n'_{p+1}}^{(p+1)} C_{\alpha'_p,\beta'_p,n'_{p+1}}^{*(p+1)}$$

        Solve eigenvalue problem $U_{(o'_p,n'_{p+1}),\eta_p}^{(i)}, \lambda_{\eta_p}^{(p)} \leftarrow \mathrm{EIG}(\rho_{(o'_p,n'_{p+1}),(o_p,\eta_{p+1})}^{(p)})$, keeping only the $D$ largest eigenvalues

        Define $M_{\alpha_{p-1},\alpha_p}^{(p)}(i_p) \leftarrow U_{o'_p,n'_{p+1},\eta_p}$, REINDEX$(n'_{p+1} \rightarrow \alpha_p, \ \eta_p \rightarrow \alpha_{p-1}, o'_p \rightarrow i_p)$

        Compute right (compressed) environment $C_{\alpha_{p-1},\beta_{p-1},\eta_p}^{(i)} \leftarrow \sum X_{\alpha_{p-1},\alpha_p}^{(i)}(i_p) O_{\beta_{p-1},\beta_p}^{(i)}(o_p,i_p) U_{o_p,\eta_{p+1},\eta_p}^{(i)*} C_{\alpha_p,\beta_p,\eta_{p+1}}^{(i+1)}$

$M_{\alpha_1}^{(1)}(i_1) \leftarrow \sum x_{\alpha_1}^{(1)}(i_1) A_{\beta_1}^{(1)}(o_1,i_1) C_{\alpha_1,\beta_1,\eta_2}^{(2)}$, REINDEX$(\eta_2 \rightarrow \alpha_1, o_1 \rightarrow i_1)$

$y \leftarrow \mathrm{QTT}(M^{(1)}, M^{(2)}, ..., M^{(L)})$ which is in right canonical form.

---

### B. Computational Cost

The primary costs arise from computing the left environments and performing the eigendecomposition of the reduced density matrices $\rho$. With physical bond dimension $d$, QTT bond dimension $D$, and QTT-O bond dimension $D_W$, the cost of tensor contraction scales like $\mathcal{O}(D^3 D_W^2 d + D^2 D_W^3 d^2)$. The cost of the eigendecomposition depends on the algorithm used. For exact eigendecomposition (which is not necessary since we are only interested in the $D$ largest eigenvalues), we must first explicitly compute the density matrix, whose cost scales like $\mathcal{O}(D^3 D_W^2 d^2 + D^2 D_W^3 d^2))$. The eigendecomposition itself scales like $\mathcal{O}((Dd)^3)$. An iterative eigensolver would be attractive, as it may allow one to avoid explicitly computing $\rho$.

### C. Extension to Comb Geometry

The algorithm introduced above is designed for the 1-D tensor train geometry. However, it can be easily extended to the comb geometry. Application of the operator along the branches uses the same procedure as above. One must then compress the branches into the spine, and then compress the spine itself. The algorithm is described in Alg. 2. Following the discussion earlier in Section II B 2, the cost of computing the environments in the comb geometry scale like $\mathcal{O}(S^4 S_W^2 + S^2 S_W^5)$ in addition to the costs of computing the environments of the branches. Here, $S$ is the size of the bonds in the QTC spine, $S_W$ is the size of the bonds in the QTC-O spine, and we assume that $S_W < S$. The cost of computing $\rho$ scales like $\mathcal{O}(S^5 S_W^2)$, and the resulting eigenvalue problem is of size $S^2$. Not only is this becoming relatively expensive, we encounter numerical issues with standard partial eigenvalue solvers (near-zero eigenvalues are not easily found).

We instead update spine tensors using the zip-up algorithm. The procedure for updating the tensors is given in Alg. 3; the algorithm otherwise remains essentially the same. The cost of updating the spine now scales ranges from $\mathcal{O}(S^4)$ to $\mathcal{O}(S^4 S_W^4)$, depending on the problem.

---

**Algorithm 2** Density Matrix Compression for Comb Geometry

---

**Input:** QTC $x$, QTC-O $A$ with $K$ branches
**Output:** QTC $y$, the low-rank approximation of $Ax$

**Build left environments**
▷ *Build branch environments* ◁
**for** spine index $k = 1, \ldots, K-1$ **do**
    Compute environments of each branch $\left\{ \left( E_{R,\text{branch}}^{(k;p)} \right) \right\}$ for $p = 1, \ldots, L$.     ▷ *The branches are QTTs of length $L$*
▷ *Build left spine environments* ◁
Left spine environments from $k \in 1 \ldots K$: $\left\{ \left( E_{L,\text{branch}}^k \right) \right\}$

**Compute compressed tensors**
**for** branch $k = K, \ldots, 1$ **do**
    ▷ *Compress branch using Alg. 1. The branch is a QTT of length $L$* ◁
    Build left branch environment $E_{L,\text{branch}}^{(k;0)}$ from spine environments (Eq. (22))
    Compute compressed branch $B^{(k)} = \text{QTT}(M^{(k;1)}, \ldots, M^{(k;L)})$ using Alg. 1, with left environments built from $\tilde{E}_{R,\text{branch}}^{(k;0)}$
    ▷ *Compress branch into spine* ◁
    Compute unitary $U_{o_1'^k,\eta_2,\eta_1}^{(k;1)}, C_{\beta_k^A,\beta_k^x,\eta_1}^{(k;1)}$ using Compress reduced density matrix step
    Update $M_{\beta_k,\alpha_{(k,1)}}^{(k;1)}(i_1^{(k)}) \leftarrow U_{o_1^{(i;1)},\eta_2,\eta_1}^{(i;1)}$, Reindex $(\eta_2 \to \alpha_{(k,1)},\ \eta_1 \to \beta_k,\ o_1'^k \to i_1^{(k)})$

    **procedure** Update Spine Tensors
        Initialize right (compressed) spine environment $\tilde{C}^{(K+1)} \leftarrow 1$
        **if** $k \neq 1$ **then**
            ▷ *Compress spine tensor* ◁
            ▷ *for $k = K$, indices with $K+1$ are dummy indices of size 1* ◁
            Compute density matrix for spine
$$\tilde{\rho}_{\eta_k',\xi_{k+1}',\eta_k,\xi_{k+1}}^{(k)} \leftarrow \sum \left( \tilde{E}_{L,\text{spine}}^{(k-1)} \right)_{\gamma_{k-1}'^x,\gamma_{k-1}'^A,\gamma_{k-1}^A,\gamma_{k-1}^x} \tilde{X}_{\gamma_{k-1}'^x,\gamma_k'^x,\beta_k'^x}^{(k)*} \tilde{A}_{\gamma_{k-1}'^A,\gamma_k'^A,\beta_k'^A}^{(k)*} \tilde{A}_{\gamma_{k-1}^A,\gamma_k^A,\beta_k^A}^{(k)} \tilde{X}_{\gamma_{k-1}^x,\gamma_k^x,\beta_k^x}^{(k)}$$
$$C_{\beta_k^A,\beta_k^x,\eta_k}^{(k;1)} C_{\beta_k'^A,\beta_k'^x,\eta_k'}^{(k;1)*} \tilde{C}_{\gamma_k^A,\gamma_k^x,\xi_{k+1}}^{(k+1)} \tilde{C}_{\gamma_k'^A,\gamma_k'^x,\xi_{k+1}'}^{(k+1)*}$$
            Solve eigenvalue problem $\tilde{U}_{(\eta_k',\xi_{k+1}'),\xi_k}^{(k)}, \lambda_{\rho_k}^{(k)} \leftarrow \text{Eig}(\tilde{\rho}_{(\eta_k',\xi_{k+1}'),(\eta_k,\xi_{k+1})}^{(k)})$
            Define $\tilde{M}_{\gamma_{k-1},\gamma_k,\beta_k}^{(b)} \leftarrow U_{\eta_k',\xi_{k+1}',\xi_k}$, Reindex$(\eta_k' \to \beta_k,\ \xi_{k+1}' \to \gamma_k,\ \xi_k \to \gamma_{k-1}))$     ▷ *Updated spine tensor*
            Define right (compressed) spine environment
$$\tilde{C}_{\gamma_{k-1}^A,\gamma_{k-1}^x,\xi_k}^k \leftarrow \sum \tilde{X}_{\gamma_{k-1}^x,\gamma_k^x,\beta_k^x}^{(k)} \tilde{A}_{\gamma_{k-1}^A,\gamma_k^A,\beta_k^A}^{(k)} \tilde{U}_{\eta_k,\xi_{k+1},\rho_k}^{(k)*} C_{\beta_k^A,\beta_k^x,\eta_k}^{(k;1)} \tilde{C}_{\gamma_k^A,\gamma_k^x,\xi_{k+1}}^{(k+1)}$$
        **else**
            ▷ *Define spine tensor* ◁
            $\tilde{M}_{\gamma_1,\beta_1}^{(1)} \leftarrow \sum \tilde{x}_{\gamma_1^x,\beta_1^x}^{(1)} \tilde{A}_{\gamma_1^A,\beta_1^A}^{(1)} C_{\beta_1^A,\beta_1^x,\eta_1}^{(1;1)} \tilde{C}_{\gamma_1^A,\gamma_1^x,\xi_2}^{(2)}$
        Define right spine environment $\tilde{E}_{R,\text{spine}}^{(k)} \leftarrow \tilde{C}^{(k)}$.

**Define new comb**
Spine $\leftarrow \text{QTT}(\tilde{M}^{(1)}, \ldots, \tilde{M}^{(K)})$
QTC $y \leftarrow \text{Comb}(\text{Spine}; (B^{(1)}, \ldots, B^{(K)}))$

---

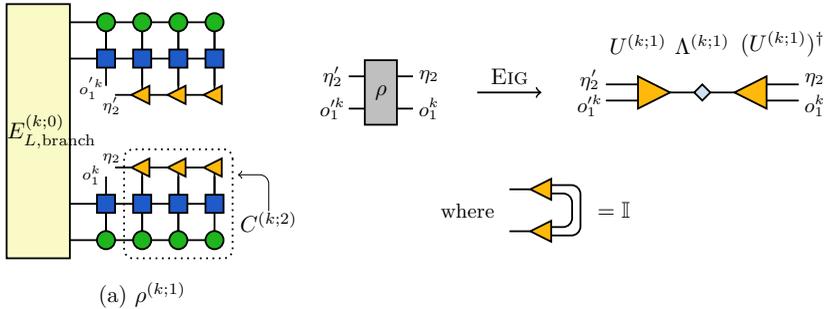

(a) $\rho^{(k;1)}$

FIG. 3. TN diagrams showing the compression and canonicalization of first tensor in the branch ($M^{(k,1)}$) into the spine. The diagram looks the same as for a normal tensor train. The orange triangles denote that the tensor is in left/right canonical form, depending on the direction the triangle is pointing in.

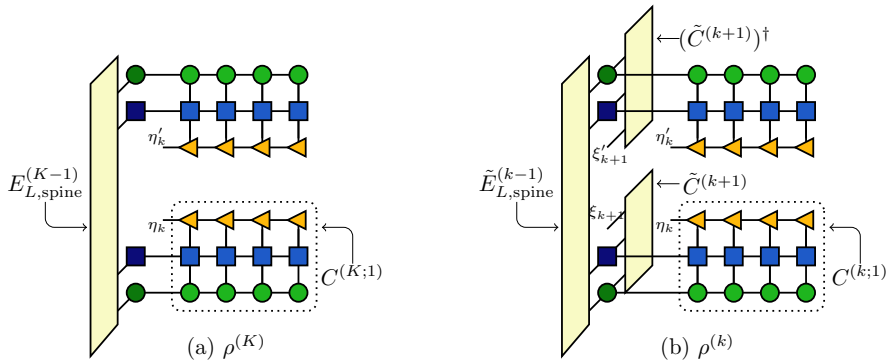

(a) $\rho^{(K)}$

(b) $\rho^{(k)}$

FIG. 4. TN diagrams showing the computation of density matrix $\rho$ for spine tensors using the density matrix algorithm (a) at site $K$ and (b) in the middle of the chain ($k = 2, ..., K - 1$). This computation is expensive so we instead use the zip-up algorithm.

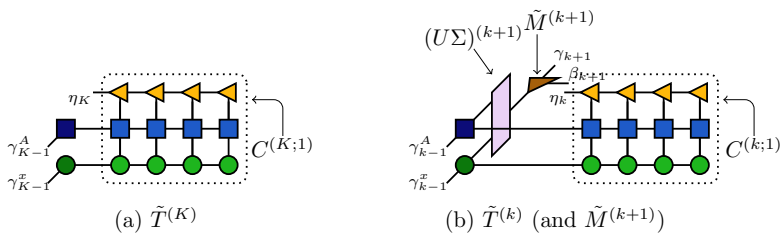

(a) $\tilde{T}^{(K)}$

(b) $\tilde{T}^{(k)}$ (and $\tilde{M}^{(k+1)}$)

FIG. 5. TN diagrams showing the computation of tensor $T^{(k)}$ for spine tensors using the zip-up algorithm (a) at site $k = K$ and (b) in the middle of the chain ($k = 2, ..., K - 1$)

.

---

**Algorithm 3** Update spine procedure based on zip-up algorithm

---

**Input:** QTC $x$, QTC-O $A$ with $K$ branches
**Output:** QTC $y$, the low-rank approximation of $Ax$
**procedure** UPDATE SPINE TENSORS(spine index $k \in (1, \ldots, K)$)

    ▷ *Zip-up algorithm for spine* ◁

    **if** $k = K$ **then**

$$\tilde{T}^{(K)}_{\gamma^A_{K-1}, \gamma^x_{K-1}, \eta_K} \leftarrow \sum \tilde{X}^{(K)}_{\gamma^x_{K-1}, \gamma^x_K, \beta^x_K} \tilde{A}^{(K)}_{\gamma^A_{K-1}, \gamma^A_K, \beta^A_K} C^{(K;1)}_{\beta^A_K, \beta^x_K, \eta_K}$$

$$U^{(K)}_{(\gamma^A_{K-1}, \gamma^x_{K-1}), \gamma_{K-1}}, \Sigma^{(K)}_{\gamma_{K-1}}, V^{(K)}_{\eta_K, \gamma_{K-1}} \leftarrow \text{SVD}\left( \tilde{T}^{(K)}_{(\gamma^A_{K-1}, \gamma^x_{K-1}), \eta_K} \right)$$

       where only singular values above a cutoff threshold ($\sim 10^{-10}$) are retained.

       Define $\tilde{M}^{(K)}_{\gamma_{K-1}, \beta_K} \leftarrow V^{(K)}_{\eta_K, \gamma_{K-1}}$, REINDEX($\eta_K \to \beta_K$)      ▷ *Updated spine tensor*

    **else if** $2 \leq k < K$ **then**

$$\tilde{T}^{(k)}_{\gamma^A_{k-1}, \gamma^x_{k-1}, \gamma_k, \eta_k} \leftarrow \sum \tilde{X}^{(k)}_{\gamma^x_{k-1}, \gamma^x_k, \beta^x_k} \tilde{A}^{(k)}_{\gamma^A_{k-1}, \gamma^A_k, \beta^A_k} C^{(k;1)}_{\beta^A_k, \beta^x_k, \eta_k} \tilde{U}^{(k+1)}_{(\gamma^A_k, \gamma^x_k), \gamma_k} \tilde{\Sigma}^{(k+1)}_{\gamma_k}$$

$$U^{(k)}_{(\gamma^A_{k-1}, \gamma^x_{k-1}), \gamma_{k-1}}, \Sigma^{(k)}_{\gamma_{k-1}}, V^{(k)}_{\gamma_k, \eta_k, \gamma_{k-1}} \leftarrow \text{SVD}\left( \tilde{T}^{(k)}_{(\gamma^A_{k-1}, \gamma^x_{k-1}), \gamma_k, \eta_k} \right)$$

       where only singular values above a cutoff threshold ($\sim 10^{-10}$) are retained.

       Define $\tilde{M}^{(k)}_{\gamma_{k-1}, \gamma_k, \beta_k} \leftarrow V^{(k)}_{\gamma_k, \eta_k, \gamma_{k-1}}$, REINDEX($\eta_k \to \beta_k$)      ▷ *Updated spine tensor*

    **else if** $k = 1$ **then**

$$\tilde{T}^{(1)}_{\gamma_1, \eta_1} \leftarrow \sum \tilde{X}^{(1)}_{\gamma^x_1, \beta^x_1} \tilde{A}^{(1)}_{\gamma^A_1, \beta^A_1} C^{(1;1)}_{\beta^A_1, \beta^x_1, \eta_1} \tilde{U}^{(2)}_{(\gamma^A_1, \gamma^x_1), \gamma_1} \tilde{\Sigma}^{(2)}_{\gamma_1}$$

       Define $\tilde{M}^{(1)}_{\gamma_1, \beta_1} \leftarrow T^{(1)}_{\gamma_1, \eta_1}$, REINDEX($\eta_1 \to \beta_1$)      ▷ *Updated spine tensor*

    If sweeping from right to left, build right spine environment $\tilde{E}^{(k)}_{R,\text{spine}}$. Else, build left spine environment.

## IV. TIME-DEPENDENT VARIATIONAL PRINCIPLE (TDVP)

Consider a unitary system, with

$$\frac{\partial}{\partial t}\psi(t) = A\psi(t) \tag{23}$$

where $A$ is an anti-Hermitian operator. According to the Dirac-Frenkel principle, the time derivative at time $t$ exists on the submanifold tangent to $u(t)$,

$$\langle v|\dot{u}(t) - Au(t)\rangle = 0, \quad \forall v \in \mathcal{T}_{u(t)}\mathcal{M}\,. \tag{24}$$

In the time-dependent variational principle (TDVP) algorithm, we evolve the tensors forward in time one by one, using equation of motions obtained from satisfying Eq. (24). Let us begin by representing the QTT for discretized $u(x)$ in right-canonical form,

$$|u(i_1, i_2, \ldots i_L; t)\rangle = \sum_{\alpha_1, \ldots \alpha_{L-1}} M^{(1)}_{\alpha_1}(i_1; t)\, B^{(2)}_{\alpha_1, \alpha_2}(i_2; t) \ldots B^{(L)}_{\alpha_{L-1}}(i_L; t) \tag{25}$$

where the tensors denoted by $B$ are right canonical; i.e., $\sum_{i_p, \alpha_p} B^{*(p)}_{\alpha_{p-1}, \alpha_p}(i_p) B^{(p)}_{\alpha_{p-1}, \alpha_p}(i_p) = \mathbb{I}$. We want to obtain an equation of motion for $M^{(1)}$. Keeping the other tensors fixed, the time derivative of $u(t; M^{(1)})$ is

$$\frac{\partial}{\partial t}|u(t; M^{(1)})\rangle = \dot{M}^{(1)}\frac{\partial}{\partial M^{(1)}}|u(t; M^{(1)})\rangle\,. \tag{26}$$

According to Eq. (24), the approximate dynamics is obtained by minimizing

$$||\dot{M}^{(1)}\frac{\partial}{\partial M^{(1)}}|u(t; M^{(1)})\rangle - A|u(t; M^{(1)})\rangle||. \tag{27}$$

Utilizing the fact that $u$ is in right canonical form, we obtain

$$\begin{aligned}
\dot{M}^{(1)}(t) &= \frac{\partial}{\partial M^{(1)*}}\langle u(t; M^{(1)})|A|u(t; M^{(1)})\rangle \\
&= \left(\frac{\partial}{\partial M^{(1)*}}\frac{\partial}{\partial M^{(1)}}\langle u(t; M^{(1)})|A|u(t; M^{(1)})\rangle\right) M^{(1)} \\
&= A^{(1)}_{\text{eff}}\, M^{(1)}.
\end{aligned} \tag{28}$$

Note that $A_{\text{eff}}$ is the original anti-Hermitian matrix projected onto the submanifold tangent to $|u(t; M^{(1)})\rangle$.

Once $M^{(1)}$ is updated to the next time step using some time integration scheme (e.g., exact integration or RK4), we prepare to shift the orthogonality center of the QTT to the next site by performing a QR decomposition on $M^{(1)}(t + dt) = A^{(1)}(t + dt)R^{(1)}(t + dt)$.

Because of the gauge degree of freedom, the tensors are not actually independent of each other, so one must propagate $R^{(1)}(t + dt)$ backwards in time before updating site 2. This is done in a similar fashion, solving

$$\dot{R}^{(1)} = -A^{(1|2)}_{\text{eff}}R^{(1)}\,, \tag{29}$$

$$A^{(1|2)}_{\text{eff}} = \left(\frac{\partial}{\partial R^{(1)*}}\frac{\partial}{\partial R^{(1)}}\langle u(t; M^{(1)})|A|u(t; R^{(1)})\rangle\right)\,. \tag{30}$$

Note the negative sign, since we are propagating backwards in time. Once $R^{(1)}(t)$ is obtained, it is contracted with the tensor at the second site, $B^{(2)}$. The orthogonality center has now been shifted to site 2, and we repeat the above steps to evolve site 2 forward in time.

In summary, the algorithm is as follows. For each site in the QTT from 1 to $L - 1$: evolve site $p$ forward in time, evolve the bond between $p$ and $p + 1$ backwards in time, and canonicalize the QTT to site $p + 1$. Finally, the last site at $p = L$ is propagated forwards in time. Note that the bond dimension of the QTT does not change as one propagates forward in time.

Updating the tensors in this sweeping fashion has a Trotter error associated with it. The scheme presented is a first-order scheme, and has error $\mathcal{O}(\Delta t)$. By sweeping from left to right and then right to left, doing time evolution with $\Delta t/2$ in each sweep, one obtains a second-order time evolution scheme with error $\mathcal{O}(\Delta t^2)$. Projecting the true

dynamics onto the tensor network manifold also introduces some error. However, in the case that a single site is updated at a time, as introduced here, the expectation value $\langle u|A|u \rangle$, and the norm of $|u \rangle$ and conserved [8, 9]. In the two-site variant of TDVP, one updates two neighboring tensors at one time. The advantage is that one can adaptively increase the rank as needed, thereby reducing some of the projection error. However, this update typically requires compression of the two tensors (so that the bond dimension remains manageable), breaking the conservation properties of TDVP.

We refer readers to Haegeman *et al.* [8] and Paeckel *et al.* [9] for more technical details.

### A. Computational Cost (TT geometry)

The cost of tangent-space methods is dominated by the calculation of $A_{\text{eff}}^{(i)}$ and the time evolution of a the tensor $M_i$, which is of size $D \times D \times d$. Recall that $D$ is the bond dimension of the QTT representation of $x$ and $d$ is the size of the physical indices. If $A$ is represented as an QTT-O with bond dimension $D_W$, then the cost of computing $A_{\text{eff}}^{(i)}$ scales like $\mathcal{O}(D^2 D_W^2 d^2 + D^4 d^2 D_W^2)$ and exact time evolution via exact diagonalization scales like $\mathcal{O}((D^2 d)^3)$. If one solved Eq. (28) using an approximate scheme such as RK4, one can avoid explicitly computing $A_{\text{eff}}^{(i)}$ and instead compute $A_{\text{eff}}^{(i)} M$ (where $M$ is a generic tensor of the appropriate size). For RK4, the cost of time evolution then scales like $\mathcal{O}(2D^3 D_W d + D^2 D_W^2 d^2)$, arising solely from tensor contraction.

### B. TDVP in Comb Geometry

The TDVP algorithm for QTCs mainly differ in the order in which tensors are propagated forward in time. We use a 1-site TDVP algorithm with the ordering proposed in Bauernfeind and Aichhorn [10] and summarized in Alg. 4:

---
**Algorithm 4** TDVP sweeping algorithm for QTC
---
**for** branch $k = 1, \ldots K - 1$ **do**
    Perform TDVP on branch starting from the free end (away from the spine), including the spine tensor
    Perform back propagation on the bond between spine tensor $\bar{M}^{(k)}$ and $\bar{M}^{(k+1)}$.
For branch $k = K$, perform TDVP on the branch, starting from spine tensor $\bar{M}^{(K)}$ and ending at the free end.

---

For a second order algorithm, the above algorithm and its exact reverse is performed, each with time step $\Delta t/2$. Also note that while not stated explicitly, the QTC must be properly canonicalized when performing TDVP. Because the sweeping algorithm is not continuous, TDVP does not automatically put the QTC in the proper canonical form (as is the case for QTTs).

### C. Computational Cost (Comb geometry)

The cost of TDVP for the branches is the same as for the TT geometry. If using exact diagonalization, the cost of calculating the effective operator scales like $\mathcal{O}(S^6)$ and the cost of exact diagonalization scales like $\mathcal{O}(S^9)$. Again, $S$ is the size of all bonds in the spine for the QTC. While small for most initial states, $S$ will generally will grow over time. We assume the bonds in the spine for the QTC-O to be small and constant, so they are ignored in the cost estimates. Unless we require $S$ to be very small, we likely would never use exact diagonalization for spine tensors because of the high cost. If one instead uses RK4, the cost scales like $\mathcal{O}(S^4)$.

## V. DENSITY MATRIX RENORMALIZATION GROUP

### A. Basic DMRG Algorithm

The density-matrix renormalization group (DMRG) algorithm is an optimization algorithm for quantized tensor networks in which each tensor (or each neighboring pair of tensors) is optimized in a sequential fashion. Typical applications include finding extreme eigenvalues [1, 11] and solving linear equations [12, 13]. In this work, we are interested in the latter application.

Let $|x \rangle$ and $|b \rangle$ be QTTs of bond dimensions $D_x$ and $D_b$, respectively, and let $A$ be a QTT-O of bond dimension $D_A$. For our application, $D_b \sim \mathcal{O}(D_x)$ and $D_A$ is a small integer value. All have physical bond dimension $d$ (2 in our

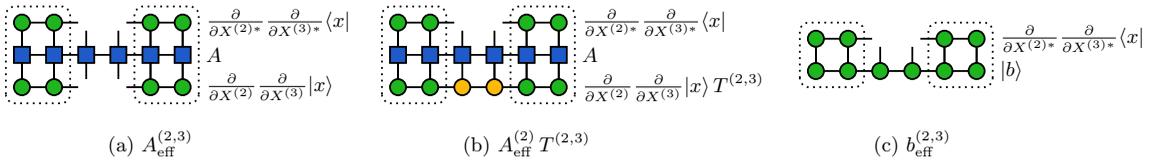

(a) $A_{\text{eff}}^{(2,3)}$       (b) $A_{\text{eff}}^{(2)} T^{(2,3)}$       (c) $b_{\text{eff}}^{(2,3)}$

FIG. 6. Tensor network diagrams for (a) $A_{\text{eff}}^{(2,3)}$, (b) $A_{\text{eff}}^{(2,3)} T^{(2,3)}$, and (c) $b_{\text{eff}}^{(2,3)}$. Portions of the TN in the dotted boxes are typically precomputed as environment tensors.

case) and are of length $L$. To solve the linear equation $A|x\rangle = |b\rangle$, each neighboring pair of tensors are to be updated in an iterative fashion with the solution to the reduced problem

$$A_{\text{eff}}^{(p,p+1)} T^{(p,p+1)} = b_{\text{eff}}^{(p,p+1)}, \tag{31}$$

where

$$T^{(p,p+1)} = \sum_{\alpha_p} X_{\alpha_{p-1},\alpha_p}^{(p)} X_{\alpha_p,\alpha_{p+1}}^{(p+1)}, \tag{32}$$

$$b_{\text{eff}}^{(p,p+1)} = \frac{\partial}{\partial T^{(p,p+1)*}} \langle x|b\rangle, \tag{33}$$

$$A_{\text{eff}}^{(p,p+1)} = \frac{\partial}{\partial T^{(p,p+1)}} \frac{\partial}{\partial T^{(p,p+1)*}} \langle x|A|x\rangle, \tag{34}$$

and $x^{(p)}$ is the $p^{\text{th}}$ tensor in $|x\rangle$. The tensor network diagram representations are shown in Fig. 6. The size of the reduced problems are $D_x^2 d^2$, and can be solved using traditional methods. We opt to solve the reduced problem using conjugate gradient descent (CGD) [14] or conjugate gradient squared (CGS) for non-Hermitian $A$ [15], so as to avoid computing $A_{\text{eff}}^{(i,i+1)}$ explicitly (see Alg. 5) [16].

The entire DMRG algorithm is summarized in Alg. 6, with three primary steps:

1. First is to build the left or right environments for $\langle x|A|x\rangle$ and for $\langle x|b\rangle$. Again, the environments are generally built recursively because each are also used to compute $A_{\text{eff}}^{(p,p+1)}$ and $b_{\text{eff}}^{(p,p+1)}$. The cost of building the environment scales like $\mathcal{O}(D_x^3 D_A d + D_x^2 D_A^2 d^2)$.

2. The second step is to update the tensors in $|x\rangle$ by solving Eq. (31) as described above, starting with sites 1 and 2 of the QTT and then sweeping left to right through the chain of tensors. In this case, one only needs to first build the right environments. After updating sites $p$ and $p+1$, one computes new left environments $E^{L,(p)}$ and $F^{L,(p)}$ as depicted in Fig. 1. It is also equally valid to sweep through the QTT tensors from right to left. In this case, one first only builds the left environments, and computes new right environments $E^{R,(p+1)}$ and $F^{R,(p+1)}$ after updating tensors $p$ and $p+1$. The cost of each sweep is dominated by the calculation of $A_{\text{eff}}^{(p,p+1)} T$, which scales like $\mathcal{O}((D_x^2 D_b + D_x D_b^2) D_A d + D_x D_b D_A^2 d^2)$ per CGD iteration.

3. At the end of each sweep, one computes the error $||b - Ax||^2$. If the error is still too large, one continues to the next iteration, sweeping in the opposite direction. By computing the error as $\langle b|b\rangle - 2\langle b|A|x\rangle + \langle x|A^\dagger A|x\rangle$, the cost typically is dominated by the last term, which scales like $\mathcal{O}(2D_x^3 D_A^2 d + 2D_x^2 D_A^3 d^2)$. One could consider a less exact measure of error to reduce the cost of this step.

## B. Block matrix DRMG

Suppose that $A$ contains some structure and is a block matrix. For example, in the case of an implicit solver for Maxwell's equations, each component of the electric field $\mathbf{E}$ and magnetic field $\mathbf{B}$ correspond to a subblock in the input $|x\rangle$ and target $|b\rangle$, and the matrix $A$ denotes couplings between the components. In this case, the linear equation to be solved is of the form

$$\sum_\xi A_{\eta,\xi} |x_\xi\rangle = |b_\eta\rangle \tag{35}$$

---

**Algorithm 5** Conjugate gradient descent

---

**Input:** tensor network $A$, tensor $b$, tensor $x$ (initial guess)
**Output:** tensor $x$
**procedure** CGD($A$, $b$, $x$)
    $\beta \leftarrow Ax$
    $r \leftarrow b - Ax$
    $d \leftarrow r$
    error $\varepsilon \leftarrow |r|$
    $i \leftarrow 1$
    **while** $i <$ MAXITER and $\varepsilon/|b| >$ CONVTOL **do**
        tensor $q \leftarrow$ CONTRACT($A$, $d$)                ▷ *Contract all tensors together*
        scalar $p \leftarrow$ CONTRACT($A$, $d$, $d^\dagger$)
        scalar $\alpha \leftarrow \varepsilon^2/p$
        tensor $x \leftarrow x + \alpha d$
        tensor $r \leftarrow r - \alpha q$
        scalar $\varepsilon_{\text{old}} \leftarrow \varepsilon$
        scalar $\varepsilon \leftarrow |r|$                      ▷ *new error is the norm of the new residual*
        scalar $\beta \leftarrow \varepsilon^2/\varepsilon_{\text{old}}^2$
        tensor $d \leftarrow r + \beta d$
        int $i \leftarrow i + 1$
    **return** $x$

---

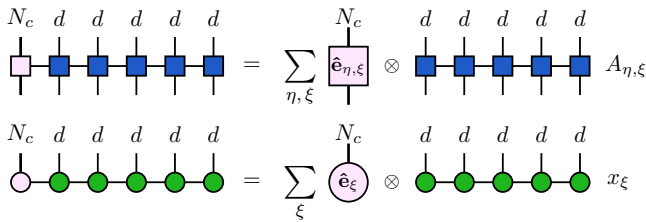

FIG. 7. Expanded QTT-O (top) and QTT (bottom), which are originally of length $L = 5$ and physical bond $d$. They are generated by summing together QTTs multiplied by an additional tensor that indexes the QTTs position in the block matrix or vector. Matrix $\hat{e}_{\eta,\xi}$ is an $N_c \times N_c$ matrix with all elements set to zero except for the $(\eta, \xi)^{\text{th}}$ element which is set to one. Vector $\eta_\xi$ is a unit vector of length $M$ of where the $\xi^{\text{th}}$ element is one and otherwise is zero. If the QTTs are originally of bond dimension $D$, the bond dimension of the expanded QTT (without compression) is $N_c^2 D$ for the QTT-Operator and $N_c D$ for the QTT-State.

where $\eta$ and $\xi$ index the position of each submatrix.

Similar to what is done in classical solvers by appending vectors and matrices together and expanding the problem space, one can combine the QTTs for each field by adding an extra tensor which denotes which field component is being operated on (see Fig. 7). The size of the physical bond dimension would be the number of field components $N_c$ (six in the case of Maxwell's equations). One then solves the system of equations using DMRG as described in the previous section. By construction, the optimization of a single site in the tensor train updates all field components. While the implementation is straightforward, the calculation can be costly. If the QTTs of each field component is of bond dimension $D$, then it is likely that the bond dimension of the composite QTT will need to be larger than $D$ in order to capture all of the original information.

Alternatively, one can perform a modified DMRG algorithm, in which one updates the tensors of each component $|x_\eta\rangle$ by solving the reduced problem

$$\sum_\xi (A_{\eta,\xi})_{\text{eff}}^{(p,p+1)} T_\xi^{(p,p+1)} = (b_\eta)_{\text{eff}}^{(p,p+1)}, \tag{36}$$

---

**Algorithm 6** Two-site DMRG with conjugate gradient descent

---

**Input:** QTT $b$ and initial guess QTT $x$, QTT-Operator $A$. QTTs are of $L$ sites, integer bond dimension $D$
**Output:** the solution QTT-State $x$ of bond dimension $D$

Define environment $E_L^{(0)} \leftarrow 1$, $E_R^{(L+1)} \leftarrow 1$
**Build left environments**
**for** site $p = 1, ..., L-1$ **do**
⌊ Compute left environment $E_L^{(p)} \leftarrow \text{CONTRACT}(E_L^{(p-1)}, x^{(p)}, A^{(p)}, x^{(p)*})$

Define environment $F_L^{(0)} \leftarrow 1$, $F_R^{(L+1)} \leftarrow 1$
**for** site $p = 1, ..., L-1$ **do**
⌊ Compute left environment $F_L^{(p)} \leftarrow \text{CONTRACT}(F_L^{(p-1)}, b^{(p)}, x^{(p)*})$

**while** error $\varepsilon < \text{CONVTOL}$ and iteration $i < \text{MAXITER}$ **do**
  ▷ *Begin DMRG sweeps (right to left)* ◁
  **for** site $p = L-1, ..., 1$ **do**

    $A_{\text{eff}}^{(p,p+1)} \leftarrow (E_L^{(p-1)}, E_R^{(p+2)}, A^{(p)}, A^{(p+1)})$
    $b_{\text{eff}}^{(p,p+1)} \leftarrow \text{CONTRACT}(F_L^{(p-1)}, F_R^{(p+2)}, b^{(p)}, b^{(p+1)})$
    $T^{(p,p+1)} \leftarrow \text{CONTRACT}(x^{(p)}, x^{(p+1)})$

    $T^{(p,p+1)} \leftarrow \text{CONJUGATE GRADIENT DESCENT}(A_{\text{eff}}^{(p,p+1)}, b_{\text{eff}}^{(p,p+1)}, T^{(p,p+1)})$
    $U_{(\alpha_{p-1}, i_p), \beta}, S_\beta, V_{\beta, (\alpha_{p+1}, i_{p+1})} = \text{SVD}\left(T_{(\alpha, i_{p-1}), (\alpha_{p+1}, i_{p+1})}^{(p,p+1)}\right)$ keeping $D$ largest singular values

      ▷ *Put in right canonical form* ◁
    $x_{\alpha_p, \alpha_{p+1}}^{(p+1)}(i_{p+1}) \leftarrow V_{\beta, \alpha_{p+1}, i_{p+1}}$
    $x_{\alpha_{p-1}, \alpha_p}^{(p)}(i_p) \leftarrow U_{\alpha_{p-1}, i_p, \beta} S_\beta$

      ▷ *Update left environments* ◁
    $E_R^{(p+1)} \leftarrow \text{CONTRACT}(E_R^{(p+2)}, A^{(p+1)}, x^{(p+1)}, x^{(p+1)*})$
    ⌊ $F_R^{(p+1)} \leftarrow \text{CONTRACT}(F_R^{(p+2)}, b^{(p+1)}, x^{(p+1)*})$

  Measure error $\varepsilon \leftarrow \text{ERROR}(A, b, x)$
  $i \leftarrow i + 1$

  **if** $\varepsilon < \text{CONVTOL}$ **then**
  ⌊ break

  ▷ *Begin DMRG sweeps (left to right)* ◁
  **for** site $p = 1, ..., L-1$ **do**

    $A_{\text{eff}}^{(p,p+1)} \leftarrow (E_L^{(p-1)}, E_R^{(p+2)}, A^{(p)}, A^{(p+1)})$
    $b_{\text{eff}}^{(p,p+1)} \leftarrow \text{CONTRACT}(F_L^{(p-1)}, F_R^{(p+2)}, b^{(p)}, b^{(p+1)})$
    $T^{(p,p+1)} \leftarrow \text{CONTRACT}(x^{(p)}, x^{(p+1)})$

    $T^{(i,i+1)} \leftarrow \text{CONJUGATE GRADIENT DESCENT}(A_{\text{eff}}^{(i,i+1)}, b_{\text{eff}}^{(i,i+1)}, T^{(i,i+1)})$
    $U_{(\alpha_{p-1}, i_p), \beta}, S_\beta, V_{\beta, (\alpha_{p+1}, i_{p+1})} = \text{SVD}\left(T_{(\alpha_{p-1}, i_p), (\alpha_{p+1}, i_{p+1})}^{(i,i+1)}\right)$ keeping $D$ largest singular values

      ▷ *Put in left canonical form* ◁
    $x_{\alpha_p, \alpha_{p+1}}^{(p+1)}(i_{p+1}) \leftarrow S_\beta V_{\beta, \alpha_{p+1}, i_{p+1}}$
    $x_{\alpha_{p-1}, \alpha_p}^{(p)}(i_p) \leftarrow U_{\alpha_{p-1}, i_p, \beta}$

      ▷ *Update right environments* ◁
    $E_L^{(p)} \leftarrow \text{CONTRACT}(E_L^{(p-1)}, A^{(p)}, x^{(p)}, x^{(p)*})$
    ⌊ $F_L^{(p)} \leftarrow \text{CONTRACT}(F_R^{(p-1)}, b^{(p)}, x^{(p)*})$

  Measure error $\varepsilon \leftarrow \text{ERROR}(A, b, x)$
⌊ $i \leftarrow i + 1$

where

$$T_\xi^{(p,p+1)} = \sum_{\alpha_p} (X_\xi)_{\alpha_{p-1},\alpha_p}^{(p)} (X_\xi)_{\alpha_p,\alpha_{p+1}}^{(p+1)} , \tag{37}$$

$$(b_\eta)_{\text{eff}}^{(p,p+1)} = \frac{\partial}{\partial T_\eta^{(p,p+1)*}} \langle x_\eta | b_\eta \rangle , \tag{38}$$

$$(A_{\eta,\xi})_{\text{eff}}^{(p,p+1)} = \frac{\partial}{\partial T_\xi^{(p,p+1)}} \frac{\partial}{\partial T_\eta^{(p,p+1)*}} \langle x_\eta | A_{\eta,\xi} | x_\xi \rangle . \tag{39}$$

The modified DMRG algorithm for solving the linear equation in block form (summarized in Alg. 7) is similar to the original DMRG algorithm, though additional bookkeeping is needed. Notice that we decided to simultaneously update the $p^{\text{th}}$ tensor in the QTTs for each component in a single optimization step. The primary procedural differences are in performing CGD of the reduced problem (see Alg. 8) and computing the error,

$$\varepsilon = \sum_\eta \left( \langle b_\eta | b_\eta \rangle - 2 \sum_\xi \langle b_\eta | A_{\eta,\xi} | x_\xi \rangle + \sum_{\xi,\beta} \langle x_\xi | A_{\eta,\xi}^\dagger A_{\eta,\beta} | x_\beta \rangle \right) .$$

Computationally, it is often advantageous to utilize the block form because the bond dimension needed to represent each subvector $|x_\xi\rangle$ is likely smaller than that required for the full vector.

---

**Algorithm 7** DMRG utilizing block matrices

---

**Input:** QTT-Os $\{A_{\eta,\xi}\}$, QTTs $\{x_\xi\}$ (initial guess), QTTs $\{\beta_\xi\}$, number of components $M$
**Output:** QTTs $\{x_\xi\}$

**for all** $\xi, \eta \in \{1, \dots, M\}$ **do**
    BUILD RIGHT ENVIRONMENTS($A_{\eta,\xi}, x_\xi$)
    BUILD RIGHT ENVIRONMENTS($b_\eta, \xi_\eta$)

**while** error $\varepsilon <$ CONVTOL and iteration $i <$ MAXITER **do**

    ▷ *Sweeping left to right* ◁
    **for** site $p = 1, \dots, L-1$ **do**
        **procedure** SOLVE REDUCED($p$)
            Obtain tensor network for $(A_{\eta,\xi})_{\text{eff}}^{(p,p+1)}$ for all $\eta, \xi$
            Compute $(b_\eta)_{\text{eff}}^{(p,p+1)}$ for all $\eta$
            Compute $T_\eta^{(p,p+1)}$ for all $\eta$
            Solve for $\{T_\eta\} \leftarrow$ BLOCKCGD($\{A_{\eta,\xi}\}, \{T_\eta\}, \{b_\eta\}$)       ▷ *(See Alg. 8)*
        **for** component $\eta = 1, \dots, M$ **do**
            Compute $U_{\alpha_{p-1},i_p,\beta} S_\beta V_{\beta,\alpha_{p+1},i_{p+1}}^\dagger \leftarrow$ SVD $\left( \left( T_\eta^{(p,p+1)} \right)_{(\alpha_{p-1},i_p),(\alpha_{p+1},i_{p+1})} \right)$ keeping the $D$ largest singular values
            Update $x_\eta^{(p)} \leftarrow U_{\alpha_{p-1},i_p,\beta}$ and $x_\eta^{(p+1)} \leftarrow V_{\beta,\alpha_{p+1},i_{p+1}} S_\beta$
            Extend left environments with updated $x_\eta^{(p)}$
    Compute error $\varepsilon$. If converged, return $\{x_\xi\}$

    ▷ *Sweeping right to left* ◁
    **for** $p = L-1, \dots, 1$ **do**
        $T^{(i,i+1)} \leftarrow$ SOLVE REDUCED($p$)
        **for** component $\eta = 1, \dots, M$ **do**
            Compute $U_{\alpha_{p-1},i_p,\beta} S_\beta V_{\beta,\alpha_{p+1},i_{p+1}}^\dagger \leftarrow$ SVD $\left( \left( T_\eta^{(p,p+1)} \right)_{(\alpha_{p-1},i_p),(\alpha_{p+1},i_{p+1})} \right)$ keeping the $D$ largest singular values
            Update $x_\eta^{(p)} \leftarrow U_{\alpha_{p-1},i_p,\beta} S_\beta$ and $x_\eta^{(p+1)} \leftarrow V_{\beta,\alpha_{p+1},i_{p+1}}$
            Extend right environments with updated $x_\eta^{(p+1)}$
    Compute error. If converged, return $\{x_\xi\}$

---

---

**Algorithm 8** Conjugate gradient descent with block matrices

---

**Input:** list of tensor networks $A_{\eta,\xi}$, kust if tensors $b_\eta$, list of tensors $x_\xi$ (initial guess)
**Output:** tensor $x$
**procedure** BLOCKCGD($A_{\eta,\xi}$, $b_\eta$, $x_\xi$)
    $\beta_\eta \leftarrow \sum_\xi A_{\eta,\xi} x_\xi$
    $r_\eta \leftarrow b_\eta - \beta_\eta$
    $d_\eta \leftarrow r_\eta$
    error $\varepsilon \leftarrow \sqrt{\sum_\eta |r_\eta|^2}$
    $i \leftarrow 1$
    **while** $i <$ MAXITER and $\varepsilon/|b| >$ CONVTOL **do**
        tensor $q_\eta \leftarrow \sum_\xi$ CONTRACT($A_{\eta,\xi}, d_\xi$)
        scalar $p \leftarrow \sum_{\eta,\xi}$ CONTRACT($A_{\eta,\xi}, d_\eta, d_\xi^\dagger$)
        scalar $\alpha \leftarrow \varepsilon^2/p$
        tensor $x_\eta \leftarrow x_\eta + \alpha d_\eta$
        tensor $r_\eta \leftarrow r_\eta - \alpha q_\eta$
        scalar $\varepsilon_{\text{old}} \leftarrow \varepsilon$
        scalar error $\varepsilon \leftarrow \sqrt{\sum_\eta |r_\eta|^2}$         ▷ *new error is the norm of the new residual*
        scalar $\beta \leftarrow \varepsilon^2/\varepsilon_{\text{old}}^2$
        tensor $d_\eta \leftarrow r_\eta + \beta d_\eta$
        int $i \leftarrow i + 1$
    **return** $x$

---

## C. Extension to Comb Geometry

In this work, we do not extend this DMRG solver to the comb geometry, since the problems that must be solved are at most 2-D problems. In this case, the comb geometry is essentially a tensor train.

## VI. ADDITIONAL RESULTS

### A. Orszag-Tang vortex

#### 1. Results for calculations with QTC geometry

The electric and magnetic (EM) fields for simulations of the Orszag-Tang vortex performed with bond dimension $D = 64$ and time step $\Delta t = 0.001\Omega_{c,p}^{-1}$ are shown in Fig. 8. As mentioned in the main text, the electric fields appear to be the primary source of noise.

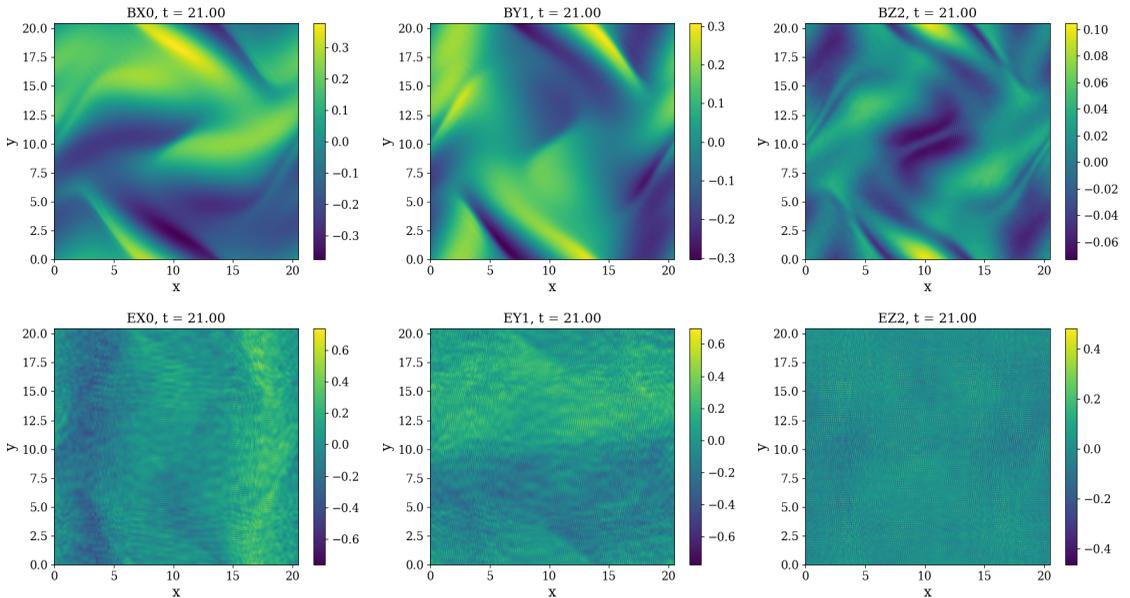

FIG. 8. Each component of the magnetic (top) and electric (bottom) fields at a time of $21\Omega_{c,p}^{-1}$, computed with a time step of $\Delta t = 0.001\Omega_{c,p}^{-1}$ and bond dimension $D = 64$.

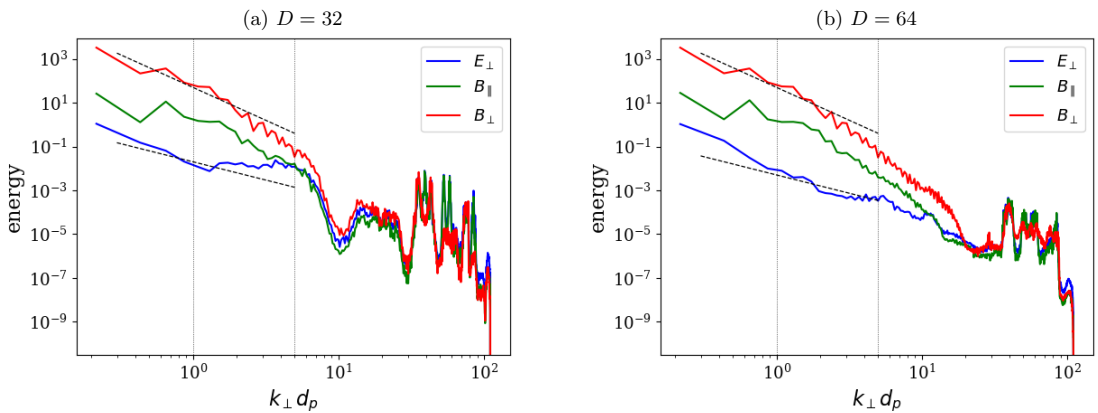

FIG. 9. Energy spectrum at a time of $21\Omega_{c,p}^{-1}$, computed with a time step of $\Delta t = 0.001\Omega_{c,p}^{-1}$. Dashed black lines show power laws of -3 and -5/3.

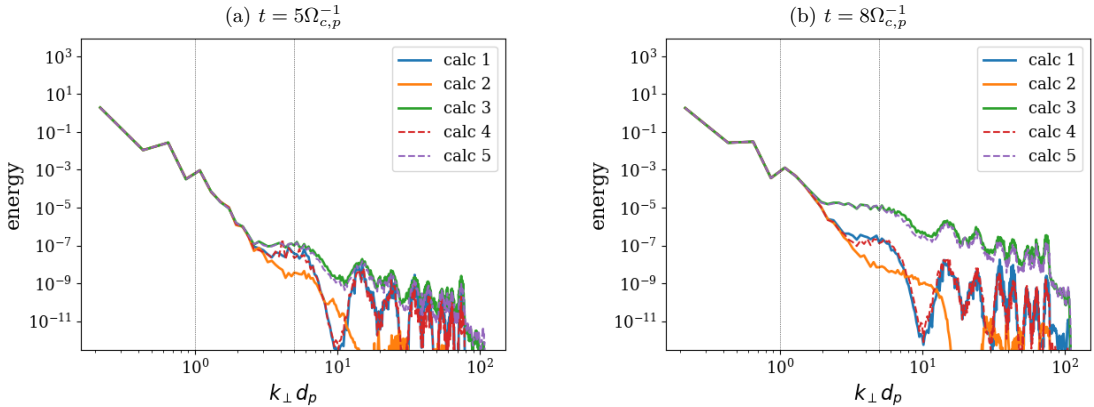

FIG. 10. Energy spectrum of the electric field at times of (a) $5\Omega_{c,p}^{-1}$ and (b) $7\Omega_{c,p}^{-1}$, computed with a time step of $\Delta t = 0.001\Omega_{c,p}^{-1}$. Calculations are as follows; (1) calculations with $D = 32$, (2) calculations with $D = 64$, (3) calculations with $D_f = 32$ for the distribution functions and $D_{EM} = 128$ for the EM fields, (4) calculations with $D = 32$ but the DMRG solver is performed with bond dimension 128, (5) calculations with $D_f = 32$ and $D_{EM} = 128$ but the electric field is compressed to bond dimension 32 before measuring the energy spectrum.

Fig. 9 shows the energy spectra of the EM fields for calculations with $D = 32$ and $D = 64$ at time $t = 21\Omega_{c,p}^{-1}$. Both calculations show similar spectra for $k_\perp d_e < 1$. (Note that given the amount of noise in the electric field, its spectrum is not accurate.) However, the spectra are noticeably different for $k_\perp d_e > 1$. Contrary to expectations, the $D = 64$ result does not show a significant steepening in the magnetic energy spectra. The spectra also have noticeable dips at $k_\perp d_p = 10$ for the $D = 32$ calculation and $k_\perp d_p = 20$ for the $D = 64$ calculation, which exactly correspond to the effective grid resolutions for their respective bond dimensions. This dip seems to limit the resolution that the EM fields can achieve, again explaining the observation mentioned in the main text that the features in $J_z$ were broader for the $D = 32$ calculation than for $D = 64$. This would suggest that the bond dimension and its associated effective grid resolution must be large enough to resolve the desired features in the simulation. This is worrying, since in the event that one would like to consider larger scale separation between the box size and the kinetic scales, one would need to use larger bond dimension such that the effective grid can fully resolve the same fine structures.

However, as argued in the main text, this observation is likely a result of the numerical noise. This noise may be introduced through low-rank approximation as well as the Crank-Nicolson solver, and does not necessarily reflect the efficiency of the QTT for representing the true solution. Fig. 10 shows the energy spectra of $E_z$ obtained with different bond dimensions for the EM fields at times of (a) $t = 5\Omega_{c,p}^{-1}$ and (b) $t = 7\Omega_{c,p}^{-1}$. Solid lines are results for (1) $D = 32$ and (2) $D = 64$ computed using the same bond dimension for the distribution functions, and (3) $D = 128$ but computed using bond dimension 32 for the distribution functions. Complementary to results shown in the main text, the flat spectrum for the third calculation is indicative of numerical noise. Because the Crank-Nicolson solver and the Vlasov equation have no dissipation, any numerical noise that is introduced continues to build. Dashed lines are results for (4) $D = 32$ but the QTT is compressed after solving Crank-Nicolson using DMRG with bond dimension 128 and (5) the same as (3) but the QTT of the $E_z$ field is compressed to $D = 32$ when constructing this plot. Results from (4) closely match results from (1), suggesting that the dip does not result from poor convergence of the DMRG solver but instead results from compression of the QTT to smaller rank. However, because results for (5) match those from (3), it is not solely the compression of the fields that generates the dip in the spectra. It is instead a cumulative effect from performing the low-rank approximation at each time step. While the observations of an "effective grid resolution" certainly warrant further investigation, it is not necessarily the case that finite bond dimension will limit the maximum resolution that can be achieved.

### 2. *Entanglement entropy for different QTN ansätzë*

In Fig. 11, we measure the entanglement entropy at each bond for three different geometries:

- the QTC geometry with flipped binary mapping for each axis (presented in the main paper).

- QTT-SFB: QTT geometry with sequential ordering, with binary mapping for position axes and flipped binary mapping for velocity axes

- QTT-PFM: QTT geometry with interleaved ordering, with forward binary mapping for position axes and mirror mapping for velocity axes

As expected, of the three geometries studied, QTT-SFB on average has the largest EE. Interestingly, while the distribution function for the QTT-PFM geometry has the lowest EE, the EE in the electric field grows much faster than in the other two cases.

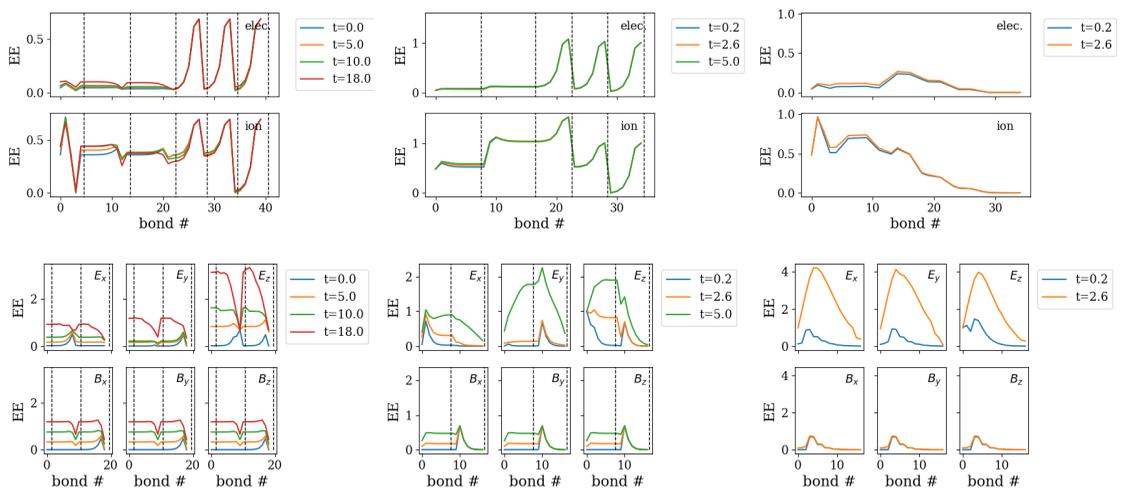

FIG. 11. Entanglement entropy measured at each bond in the QTN for the electron and ion distributions (top) and the electric and magnetic fields (bottom), obtained using (left) the QTC ansatz with $D = 64$, (middle) the QTT-SFB ansatz with $D = 128$, and (right) the QTT-PFM ansatz with $D = 32$. Calculations were performed with a time step of $\Delta t = 0.002$. Time is in units of $\Omega_{c,p}^{-1}$.

### 3. Calculations with uncentered Crank-Nicolson

Single-step time integration schemes can generally be written as

$$\frac{\phi_{n+1} - \phi_n}{\Delta t} = (1-\alpha)F(\phi_n, t_n) + \alpha F(\phi_{n+1}, t_{n+1}),$$ (40)

where $\phi_n$ is the solution and $F(\phi_n, t_n)$ is the time derivative of $\phi$ at time $t_n$. In the case that $\alpha = 1/2$, one obtains the Crank-Nicolson scheme. The time integration scheme is stable for $\alpha \geq 0.5$ [17].

In Fig. 12, we compare the contour plots and spectra of the electric fields obtained using (1) the original Crank-Nicolson scheme and (2) Eq. (40) with $\alpha = 0.6$ to perform integration of Maxwell's equations. Unlike the Crank-Nicolson scheme, the un-centered scheme introduces numerical dissipation, which visibly helps dampen the noise in the electric field. However, additional work to determine how one can efficiently and accurately reduce noise even further is required.

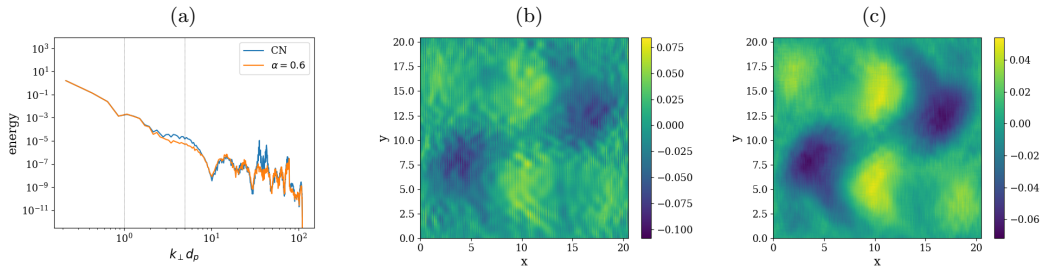

FIG. 12. Plots comparing results using different solvers for updating Maxwell's equations, obtained using bond dimension $D = 32$ and time step $\Delta t = 0.002\Omega_{c,p}^{-1}$. Plots are at a time of $t = 12\Omega_{c,p}^{-1}$. (a) Energy spectrum of the perpendicular electric field. (b) Contour plot of $E_z$ computed using a Crank-Nicolson solver (Eq. (40) with $\alpha = 0.5$). (c) Contour plot of $E_z$ computed using an un-centered solver (Eq. (40)) with $\alpha = 0.6$.

### 4. Observed scalings of cost with respect to QTN bond dimension

The wall-time of the real-space and velocity-space advection steps are measured with respect to QTN bond dimension $D$ for the Orszag-Tang problem. (Since the velocity-space advection step is performed twice per actual time step, the plotted value is those two times averaged together.) The wall-times are measured at various time steps and appear to converge by 1000 time steps. The observed costs are roughly $\mathcal{O}(D^{2.5})$ for computing the advection term in real-space and $\mathcal{O}(D^4)$ for the advection term in velocity-space, which is consistent with theoretical expectations.

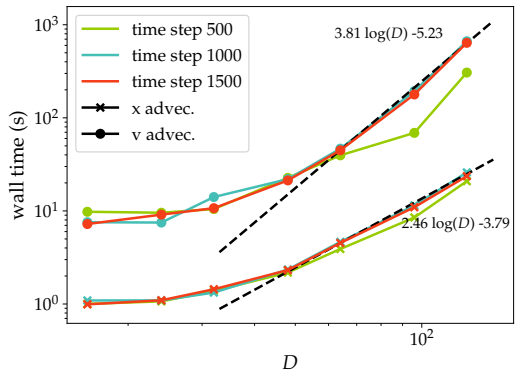

FIG. 13. Plot of wall times for the real-space advection step (denoted using 'x' markers) and velocity-space advection step (denoted using circle markers) computed with QTCs of various bond dimension $D$. Wall times are measured at the specified time steps of the Orszag-Tang problem, and the time step size is $\Delta t = 0.002\Omega_{c,p}^{-1}$. The values of $D$ measured are 16, 24, 32, 48, 64, 96, and 128. Dashed black lines are fitted lines, demonstrating the computational scalings for each calculation.

## B. GEM reconnection problem

Figs. 14 and 15 show the magnetic field, electric field, and distribution function cross sections at times of $25\Omega_{c,p}^{-1}$ and $40\Omega_{c,p}^{-1}$. The simulations were performed with with $D = 64$ and time step $0.005\Omega_{c,p}^{-1}$. At $t = 25\Omega_{c,p}^{-1}$, some numerical noise is already evident in the electric field components.

Fig. 16 shows the spectra of the electric and magnetic (EM) fields at those two times, for calculations performed with $D = 32$ and $D = 64$. At $t = 25\Omega_{c,p}^{-1}$, the spectra for the two bond dimensions look similar, and show the expected steepening at $k_\perp d_e > 1$. For $d_p^{-1} < k_\perp < d_e^{-1}$, $B_\perp$ roughly obeys a $k^{-3}$ power law while $E_\perp$ and $B_\parallel$ appear to fit a $k^{-5/3}$ power law; for $k_\perp d_e > 1$, the decay in magnetic energy roughly follows $k^{-8}$. However, at $t = 40\Omega_{c,p}^{-1}$, the spectra for $D = 32$ appear much flatter, which is due to the fields being dominated by noise.

Following the discussion in the previous section, since the EM fields are represented as QTTs with rank $D = 128$, we might expect to observe a dip in the spectra at around $k_\perp d_p = 40$. Though not as pronounced as in the Orszag-Tang simulations, it does appear as if there might be a dip instead of noise at that length scale.

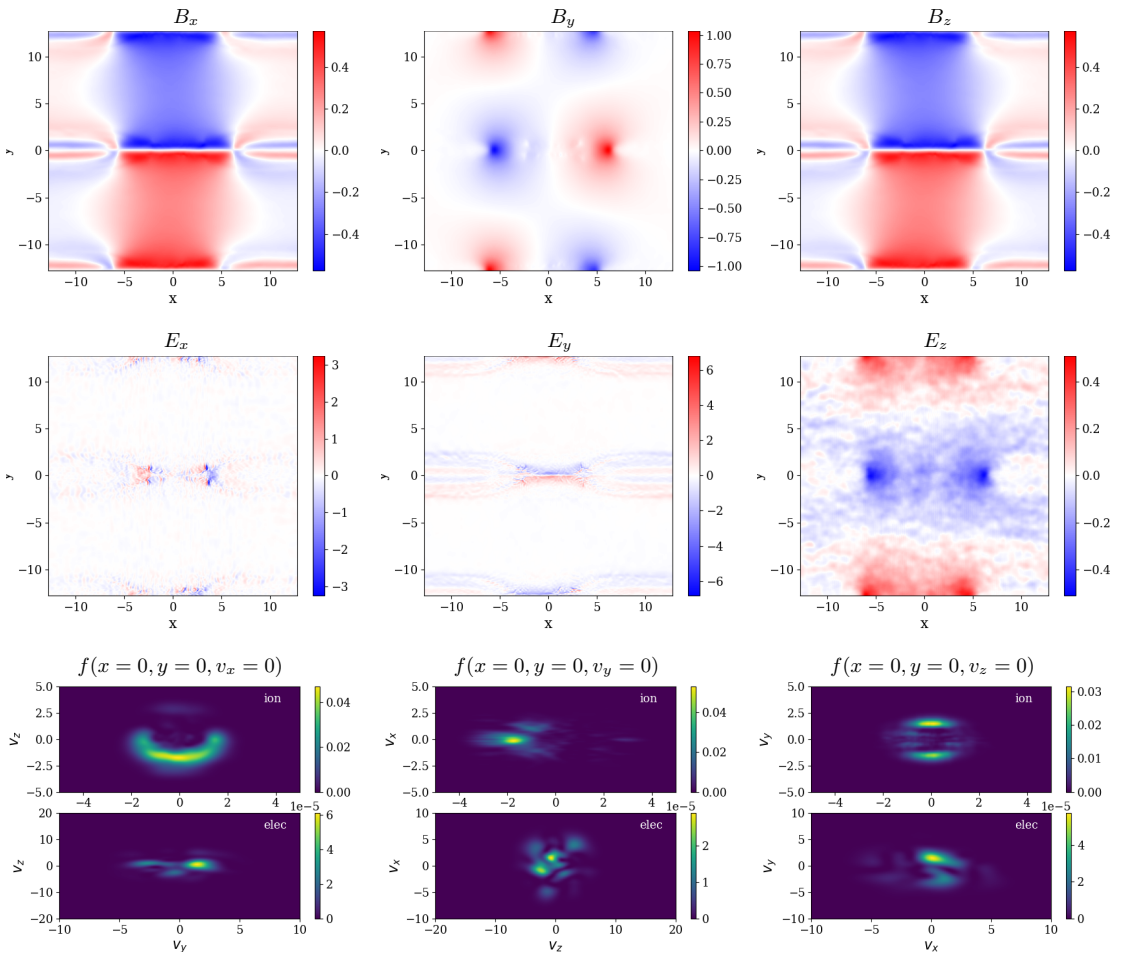

FIG. 14. Each component of the magnetic (top) and electric (center) fields at a time of $25\Omega_{c,p}^{-1}$, computed with a time step of $\Delta t = 0.005$ and bond dimension $D = 64$. At the bottom are plots of the distribution function at the X-point, showing the cross section at zero for the axis not plotted (ordered as $v_x, v_y, v_z$ from left to right).

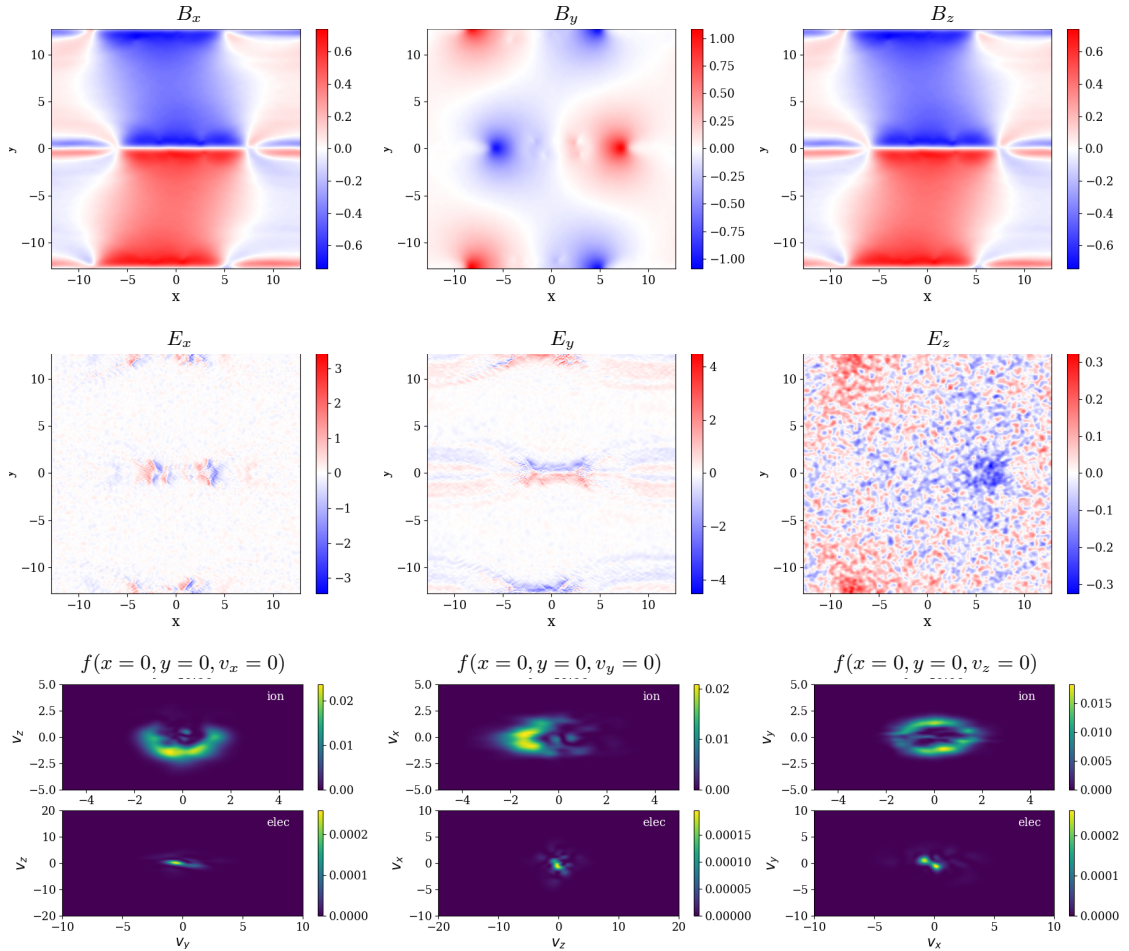

FIG. 15. Each component of the magnetic (top) and electric (middle) fields at a time of $40\Omega_{c,p}^{-1}$, computed with a time step of $\Delta t = 0.005\Omega_{c,p}^{-1}$ and bond dimension $D = 64$. At the bottom are plots of the distribution function at the X-point, showing the cross section at zero for the axis not plotted (ordered as $v_x, v_y, v_z$ from left to right). We observe that the electric fields become dominated by noise.

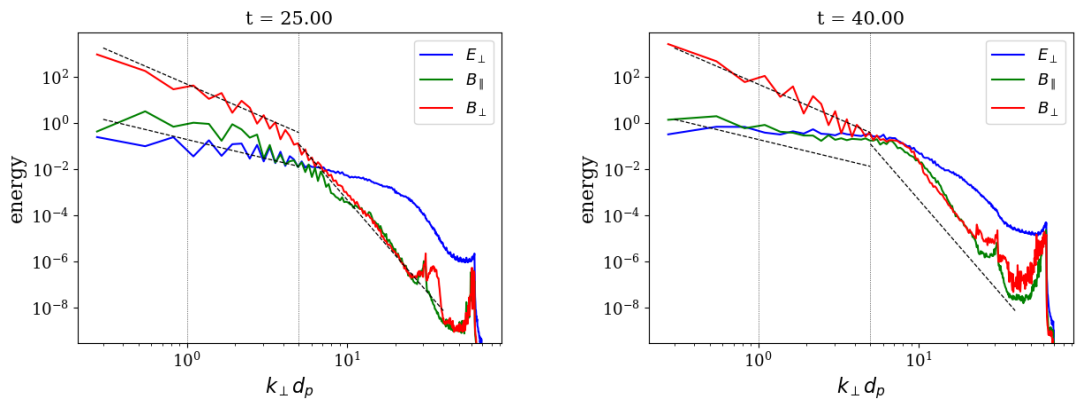

(a) $D = 32$

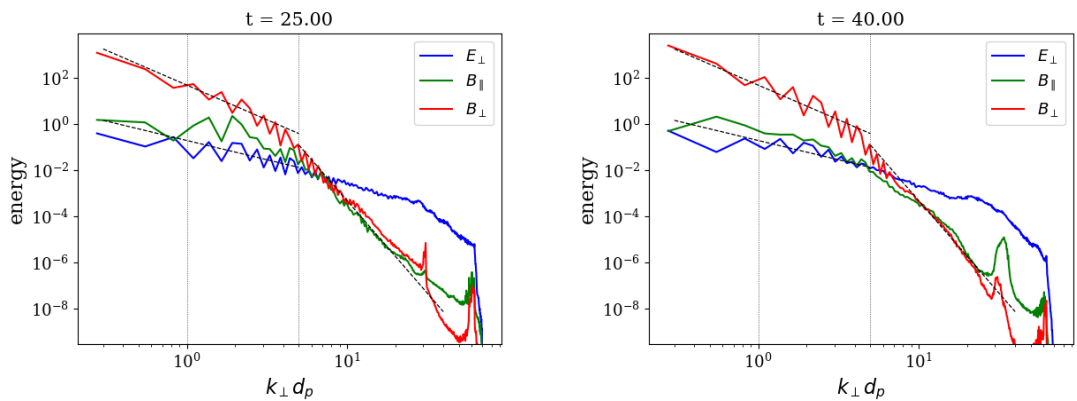

(b) $D = 64$

FIG. 16. Energy spectrum at a time of $25\Omega_{c,p}^{-1}$ and $40\Omega_{c,p}^{-1}$, computed with a time step of $\Delta t = 0.005\Omega_{c,p}^{-1}$. The thin vertical dotted lines depict $k = d_p^{-1}$ and $k = d_e^{-1}$. The sloped dashed black lines show power laws of -3, -5/3, and -8. The $D = 64$ calculations show match expectations reasonably well. In contrast, for the $D = 32$ results, while the spectra at earlier times appear within expectation, at later times the results show much flatter spectra. This is due to the noise in the electric field dominating the dynamics.

---



\end{document}